\documentclass[PhD]{iistdiss}
\usepackage{enumerate}
\usepackage{float}
\usepackage{times}
\usepackage{t1enc}
\usepackage{indentfirst}
\usepackage{graphicx}
\usepackage{epstopdf}
\usepackage[hypertex]{hyperref} 
\usepackage{amsmath} 
\usepackage{amssymb}
\usepackage{physics}
\newcommand{\q}{\quad}

\let\textgrave\`
\newcommand{\bea}{\begin{eqnarray}}
\newcommand{\eea}{\end{eqnarray}}
\newcommand{\be}{\begin{equation}}
\newcommand{\ee}{\end{equation}}
\newcommand{\aver}[1]{\left\langle{#1}\right\rangle}

\newcommand{\modu}[1]{\left|{#1}\right|}
\newcommand{\inner}[2]{\left\langle{#1}|{#2}\right\rangle}
\newcommand{\trev}{T_{\rm rev}}
\begin{document}

\title{AN INVESTIGATION OF NONCLASSICAL PROPERTIES OF LIGHT USING AN OPTICAL TOMOGRAM}

\author{ROHITH M}

\date{FEBRUARY, 2016}
\department{Physics}

\maketitle
\dedication
\begin{center}
{\it Dedicated with extreme affection and gratitude to Amma}
\end{center}
\certificate
{
\vspace*{0.5in}

\noindent This is to certify that the thesis entitled {\bf An Investigation of Nonclassical Properties of Light Using an Optical Tomogram}, submitted by {\bf Mr. Rohith M} to the Indian Institute of Space Science and Technology, Thiruvananthapuram, in partial fulfillment for the award of the degree of {\bf Doctor of Philosophy} is a {\it bona fide} record of research work carried out by him under my supervision.  The contents of this thesis, in full or in parts, have not been submitted to any other Institution or University for the award of any degree or diploma.

\vspace*{1.0in}

\hspace*{-0.25in}
\flushleft{
\parbox{2in}{
\noindent{Dr. Sudheesh Chethil}\\
Supervisor\\
Department of Physics
}
}

\vspace*{1.0in}
\noindent Thiruvananthapuram\hfill Counter signature of HOD with seal\\
February, 2016
}
\declaration
\vspace*{0.5in}

\noindent I declare that this thesis entitled {\bf An Investigation of Nonclassical Properties of Light Using an Optical Tomogram} submitted in fulfillment of the degree of {\bf Doctor of Philosophy} is a record of original work carried out by me under the supervision of {\bf Dr. Sudheesh Chethil}, and has not formed the basis for the award of any other degree or diploma, in this or any other Institution or University. In keeping with the ethical practice in reporting scientific information, due acknowledgements have been made wherever the findings of others have been cited.

\vspace*{1.5in}
\hspace*{4in} 
\parbox{3in}{
\noindent  Rohith M \\
\noindent  (SC11D020) 
}  

\vspace*{0.25in}
\begin{flushleft}
Thiruvananthapuram- 695 547\\
11/02/2016
\end{flushleft}
  
\setlength\parindent{1.5cm}
\acknowledgements
I take this opportunity to express my sincere gratitude to my research supervisor, Dr. Sudheesh Chethil, although a few words cannot do justice to the pivotal role he has had in my research career. I have greatly benefited from his in-depth subject knowledge and programming skills. His advices on both research as well as on my career have been priceless.

I thank my Doctoral Committee members, Dr. Prasanta K. Panigrahi, Dr. S. Lakshmi Bala,  Dr. C. V. Anil kumar, Dr. C. S. Narayanamurthy, and Dr. S. Murugesh, for the effort they have made in reviewing this work, giving valuable suggestions and for the constructive criticisms at different stages of my research. I thank the cooperation and inspiring words of all the teaching and non-teaching staffs and the research scholars of the Department of Physics, during the course of my research. 
I also place on record my sincere thanks to Mr. R. Rajeev for spending his time on the topic of entanglement dynamics of the quantum states. It is a great pleasure to thank the former Chairman of the Board of Management, IIST, Dr. K. Radhakrishnan for allowing me to continue my Ph. D., when I got a job. Special thanks to Sri. Aryadan Muhammed, Minister for Electricity and Transport, Goverment of Kerala, for the support given to me during the course of this work. 
 
I express my overwhelming gratitude to my family for their love and support. Words cannot express how grateful I am to my mother, for all the efforts she have made to make me in this platform. Many friends have helped me stay sane through these difficult years. Their love, support and care helped me overcome setbacks and stay focused on my research. I would like to thank Chithra Sharma for providing some of the reference materials whenever I needed them. Finally, I would like to thank all my teachers, in particular, Dr. V. M. Bannur, Mr. Thomas K. George, Mrs. Leela, Mr. Haridasan, and Mrs. Yamuna, for their inspiring words and unconditional support.

\vspace*{0.25in}
\hfill Rohith M
\abstract
\setlength{\parskip}{0 ex}
\linespread{1}\selectfont
By analyzing the optical tomogram of a linear superposition of coherent states, we show that distinctive signatures of the macroscopic superposition states are displayed directly in the optical tomograms of the states. We also study the effect of decoherence on the optical tomograms of the macroscopic superposition states. We consider the amplitude decay and phase damping models of decoherence  and  show the direct manifestations of decoherence in  the optical tomogram. Since the wave packet fractional revivals are associated with the generation of macroscopic superposition states, these signatures help in visualizing the revivals and fractional revivals occurring in a nonlinear medium directly in the optical tomogram of the time-evolved state. We have investigated the optical tomogram of the time-evolved state obtained by the evolution of an initial wave packet, corresponds to an ideal cohrent state, in a Kerr-like medium. We found that the optical tomogram of the time-evolved state at the instants of fractional revivals  show structures with sinusoidal strands. At $k$-subpacket fractional revival time, the optical tomogram of the time-evolved state shows a structure with $k$ sinusoidal strands. The sinusoidal strands are completely absent when the initial wave packet collapses during the evolution in the medium.  The structures with sinusoidal strands are not lost when the interaction of the system with its external environment is for a short time.

Using a class of initial superposed wave packets evolving in the Kerr-like medium, we further show that the condition for the occurrence of fractional revival phenomenon depends on the number of wave packets composing the initial superposition state. The initial state considered for this purpose is  the superposed coherent states which are the eigenstates of the powers of annihilation operator. Analyzes based on the expectation values of observables, R\'{e}nyi  uncertainty relation and Wigner function are also used to support our findings. For an initial superposed coherent states, the number of sinusoidal strands in the optical tomogram of the time-evolved state at $k$-subpacket fractional revival is $k$ times the number of sinusoidal strands present in the optical tomogram of the initial state.

In the case of a two-mode electromagnetic field, we investigate the entanglement of the state directly using the optical tomogram. We study the optical tomograms of maximally entangled states generated at the output modes of a beam splitter. We take even and odd coherent states in one of the input modes and  vacuum state in the other input mode of the beam splitter. We have shown that the signatures of entanglement can be observed directly in  the single-mode optical tomogram of the state without reconstructing the density matrix of the system.  Two distinct types of optical tomograms are observed  in any one  of the output modes of the beam splitter  based  on the quadrature measurement in the other output mode if the output modes are entangled.  The different features shown by the optical tomograms  are verified by investigating the photon number statistics of the corresponding state. We also analyze the effect of decoherence on the optical tomograms of the entangled states. 

Further, we examine the optical tomograms of the entangled states generated using a beam splitter with a Kerr medium placed into one of its  input modes. The entanglement dynamics of the initial coherent state captures the signatures of revival and fractional revivals. The dynamics of entanglement using von Neumann entropy plot shows local minima at the instants of fractional revivals. The maximum amount of entanglement  is obtained at the instants of collapses of wave packets during the evolution in the medium. The maximum value of entanglement increases with an increase in the field strength. We have found the signatures of entanglement in the optical tomogram for the entangled states generated at the instants of two- and three-subpacket fractional revival times. 
\clearpage

\begin{singlespace}
\tableofcontents
\newpage
\addcontentsline{toc}{chapter}{LIST OF FIGURES}
\listoffigures
\end{singlespace}
\linespread{1.5}\selectfont
\abbreviations

\noindent 
\begin{tabbing}
xxxxxxxxxxx \= xxxxxxxxxxxxxxxxxxxxxxxxxxxxxxxxxxxxxxxxxxxxxxxx \kill
MaxLik \> Maximum-likelihood reconstruction method\\
MaxEnt \> Maximum-entropy reconstruction method\\
FWHM \> Full Width at Half Maximum\\
EEP \> Entropic Entanglement Potential\\
2D \> Two Dimensions
\end{tabbing}

\pagebreak

\chapter*{\centerline{NOTATION}}
\addcontentsline{toc}{chapter}{NOTATION}

\begin{singlespace}
\begin{tabbing}
xxxxxxxxxxxxxxxx \= xxxxxxxxxxxxxxxxxxxxxxxxxxxxxxxxxxxxxxxxxxxxxxxx \kill
$\ket{\alpha}$ \> coherent state of single-mode electromagnetic field\\
$\delta$ \> argument of $\alpha$ ($\alpha=\modu{\alpha}e^{i\delta}$) \\
$\modu{\alpha}^2$ \> mean photon number in the coherent state $\ket{\alpha}$\\
$\alpha$, $\beta$ \> complex numbers labelling coherent states \\
$\hat{X}_\theta$ \> rotated quadrature operator\\
$\theta$ \> phase of the local oscillator in homodyne detection arrangement\\
$\ket{X_\theta,\theta}$ \> eigenvector of $\hat{X}_\theta$\\
$X_\theta$ \> eigenvalue of $\hat{X}_\theta$\\
$\psi_\alpha(X_\theta,\theta)$ \> quadrature representation of the coherent state $\ket{\alpha}$ \\
$\omega\left(X_\theta, \theta\right)$ \> optical tomogram of a single-mode electromagnetic field\\
$\ket{\psi_{l,h}}$ \> state vector for a linear superposition of $l$ coherent states\\
$\rho$ \> density matrix \\
$e_j$, $e_j^\dag$ \> ladder operators for the $j^{\rm th}$ environment mode\\
$\hbar$ \> Plank's constant\\
$H_{\rm amp}$ \> interaction Hamiltonian for the single-mode field \\
\> in amplitude decay model\\
$\gamma$ \> decay rate in amplitude decay model\\
$\gamma\tau$ \> scaled time in amplitude decay model\\
$H_n$ \> Hermite polynomial of order $n$\\
$H_{\rm ph}$ \> interaction Hamiltonian for the single-mode field \\
\> in phase damping  model\\
$\kappa$ \> decay rate in phase damping model\\
$\kappa\tau$ \> scaled time in phase damping model\\
$H$ \> Hamiltonian for a single-mode field propagating in a Kerr medium\\
$a$, $a^\dag$ \> ladder operators for the single-mode electromagnetic field\\
${\rm {\bf N}}$ \> photon number operator $a^\dag a$\\
$\ket{n}$ \> $n$-photon Fock state\\
$\chi$ \> a positive constant in the Hamiltonian $H$\\
$U(t)$ \> unitary time evolution operator\\
$\hat{x}$, $\hat{p}$ \> position and momentum operators\\
$\ket{\psi_k}$ \> state vector a time $t=\pi/k\chi$  \\
$W^{(k)}(\beta)$ \> Wigner function of the state $\ket{\psi_k}$\\
$R_f^{(\zeta)}$ \> R\textgrave{e}nyientropy of order $\zeta$ with probability density $f$\\
$\rho^{(k)}$ \> density matrix of the state $\ket{\psi_k}$\\
$\omega^{(k)}\left(X_\theta, \theta\right)$ \> optical tomogram of the state $\ket{\psi_k}$.\\
$U_{BS}$ \> unitary operator corresponding to beam splitter transformation\\
$\rho_{cd}$ \> density matrix for the state at the output of the beam splitter\\
$\omega\left(X_{\theta_1},\theta _1; X_{\theta_2},\theta _2\right)$ \> optical tomogram of a two-mode state\\
$\omega_i\left(X_{\theta_i},\theta _i\right)$ \> optical tomogram for a subsystem\\
$\rho_k$ \> density matrix for a subsystem\\
$E$ \> entanglement\\
$Q$ \> Mandel's $Q$ parameter\\
$E_N$ \> logarithmic negativity\\
$H_{\rm amp}^{(2)}$ \> interaction Hamiltonian for the two-mode field\\
\> in amplitude decay model\\
$H_{\rm ph}^{(2)}$ \> interaction Hamiltonian for the two-mode field\\
\> in phase damping  model\\

\textbf{Subscripts}\\
rev   \> revival\\
BS   \> beam splitter\\
ss   \> separable state\\
amp  \> amplitude decay model\\
ph   \> phase damping model\\
\end{tabbing}
\end{singlespace}
\pagebreak
\clearpage


\pagenumbering{arabic}
\chapter{INTRODUCTION}\label{Ch_Introduction}
\thispagestyle{plain}

The state of a quantum mechanical system is represented by its state vector $\ket{\psi}$ or the density matrix $\rho$ in the appropriate Hilbert space. Writing the long and often complicated expressions for the wave vectors or the density matrices of infinite dimensional bosonic systems, such as the electromagnetic radiation states, is tiresome, with the added disadvantage of losing sight of the key characteristics that are crucial to extract information. Alternatively, one can use the phase space formulation of the quantum mechanics \citep{Weyl1931,Wigner1932,Groenewold1946,Moyal1949}, in which   the quantum states are represented by its quasiprobability distributions in the phase space. There are mainly three types of quasiprobability distributions: Glauber-Sudarshan $P$ function, Husimi $Q$ function, and the Wigner function \citep{Barnett1997}. These are joint probability distributions of position and momentum variables. The quasiprobability distribution is a real-valued function and is similar to a true probability distribution for the field amplitude, in the sense that, they are normalized and the moments of the products of creation and annihilation operators can be obtained by evaluating an integral weighted by the quasiprobability distribution. However, they are not always positive and thus interpreting  it as a probability distribution is not always possible. 

The quasiprobability distributions are very convenient for understanding various features of the quantum states of electromagnetic fields. For example, if the $P$ function of the field is non-negative everywhere in phase space, it is a classical state of the electromagnetic field. Another important class of radiation field is the nonclassical states, which are quantum states whose characteristic properties cannot be explained by the principles of classical electrodynamics. Such a state can be described in terms of quantized  electromagnetic fields. The nonclassical states of light can be spotted easily from its quasiprobability distributions. A quantum state is said to be nonclassical if its $P$ function is negative somewhere in phase space or more singular than a Dirac delta function. The quantum mechanical phenomena such as, quadrature squeezing, entanglement, oscillations in the photon number distributions, photon antibunching, and fractional revivals, are associated with the nonclassicality of the quantum state. The nonclassical states of light are potential candidates for various applications in quantum information and quantum computation. The experimental characterization of the nonclassical states plays a significant role in understanding the fundamental features associated with the states.

None of these quantities, that is, state vector, density matrix, or quasiprobability distributions, are directly measurable. 
Measurements of a suitable quorum of observables can describe the quantum state completely \citep{Fano1957}. For optical fields, this can be achieved by measuring the rotated quadrature operator of the electromagnetic field \citep{Bertrand1987,Vogel1989}. It has been shown that there exists a one-to-one correspondence between the quasiprobability distribution and the probability distribution of rotated quadrature phases of the field \citep{Bertrand1987,Vogel1989}. The probability distribution of the rotated quadrature operator of the electromagnetic field is called an optical tomogram. The optical tomogram contains all the information about the system, and can serve as an alternative representation of the quantum system. In fact, an alternative formulation of quantum mechanics in which the quantum states are described by tomographic probability distributions  was discussed in \citep{Ibort2009}. In experiments, a series of homodyne measurements of the rotated quadrature operator of the field are done on an ensemble of identically prepared systems and generate the optical tomogram of the state \citep{Leonhardt1997}. The first experimental observation of squeezed state of light, by measuring the quadrature amplitude distribution using the balanced homodyne detection arrangement, has been done in \citep{Smithey1993}.  Thereafter, many nonclassical states of light have been characterized by optical homodyne tomography. A  review of continuous-variable optical quantum state tomography, including a list of the optical quantum states characterized by the same, is given in \citep{Lvovsky2009}.

It is a usual practice in experiments to reconstruct the density matrix or the quasiprobability distributions of the state from the optical tomogram and study its nonclassical properties. The mathematical methods for the reconstruction process can be divided into two categories \citep{Lvovsky2009}. First one is the inverse linear transform methods, which include the inverse Radon transform method and the pattern function method. 
These methods work well only in the limit of a very large number of data and very precise measurements, and are rarely used in experiments. The second one is the methods of statistical inference. This includes maximum-likelihood (MaxLik) reconstruction and maximum-entropy (MaxEnt) reconstruction method.  The reconstructed quasiprobability distributions provide a convenient way to visualize the different nonclassical features of the state in phase space.

It should be emphasized that no reconstruction process  is perfect due to systematic and statistical errors in the estimation of measured statistical distribution. In fact, the original errors of the experimental data can propagate during the process of reconstruction of density matrix or the quasiprobability distribution of the state. Therefore, the reconstruction of the quantum state from an optical tomogram can lead to the loss of information about the actual state prepared in an experiment. An experimental tomogram contains the complete information about the quantum state only in the limit of a very large number of experimental runs  and very precise measurements. An attempt to increase the number of experimental runs can reduce the statistical errors in the measurements, but it increases the systematic errors associated with the experiment. If one can extract the characteristic properties of the quantum state directly from its optical tomogram avoiding the intermediate reconstruction of density matrix or the quasiprobability distribution of the state, more comprehensive will be the information about the state prepared in the experiment and thus highly sophisticated quantum mechanical phenomena can be investigated with high accuracy. It has been shown that 
the physical properties of quantum states can be derived directly using optical tomogram without the intermediate calculation of the density matrix or the quasiprobability distribution of the state,  and  the tomographic approach can be used to estimate the  errors in the histograms of experimentally obtained quadrature values \citep{Bellini2012}. We explore this idea further to study theoretically the nonclassical properties of light directly from the optical tomogram of the state.

Optical tomograms of several nonclassical states of light have been investigated theoretically  in the literature \citep{Filippov2011,Korennoy2011,Miranowicz2014,Manko2012b}. Optical tomogram of a quantum state can be evaluated by a suitable transformation in the symplectic tomogram of the state  \citep{Mancini1995,Ariano1996,Mancini1997}.  The main advantages of studying the optical tomogram of the quantum state are the following. Theoretically calculated  optical tomogram can be used to compare and verify the experimentally measured optical tomogram of the corresponding state. Also, the optical tomogram is convenient for understanding the effect of environment-induced decoherence on the quantum state \citep{Rohith2015}. The correctness of measured tomogram can be checked using the properties of the tomogram, like uncertainty relations \citep{Nicola2006}, tomographic entropic inequalities \citep{Manko2009}, purity constraints \citep{Manko2011}, etc. 

Various kinds of optical macroscopic superposition states have been investigated theoretically \citep{Buzek1995} and have been characterized by continuous-variable optical homodyne tomography \citep{Lvovsky2009}. 
An investigation of the symplectic tomogram of the even and odd coherent states, which are superpositions of two coherent states, have been discussed in \citep{Mancini1996}. Due to quantum interference, the properties of the macroscopic superposition states are different from the properties of the constituent states, as well as from the incoherent superposition or statistical mixture of the constituent states. The nonclassical properties of a macroscopic superposition state, such as quadrature squeezing, entanglement, fractional revival, and oscillations in the photon number distributions, are useful for several technological applications ranging from gravitational wave detectors to quantum computation.
We have investigated the optical tomograms of the macroscopic superposition states of light and  found that distinctive signatures of the superposition states are captured directly in the optical tomogram, enabling the selective identification of the macroscopic superposition states. 

Macroscopic superposition states are sensitive to interactions with the external environment which leads to the decoherence of the state. The effect of decoherence on the quasiprobability distribution of the optical macroscopic superposition states have been investigated in detail \citep{Milburn1986a,Milburn1986b,Daniel1989}. We have found the manifestations of environment-induced decoherence directly in the optical tomogram of the state. The macroscopic superposition states can be generated at the instants of fractional revival, which is a nonclassical phenomenon, during the time evolution of an initial wave packet in a nonlinear medium \citep{Yurke1986,Miranowicz1990,Paprzycka1992,Tara1993}. The fractional revival phenomenon has been investigated both theoretically and experimentally in a wide class of systems \citep{Robinett2004}. 
The superposed wave packets generated by the evolution of an initial coherent state \citep{Glauber1963} in a nonlinear medium have potential applications in quantum cloning with continuous-variables \citep{Cerf2000}.  Two- and four-superposition states generated at fractional revival instances are useful for implementing the one- and two-bit logic gates \citep{Shapiro2003}. Recent experimental observation of multicomponent Schr\"{o}dinger cat states using single-photon Kerr effect opens up new directions for continuous-variable quantum computation \citep{Kirchmair2013}. Identifying the signatures of fractional revivals is an important aspect in the study of macroscopic superposition states.

The time evolution of an initial wave packet $\ket{\psi(0)}$ in a nonlinear medium can exhibit revivals and fractional revivals at specific instants of time. A revival of a well-localized initial wave packet occurs when it evolves in time to a wave packet that reproduces the initial waveform. The characteristic time scale over which this phenomenon happens is called  the revival time $\trev$. At revival time, the autocorrelation function $A(t)=\modu{\inner{\psi(0)}{\psi(t)}}^2$ return to its initial value of unity. Within the characteristic time scale $\trev$, the wave packet may split into a number of scaled copies of the initial state at specific instants during the evolution. This is known as the fractional revival of an initial wave packet \citep{Averbukh1989}.  A $k$-subpacket fractional revival occurs when the initial wave packet splits into a superposition of $k$ wave packets of the initial form. The revivals and fractional revivals have been observed experimentally in a variety of quantum systems such as Rydberg atomic wave packets, molecular vibrational states, Bose-Einstein condensates, and so forth \citep{Rempe1987, Yeazell1990, Yeazell1991, Meacher1991, Vrakking1996, Greiner2002, Matsukevich2006}.

There are various methods in the literature to  identify  and analyze the fractional revivals. The distinctive signatures of the different fractional revivals of a suitably prepared initial wave packet are displayed in the mean values and higher moments of appropriate observables \citep{Sudheesh2004}. The entropy associated with the phase  distribution \citep{Vaccaro1995}, Wehrl's entropy \citep{Jex1994}, and  R\'{e}nyi entropy \citep{Romera2007,Romera2008,Santos2010}  can be used  to study the formation of macroscopic quantum superposition states.  The Wigner function or Husimi $Q$ function plots can also be used to visualize the revivals and  fractional revivals in phase space. The characterization of the time-evolved state in a nonlinear medium can be performed using optical homodyne tomography. The formation of different macroscopic superposition states at the instants of fractional revivals can be traced directly using the optical tomogram of the time-evolved state. We have shown that the signature of revivals and fractional revivals are captured directly in the optical tomogram of the time-evolved state in a nonlinear medium \citep{Rohith2015}.

Another important challenge for the experimental generation of macroscopic superposition state and using it in different technological applications is to handle the environment-induced decoherence of the state arising from the interaction of the system with its external environment \citep{Schlosshauer2005}. The macroscopic superposition states are sensitive to interaction with its environment in an actual experimental setting, and this interaction can even destroy the states generated. The decoherence of the superposition state lead to the leakage of its quantum properties, that is, a transition from quantum to classical world \citep{Zurek1991,Zurek2003}. The external environment can be modelled as a collection of an infinite number of harmonic oscillators. Depending on the type of interaction between the system and the environment, the decoherence of the quantum state can occur at least in two ways: The first one is due to the photon absorption by the environment, also known as amplitude decay, and the second one is due to the phase damping.  Physically, the amplitude damping corresponds to the decay of photons from the system to the environment. The phase damping model corresponds to the scattering losses, in which the number of photons in the system remains unchanged while the environment quanta can be emitted or absorbed. There is no energy transfer between the system and environment, only the phase of the system is changed. These two models are well described by master equations \citep{Gardiner1991}. We have analyzed the effect of decoherence on the optical tomogram of the superposition states. We have shown the manifestation of the environment-induced decoherence directly in the optical tomogram of the macroscopic superposition states \citep{Rohith2015}. The analytical expressions for the optical tomograms of the states in the presence of decoherence (with suitable decay rate), given in this thesis, provide the possibility of direct comparison and verification of the optical tomograms obtained by the homodyne measurements of photonic states.

Most of the theoretical investigations on the revivals and fractional revival phenomena deal with the evolution of an initial single wave packet \citep{Yurke1986,Miranowicz1990,Tara1993,Sudheesh2004,Sudheesh2005a,Sudheesh2005b}. The coherent state \citep{Glauber1963} and photon-added coherent state \citep{Agarwal1991,Zavatta2004} are the examples for such single wave packet. The fractional revival occurs at the same instants of time for both of these initial states \citep{Sudheesh2004,Sudheesh2005b}.  We have shown that the conditions for the occurrence of fractional revivals depend on the number of subpackets composing the initial superposition state \citep{Rohith2014}.

Another important nonclassical feature of the electromagnetic field is the entanglement. Quantum entanglement plays a crucial role in quantum information and quantum computing. It has been a key resource for quantum information processing. After the celebrated EPR paper \citep{Einstein1935}, a tremendous amount of work has been done in the field of quantum entanglement \citep{Horodecki2009}. In most of the quantum information processes, such as quantum teleportation \citep{Bennett1993}, quantum cryptography \citep{Gisin2002}, superdense coding \citep{Bennett1992}, and quantum metrology \citep{Giovannetti2006}, the systems are prepared initially in an entangled state. Much attention is devoted to the discussion of entanglement properties of continuous-variable systems, for their great practical relevance in applications to quantum optics and quantum information \citep{Adesso2007}.

Various devices have been proposed and realized experimentally to generate quantum entanglement. The simplest one is by using a quantum mechanical beam splitter. A beam splitter generates entangled states if one of the input fields are nonclassical \citep{Kim2002}. It has been shown that a standard nonlinear optics interaction, arising from  a Kerr nonlinearity, followed by a simple interaction with a beam splitter can produce a large amount of entanglement in an arbitrarily short time \citep{vanEnk2003,vanEnk2005}. For an initial coherent state with large field strength, such an arrangement generates maximally entangled states in $k$ dimensions at the instants of $k$-subpacket fractional revivals. These states are referred as the {\it multidimensional entangled coherent states} and are  useful in quantum teleportation protocols \citep{vanEnk2003}. We have found that, for an initial coherent state with a finite field strength, the maximum amount of entanglement is obtained at the instants of collapses of the wave packet during the evolution in the Kerr medium \citep{Rohith2016b}. Once the entangled states are created in an experiment, it is important to characterize the state of the system precisely. The optical homodyne tomography can serve as an efficient technique to measure and reconstruct the state of entangled optical fields. Two homodyne detection arrangements, one for each mode, can be used to characterize the two-mode entangled states of light. Various kinds of entangled states of light  have been characterized recently \citep{DAuria2009,Yao2012,Lvovsky2013,Morin2014}. In  experiments, the two-mode density matrix of the system has been reconstructed from the optical tomogram, and the amount of entanglement is calculated using the reconstructed density matrix.

We have investigated theoretically the optical tomograms of the maximally entangled states generated at the output of a beam splitter. A conditional measurement on one of the modes of entangled states may change the state in the other mode due to entanglement, and such changes may show up in the optical tomogram of the state. This property can be explored to find the signatures of entanglement in the optical tomogram of the state, without reconstructing the density matrix of the state.  We have shown that for the entangled two-mode states, the optical tomogram in one of the mode shows different features when upon changing  the parameters  associated with the other mode. These signatures will help in determining whether the state is entangled or not just by looking at the optical tomogram in one of the modes \citep{Rohith2016a}. The different features shown by the state in one of the modes, upon changing the parameters in the other mode, are verified by studying the photon number statistics of the state.  
Since our calculations are based on the optical tomogram in only one of the two modes, it not only avoids the computational complexity of finding the two-mode density matrix or the quasiprobability distribution of the state, but it also reduces the number of homodyne measurements to be performed to determine whether the state is  entangled or not.

The robustness of the entangled state is an important factor for using such states in quantum information protocols. The interaction with its external environment leaves the system in a mixed state, and the decay of entanglement of the state can be quantified using the logarithmic negativity \citep{Vidal2002}. The decoherence of the  multidimensional entangled coherent states using a fictitious beam splitter model is described in \citep{vanEnk2005}. The two-mode density matrix of the entangled state at the output of the beam splitter in the presence decoherence is calculated by solving the master equations in amplitude and phase damping models. We have found the manifestations of decoherence of the entangled state at the output of the beam splitter directly in the optical tomogram \citep{Rohith2016a}.

A summary of the contents of the rest of this thesis is as follows:

{\bf Chapter~\ref{ch_OT_PSI_lh}} describes the manner in which the distinctive signatures of a macroscopic superposition state are displayed in the optical tomogram of the state. This help in selectively identifying a macroscopic superposition state directly from the optical tomogram. The state considered is a linear superposition of coherent states. The effect of decoherence on the optical tomograms of superposed coherent states are investigated using the zero-temperature master equations corresponding to amplitude decay and phase damping models of decoherence.
  
  {\bf Chapter~\ref{Ch_CSevolution}} discusses how signatures of revivals and fractional revivals are captured in the optical tomogram of the time-evolved state. The model Hamiltonian considered is that of   a single-mode field propagating in a Kerr-like medium. The initial state considered is a coherent state. We have found the manifestations of amplitude decay and phase damping models of decoherence on the optical tomogram of the states at the instants of fractional revivals \citep{Rohith2015}.

  {\bf Chapter~\ref{Ch_SCSevolution}} is concerned with the fractional revivals of initial superposed coherent states evolving in the Kerr-like medium. It illustrates the dependence of fractional revival times on the number of subpackets composing the initial superposition state. The dynamics are investigated using the optical tomogram, Wigner function, expectation value analysis, and R\'{e}nyi entropy \citep{Rohith2014}.
    
  {\bf Chapter~\ref{Ch_EntangledOT}} deals with the optical tomograms two-mode continuous-variable entangled states generated using a beam splitter. The signatures of entanglement are captured directly in the single-mode optical tomogram of the state, without reconstructing the density matrix of the system. We also present the analysis of the robustness of entangled states generated at the output of a beam splitter, using the zero-temperature master equations corresponding to amplitude and phase damping models of decoherence \citep{Rohith2016a}.
  
 {\bf Chapter~\ref{Ch_EntanglementDynamics}} describes the optical tomograms of entangled states generated in a beam splitter with the Kerr medium placed into one of its input arms. The entanglement dynamics of the initial coherent state shows an arbitrarily large amount of entanglement at the instants of collapses during the evolution in the medium \citep{Rohith2016b}. The signatures of entanglement in the optical tomogram of the entangled states generated at the instants of two- and three-subpacket fractional revival times are discussed.
     
  {\bf Chapter~\ref{Ch_Conclusion}} concludes the thesis with some brief remarks and a list of open problems for further research.
\chapter{OPTICAL TOMOGRAM OF A MACROSCOPIC SUPERPOSITION STATE}\label{ch_OT_PSI_lh}
\thispagestyle{plain}
\section{Introduction}  
In this chapter, we discuss the manner in which the distinctive signatures of a macroscopic superposition state are displayed in the optical tomogram of the state. Such an investigation can help the identification of the macroscopic superposition states directly from the optical tomogram. The most natural candidate for this purpose would be a linear superposition of coherent states. For completeness, we write down the definition of the coherent state. A coherent state $\ket{\alpha}$  is the eigenstate of the annihilation operator $a$ with eigenvalue $\alpha$,
  \begin{equation}
  a\ket{\alpha}=\alpha\ket{\alpha},
  \end{equation}
where $\alpha$ is a complex number. Let $\alpha=\sqrt{\modu{\alpha}^2}\,\exp(i\,\delta)$, where $\modu{\alpha}^2$ is the mean number of photons in the coherent state $\ket{\alpha}$ and $\delta$ is the argument of $\alpha$. We consider a linear superposition of $l$ coherent states of the form \citep{Peng1990,Napoli1999}
\begin{equation}
\ket{\psi_{l,h}}=N_{l,h}\sum_{r=0}^{l-1} e^{-i 2 \pi r h/l}\ket{\alpha e^{i 2 \pi r /l}}, \label{ch2psi_lh}
\end{equation}
where the normalization constant
\bea
N_{l,h}=\frac{1}{\sqrt{l}} \modu{\sum_{r=0}^{l-1} e^{-i2 \pi r h/l} e^{-\modu{\alpha}^2 (1-e^{i 2 \pi r /l})}}^{-1/2},
\eea
and $h=0,1,...,(l-1)$. If we set $l=1$ and  $h=0$ in the above  equation we retrieve the coherent state $\ket{\alpha}$.  For $l=2$, we get  two states which correspond to $h=0$ and $h=1$ and they are called even and odd coherent states, respectively \citep{Dodonov1974}. The states $\ket{\psi_{l,h}}$ with $h=0$ and $h=1$ are called even and odd coherent states of order $l$, respectively \citep{Napoli1999}. The states $\ket{\psi_{l,h}}$ for a given $l$ with $h=0,1,2,\dots (l-1)$ are orthonormalized eigenstates of the powers of annihilation operator $a^l$ with eigenvalue $\alpha^l$. 
The nonclassical properties of the state $\ket{\psi_{l,h}}$ was discussed in \citep{Buzek1992,Sun1992}. We study the optical tomogram of the superposed coherent state $\ket{\psi_{l,h}}$ and discuss the manifestations of environment-induced decoherence on the optical tomogram of the state. In the next section, we give a brief outline of the calculation of the optical tomogram of a quantum state and its general properties. 
 
\section{Optical tomogram of a quantum state}
\label{representation}
Consider the homodyne quadrature operator 
\begin{equation}
\hat{X}_{\theta}= \frac{1}{\sqrt{2}}\left(a\, e^{-i\theta}+a^\dag  e^{i\theta}\right),\label{ch2QuadratureOperator}
\end{equation}
where $\theta$ is the phase of the local oscillator in the homodyne detection setup \citep{Leonhardt1997} and  $a$ and $a^\dag$ are the photon annihilation and creation operators of the single-mode electromagnetic field, respectively. The phase of the local oscillator varies in the domain $0\leq \theta \leq 2\pi$. 
The optical tomogram $\omega\left(X_{\theta},\theta\right)$ of a quantum state with density matrix $\rho$  can be calculated by the following expression \citep{Vogel1989,Lvovsky2009}:
\be
\omega\left(X_{\theta},\theta\right)=\bra{X_{\theta},\theta}\rho\ket{X_{\theta},\theta},
\label{ch2opt_tomo_def}
\ee
where
\be
\ket{X_{\theta},\theta}=\frac{1}{\pi^{1/4}} \exp\left[-\frac{{X_{\theta}}^2}{2}-\frac{1}{2} e^{i\,2\theta} {a^\dag}^2+\sqrt{2}\, e^{i\,\theta} X_{\theta}\, a^\dag\right]\ket{0}\nonumber
\ee
is the eigenvector of the Hermitian operator $\hat{X}_{\theta}$ with eigenvalue $X_{\theta}$ \citep{Barnett1997}.
 For a pure quantum state with wave vector $\ket{\psi}$, the expression (\ref{ch2opt_tomo_def}) can be rewritten as
\be
\omega(X_{\theta},\theta)=\modu{\bra{X_{\theta},\theta}\psi\rangle}^2.\label{ch2opt_tomo_def_purestate}
\ee
The normalization condition of the optical tomogram $\omega (X_{\theta},\theta)$ is given by
\be
\int dX_\theta\, \omega (X_{\theta},\theta)=1.
\ee
The optical tomogram $\omega (X_{\theta},\theta)$ of a quantum state is non-negative 
\be
\omega (X_{\theta},\theta)\geq 0,
\ee
and has the following symmetry property:  
\be
\omega (X_{\theta},\theta+\pi)=\omega (-X_{\theta},\theta).
\ee
In the subsequent section, we use the Eq.~(\ref{ch2opt_tomo_def_purestate}) to evaluate the optical tomogram of a macroscopic superposition state.

\section{Optical tomogram of superposed coherent states}
 
The optical tomogram of the superposed coherent state $\ket{\psi_{l,h}}$ can be calculated  using the definition given in Eq.~(\ref{ch2opt_tomo_def_purestate}) as
\begin{align}
\omega_{l,h}\left(X_{\theta},\theta\right)=\modu{\bra{X_{\theta},\theta}\ket{\psi_{l,h}}}^2.
\end{align}
Substituting Eq.~(\ref{ch2psi_lh}) in the above equation, we get
\begin{align}
\omega_{l,h}\left(X_{\theta},\theta\right)=N_{l,h}^2\modu{\sum_{r=0}^{l-1} e^{-i 2 \pi r h/l}\bra{X_{\theta},\theta}\ket{\alpha e^{i 2 \pi r /l}}}^2.\label{ch2_opt_psi_lh_def}
\end{align}
The quadrature representation of the coherent state $\ket{\alpha}$, given by \citep{Barnett1997}
\begin{align}
\bra{X_{\theta},\theta}\ket{\alpha}=\frac{1}{\pi^{1/4}}\exp\left[-\frac{X_{\theta}^2}{2}-\frac{\modu{\alpha}^2}{2}-\frac{\alpha^2\,e^{-i\,2\theta}}{2}+\sqrt{2}\,\alpha\, X_{\theta} e^{-i\,\theta}\right], 
\end{align}
can be used to simplify the above expression as
\begin{equation}
\omega_{l,h}\left(X_{\theta},\theta\right)=\frac{N_{l,h}^2}{\sqrt{\pi}}\modu{\sum_{r=0}^{l-1} e^{-i 2 \pi r h/l} \exp\left[-\frac{X_{\theta}^2}{2}-\frac{\modu{\alpha}^2}{2}-\frac{\alpha_r^2\,e^{-i\,2\theta}}{2}+\sqrt{2}\,\alpha_r\, X_{\theta} e^{-i\,\theta}\right]}^2,
\label{ch2_opt_psi_lh}
\end{equation}
where $\alpha_r=\alpha\, e^{i\, 2 \pi r/l}$. Next, we analyze the optical tomogram of the superposed coherent state $\ket{\psi_{l,h}}$ for different $l$ values.

\begin{figure}[h]
\centering
\includegraphics[scale=0.6]{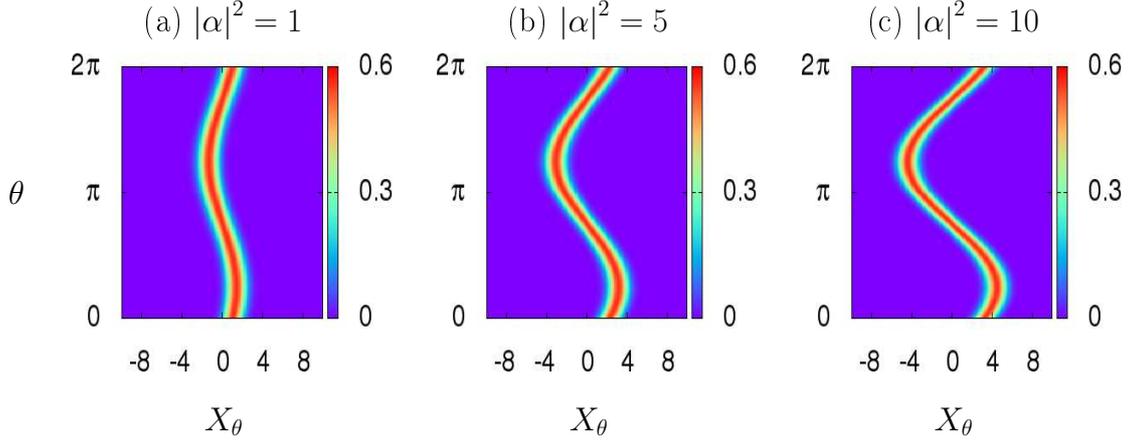}
\caption{Optical tomograms of the coherent state $\ket{\alpha}$ with $\delta=\pi/4$ for (a)  $\modu{\alpha}^2=1$, (b) $\modu{\alpha}^2=5$, and (c) $\modu{\alpha}^2=10$.}
\label{ch2fig:opt_cs}
\end{figure}
The optical tomogram of the coherent state $\ket{\alpha}$ (setting $l=1$ and $h=0$ in Eq.~(\ref{ch2_opt_psi_lh})) is calculated as 
\be
\omega_{1,0}\left(X_{\theta},\theta\right)=\frac{1}{\sqrt{\pi}} \exp\left\{-\left[X_{\theta}-\sqrt{2}\modu{\alpha}\cos(\delta-\theta)\right]^2\right\}.
\label{ch2OT_CS}
\ee
The maximum intensity of the optical tomogram $\omega_{1,0}\left(X_{\theta},\theta\right)$ is $1/\sqrt{\pi}$, which occurs along the sinusoidal path, defined by $X_{\theta}=\sqrt{2\modu{\alpha}^2}\cos(\delta-\theta)$, in the $X_{\theta}$-$\theta$ plane. Hence, the projection of the optical tomogram on the $X_{\theta}$-$\theta$ plane is a structure with a single sinusoidal strand. The optical tomograms of the coherent state $\ket{\alpha}$ with $\delta=\pi/4$  for different field strengths $\modu{\alpha}^2$ are shown in Fig.~(\ref{ch2fig:opt_cs}). 
\begin{figure}[H]
\centering
\includegraphics[scale=0.6]{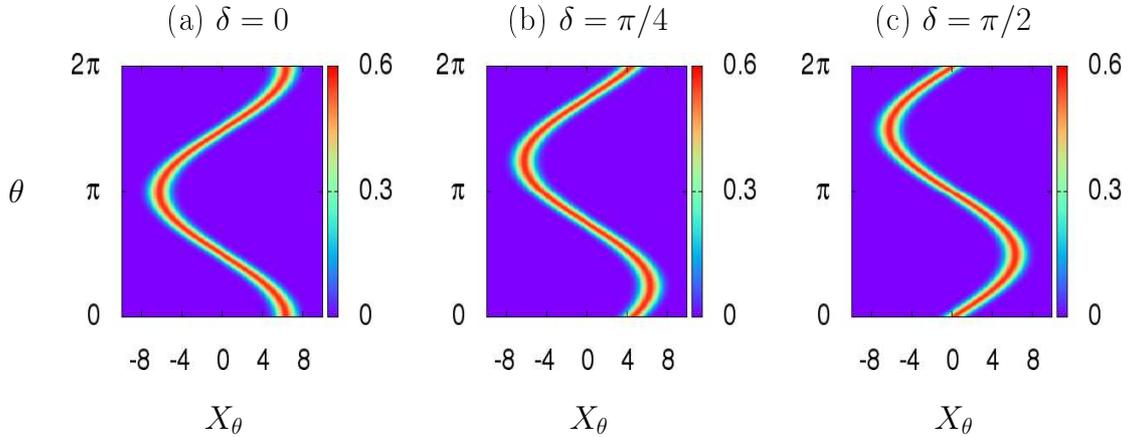}
\caption{Optical tomogram of the coherent state $\ket{\alpha}$ with field strength $\modu{\alpha}^2=20$ and (a) $\delta=0$, (b) $\delta=\pi/4$, and (c) $\delta=\pi/2$.}
\label{ch2fig:opt_csDelta}
\end{figure}
It is clear from the figures that, with an increase in the field strength $\modu{\alpha}^2$, the sinusoidal strand expands in the horizontal direction (along the direction of $X_\theta$ axis). Along the $X_{\theta}$ axis ($\theta=0$), the maximum intensity of the optical tomogram occurs at $X_{\theta}=\sqrt{2\modu{\alpha}^2}\cos\delta$. To show this feature, we have plotted the optical tomogram of a coherent state $\ket{\alpha}$ with field strength $\modu{\alpha}^2=20$ for different $\delta$ values in Fig.~\ref{ch2fig:opt_csDelta}. Along the $X_{\theta}$ axis, the maximum intensity of the optical tomograms shown in Figs.~\ref{ch2fig:opt_csDelta}(a)-\ref{ch2fig:opt_csDelta}(c), occurs at locations $X_\theta=\sqrt{40}$, $X_\theta=\sqrt{20}$, and $X_\theta=0$, respectively. In the rest of this chapter, we set the value of $\delta$ to be $\pi/4$ without loss of generality.

\begin{figure}[h]
\centering
\includegraphics[scale=0.8]{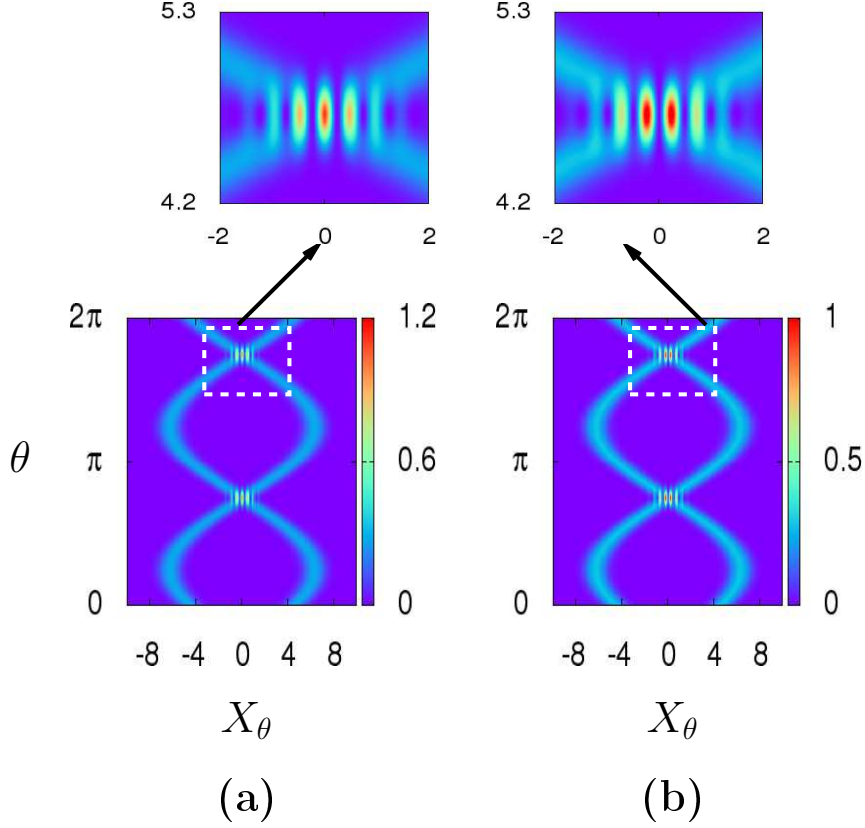}
\caption{Optical tomogram $\omega_{l,h}(X_{\theta},\theta)$ of the (a) even ($l=2$, $h=0$) and (b) odd  ($l=2$, $h=1$) coherent states with $\modu{\alpha}^2=20$.}
\label{ch2fig:opt_l2}
\end{figure}
The optical tomograms of the even and odd coherent states can be obtained by a suitable transformation in the symplectic tomograms of such states  \citep{Mancini1996}. We analyze the  optical tomograms of the even and odd coherent states and investigate the signatures of these  states directly in the optical tomogram. Figure~\ref{ch2fig:opt_l2} shows the optical tomogram of the even and odd coherent states $\ket{\psi_{2,h}}$, which is a superposition of coherent states $\ket{\alpha}$ and $\ket{-\alpha}$. For both even and odd coherent states, the optical tomogram is a structure with two sinusoidal strands in $X_{\theta}$-$\theta$ plane. Along the $X_{\theta}$ axis, the maximum intensities of the optical tomogram occur at $X_{\theta}=\sqrt{40}\cos(\pi/4)$ and $X_{\theta}=\sqrt{40}\cos(5\pi/4)$, corresponding to the sinusoidal strands of $\ket{\alpha}$ and $\ket{-\alpha}$, respectively.  The optical tomogram shows large oscillations in the regions where the two sinusoidal strands intersect because of the quantum interference between the states $\ket{\alpha}$ and $\ket{-\alpha}$. The difference between the optical tomograms of even and odd coherent states (shown in Fig.~\ref{ch2fig:opt_l2}(a) and \ref{ch2fig:opt_l2}(b), respectively)  can be observed from the variation of intensities  (note down the color box given on the right hand side of each figure) as well as from the regions of intersection of the two sinusoidal strands.   For clarity, we have shown the zoomed portions of the interference regions, on the top of each of the optical tomograms, correspond to even and odd coherent states. It is clear that there only one high-intensity spot at the interference region for the even coherent state whereas in the case of odd coherent state there are two high-intensity spots. Therefore, the optical tomograms of even and odd coherent states can be distinguished from one another directly from the optical tomogram of the states.  Here we can conclude that a structure with two sinusoidal strands in the optical tomogram of the superposed coherent states is a signature of the superposition of two coherent states.

\begin{figure}[h]
\centering
\includegraphics[scale=0.7]{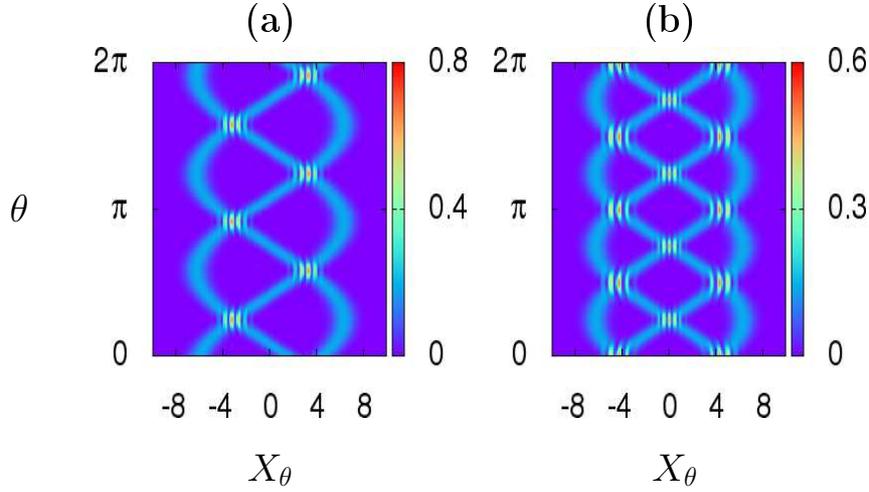}
\caption{Optical tomogram $\omega_{l,h}(X_{\theta},\theta)$ of the superposed coherent states with $\modu{\alpha}^2=20$ and $h=0$ for (a) $l=3$, and (b) $l=4$. }
\label{ch2fig:opt_l34}
\end{figure}
In Fig.~\ref{ch2fig:opt_l34}, we plot the optical tomogram of the state $\ket{\psi_{l,0}}$ for $l=3$  and $l=4$ with $\modu{\alpha}^2=20$, corresponding to a superposition of three and four coherent states, respectively. The optical tomogram of the state $\ket{\psi_{3,0}}$ shown in Fig.~\ref{ch2fig:opt_l34}(a), which is the superposition of three coherent states $\ket{\alpha}$, $\ket{\alpha\,e^{i\,2\pi/3}}$, and $\ket{\alpha\,e^{i\,4\pi/3}}$, displays a structure with three sinusoidal strands. Similarly, the optical tomogram of the state $\ket{\psi_{4,0}}$, which is the superposition of four coherent states $\ket{\alpha}$, $\ket{i\alpha}$, $\ket{-\alpha}$, and $\ket{-i\alpha}$, displays a structure with four sinusoidal strands (see Fig.~\ref{ch2fig:opt_l34}(b)).  We have repeated the above analysis for higher number of superposition ($l>4$) of coherent states and found that the optical tomogram of a superposition state composing $l$ coherent states is a structure with $l$ sinusoidal strands in the $X_{\theta}$-$\theta$ plane. The resolution of these $l$ sinusoidal strands increases with the increase in the field strength $\modu{\alpha}^2$. It is also noticed that, for a given $l$, the structures with $l$ sinusoidal strands in the optical tomograms of the states $\ket{\psi_{l,h}}$ with $h=0,\,1,\,2,\,(l-1)$ differ from one another in the regions of intersections of the sinusoidal strands.

\section{Effect of decoherence on the optical tomogram}
\label{ch2Tomogramdecoherence}
So far we have analyzed the optical tomograms of pure quantum states. However, the macroscopic superposition states are very much sensitive to noise and the interaction with the external environment.  This leads to decoherence of the quantum states prepared in an experiment. In this section, we study the effect of environment-induced decoherence on the optical tomogram of the superposition state given in Eq.~(\ref{ch2psi_lh}). The density matrix of the state $\ket{\psi_{l,h}}$ at time $\tau=0$ is given by  
\begin{equation}
\rho(\tau=0)=\ket{\psi_{l,h}}\bra{\psi_{l,h}}.
\end{equation}
The evolution of this state under decoherence, represented by $\rho(\tau)$, can be calculated using  the appropriate master equations that describe the amplitude decay and phase damping of the state.  The interaction with the external environment leaves the system in a mixed state; that is, the state $\rho(\tau)$ is a mixed state for $\tau > 0$. Next we evaluate the optical tomogram of the state ${\rho}(\tau)$ in the amplitude decay and phase damping models.

\subsection{Amplitude decay model}
In this model the interaction of the single-mode field (mode $a$) with the environment modes $e_j$ under the rotating-wave approximation can be described by the Hamiltonian \citep{Gardiner1991}
\bea
H_{\rm amp}=\sum_{j=0}^{\infty} \hbar \gamma \left(a\, {e_j}^\dag+a^\dag e_j\right),
\eea
where $\gamma$ is the coupling strength of the mode $a$ with the environment. In the  Born-Markov approximation, the density matrix $\rho$ in the interaction picture   obeys the zero-temperature master equation \citep{Walls1985}
\begin{equation}
\frac{d \rho }{d \tau}=\gamma \left(2 a \rho a^\dag-a^\dag a \rho -\rho a^\dag a  \right). \label{ch2master}
\end{equation}
The solution of this equation can be written as \citep{Walls1985}
\begin{align}
\rho(\tau)=&N_{l,h}^2\sum_{r,r^\prime=0}^{l-1} e^{-i\,2\pi h(r-r^\prime)/l}\exp\left[-\modu{\alpha}^2 \left(1-e^{i\,2\pi(r-r^\prime)/l}\right)\left(1-e^{-2\gamma \tau}\right)\right]\nonumber\\
&\times \ket{\alpha_r\,e^{-\gamma \tau}}\bra{\alpha_{r^\prime}\,e^{-\gamma \tau}}.\label{ch2master_solution}
\end{align}
It should be noted that, in the long-time limit (i.e, $\tau\rightarrow\infty$) only the vacuum state will survive under amplitude decoherence, that is,
\be
{\rho}(\tau\rightarrow\infty)=\ket{0}\bra{0}.
\label{ch2rho_tau_infty}
\ee
The optical tomograms of the superposed coherent states $\ket{\psi_{l,h}}$ in the presence of amplitude damping can be obtained by substituting Eq.~(\ref{ch2master_solution}) in Eq.~(\ref{ch2opt_tomo_def}):
\begin{align}
\omega_{l,h}(X_{\theta},\theta,\tau)=&\frac{N_{l,h}^2}{\sqrt{\pi}}\sum_{r,r^\prime=0}^{l-1} e^{-i\,2\pi h(r-r^\prime)/l}\exp\left[-\modu{\alpha}^2 \left(1-e^{i\,2\pi(r-r^\prime)/l}\right)\left(1-e^{-2\gamma \tau}\right)\right]\nonumber\\
&\times \zeta(X_{\theta},\theta,\alpha_r,\tau)\zeta^{\ast}(X_{\theta},\theta,\alpha_{r^\prime},\tau),\label{ch2opt_amplitude}
\end{align}
where 
\begin{align}
\zeta(X_{\theta},\theta,\alpha_r,\tau)=&\exp\left(-\frac{{X_{\theta}}^2}{2}-\frac{\modu{\alpha_r}^2\,e^{-2\gamma \tau}}{2}- \frac{\alpha^2_r \,e^{-2\gamma \tau}\,e^{-i{2\theta}}}{2}+\sqrt{2}\,X_{\theta}\,\alpha_r\,e^{-\gamma \tau}\, e^{-i{\theta}}\right).
\label{ch2opt_amplitude_factor}
\end{align}

\begin{figure}[!htb]
\centering
\includegraphics[scale=0.65]{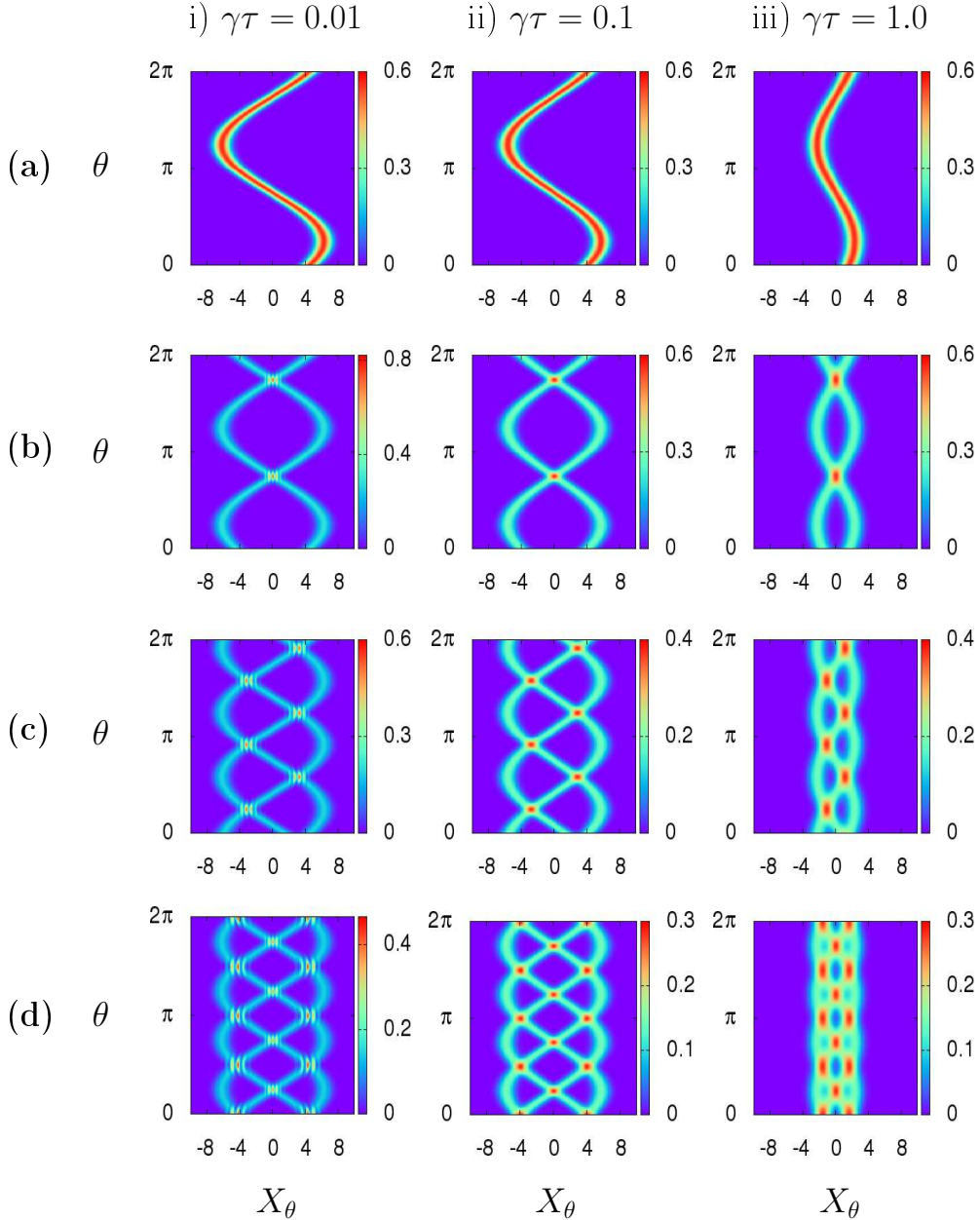}
\caption{Optical tomograms of the coherent state and superposed coherent states $\ket{\psi_{l,h}}$ in the presence of amplitude damping  for (a) $l=1$, (b) $l=2$, (c) $l=3$, and (d) $l=4$ at times (i) $\gamma \tau =0.01$, (ii) $\gamma \tau =0.1$, and (iii) $\gamma \tau =1.0$. For all these plots $h=0$ and $\modu{\alpha}^2=20$.}
\label{ch2fig:AmplitudeDamp}
\end{figure}

Next we use Eq.~(\ref{ch2opt_amplitude}) to plot the optical tomogram of superposed coherent states $\ket{\psi_{l,h}}$ in the presence of amplitude damping. For reference, we have also plotted the effect of amplitude damping on the optical tomogram of a coherent state $\ket{\alpha}$ with field strength $\modu{\alpha}^2=20$ in Fig.~\ref{ch2fig:AmplitudeDamp}(a). The short time interaction of the system with the environment do not destroy the coherence property of a coherent state, but it reduces the amplitude of the state exponentially in time $\tau$. The reduction in the amplitude of the coherent state is clearly observed in Fig.~\ref{ch2fig:AmplitudeDamp}(a), where sinusoidal strand in the optical tomogram shrinks in the horizontal direction (along the $X_{\theta}$ axis). We take the value of coupling constant $\gamma$ to be $0.01$ for all the plots in Fig.~\ref{ch2fig:AmplitudeDamp}.  In Figs.~\ref{ch2fig:AmplitudeDamp}(b)-\ref{ch2fig:AmplitudeDamp}(d)  we plot the optical tomograms of the state $\ket{\psi_{l,h}}$  in the presence of amplitude damping at different  times $\gamma\tau$ (scaled time) for $l=2,\,3$, and $4$. The structures with sinusoidal strands are not lost when  the interaction of the state with the environment is for a short duration of time  (for example, when   $\gamma\tau=0.1$). The sinusoidal strands in the optical tomogram get close together and get distorted  with an increase in   time $\gamma\tau$  and they merge for large $\gamma\tau$. Figures~\ref{ch2fig:AmplitudeDamp}(b)-\ref{ch2fig:AmplitudeDamp}(d) show the merging of the two, three, four sinusoidal strands in the optical tomogram of the states $\ket{\psi_{2,0}}$,  $\ket{\psi_{3,0}}$ and $\ket{\psi_{4,0}}$, respectively. 
The merging of the sinusoidal strands with the increase in  time $\gamma\tau$ is due to the decay of amplitude of the quantum state as a result of the photon absorption by the environment. Another important fact is that  the oscillations in the optical tomogram in the interference regions of the sinusoidal strands decrease with the increase in time $\gamma\tau$, which can be observed in Fig.~\ref{ch2fig:AmplitudeDamp}.

\begin{figure}[h]
\centering
\includegraphics[scale=0.7]{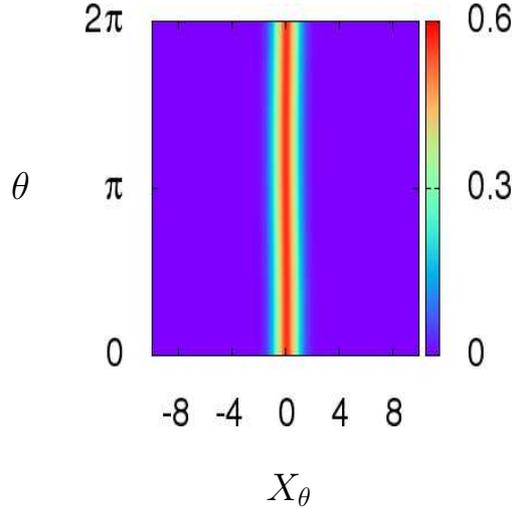}
\caption{Optical tomogram of the superposed coherent state $\ket{\psi_{4,0}}$ at long times $\gamma\tau$ in the presence of amplitude damping.}
\label{ch2fig:longtime}
\end{figure}

 All the superposition states considered above  decay to the vacuum state in the long-time limit, i.e., when $\gamma \tau\rightarrow\infty$,  and  the corresponding optical tomogram is given by  
\begin{align}
\omega_{l,h}\left(X_\theta,\theta,\tau\rightarrow\infty\right)=\frac{1}{\sqrt{\pi}}\,e^{-{X_\theta}^2}.
\label{ch2Opt_vacuum}
\end{align}
The above optical tomogram $\omega\left(X_\theta,\theta,\tau\rightarrow\infty\right)$  is a structure with a single straight strand in the $X_\theta$-$\theta$ plane.  The optical tomogram of the state $\ket{\psi_{4,0}}$ at long times $\gamma\tau$, shown in Fig.~\ref{ch2fig:longtime}, confirms this result. We repeated the analysis described above for the higher superposition $l>4$ of coherent states and found the  similar results.

\subsection{Phase damping model}
In the phase damping model, the interaction between the system (represented by the mode $a$)  and the environment modes $e_j$ can be modeled by the Hamiltonian \citep{Gardiner1991} 
\bea
H_{\rm ph}=\sum_{j=0}^{\infty} \hbar\, \kappa \left(A {e_j}^\dag+A^\dag  e_j\right),
\eea
where $A=a^\dag a$ and $\kappa$ is the coupling constant. In this case, the interaction with the environment causes no loss of energy of the system but the information about the relative phase of the energy eigenstates is lost. The Markovian dynamics of the state  $\rho$  is described by the zero-temperature master equation \citep{Walls1985}
\begin{equation}
\frac{d \rho}{d \tau}=\kappa \left(2 A \rho A^\dag-A^\dag A \rho -\rho A^\dag A  \right). \label{ch2master_phase}
\end{equation}
The solution of this equation can be written in the Fock basis as 
\begin{equation}
\rho(\tau)=\sum_{n,n^\prime=0}^{\infty} {\rho}_{n,n^\prime}(\tau) \ket{n}\bra{n^\prime}, \label{ch2decoh_densitymatrix}
\end{equation}
where the density-matrix elements ${\rho}_{n,n^\prime}(\tau)$ for an arbitrary initial state ${\rho}(\tau=0)$ are given by \citep{Gardiner1991}
\bea
{\rho}_{n,n^\prime}(\tau)=\exp\left[-\kappa\left(n-n^\prime\right)^2\,\tau\right]{\rho}_{n,n^\prime}(\tau=0).
\label{ch2master_ph_solution}
\eea
Note that the diagonal matrix elements  do not decay due to phase damping. Using Eq.~(\ref{ch2master_ph_solution}) we calculate the matrix elements of ${\rho}(\tau)$ in the presence of phase damping for the initial superposed coherent states $\ket{\psi_{l,h}}$ as
\begin{align}
{\rho}_{n,n^\prime}(\tau)=&\frac{l^2\,N_{l,h}^2
e^{-\left(n-n^\prime\right)^2\kappa\tau-\modu{\alpha}^2}\alpha^{n}
{\alpha^\ast}^{n^\prime}}{\sqrt{n!\,n^\prime!}}\delta_{\left[(n-h)/l\right],(n-h)/l}\,\delta_{\left[(n^\prime-h)/l\right],(n^\prime-h)/l},
\label{ch2PhDec_density_elements_ECS}
\end{align}
where $\delta$ is  Kronecker delta function, and $\left[x\right]$ is integer part of $x$. The optical tomogram of the state $\rho(\tau)$, using Eq.~(\ref{ch2opt_tomo_def}),  takes the form
\begin{equation}
\omega_{l,h}\left(X_\theta, \theta,\tau\right)=\sum_{n,n^\prime=0}^{\infty} {\rho}_{n,n^\prime}(\tau) \bra{X_\theta, \theta} n\rangle \langle n^\prime\ket{X_\theta, \theta}. \label{ch2omega_decoh}
\end{equation}
The expression (\ref{ch2omega_decoh}) for the optical tomogram has been simplified to 
\begin{equation}
\omega_{l,h}\left(X_\theta, \theta,\tau\right)=\frac{e^{{-X_\theta}^2}}{\sqrt{\pi}}\sum_{n,n^\prime=0}^{\infty} {\rho}_{n,n^\prime}(\tau) \frac{H_n(X_\theta)\, H_{n^\prime}(X_\theta)}{2^{(n+n^\prime)/2}\,\sqrt{n!\,n^\prime!}} e^{-i\,(n-n^\prime)\theta},
\label{ch2omega_Tau}
\end{equation}
where we have used the quadrature representation $\bra{X_\theta, \theta} n\rangle$ of the Fock state $\ket{n}$,
\begin{equation}
\bra{X_\theta, \theta} n\rangle=\frac{1}{\pi^{1/4}\,2^{n/2}}\frac{e^{{-X_\theta}^2/2}}{\sqrt{n!}} H_n(X_\theta) e^{-i\,n\,\theta}.\nonumber
\end{equation}
In the above equations $H_n(\cdot)$ denotes the Hermite polynomial of order $n$. Substituting Eq.~(\ref{ch2PhDec_density_elements_ECS}) in Eq.~(\ref{ch2omega_Tau}) we get the optical tomogram of the state $\ket{\psi_{l,h}}$ under phase damping as
\begin{align}
\omega_{l,h}\left(X_\theta, \theta,\tau\right)=&\frac{l^2 N_{l,h}^2\,\exp\left[-X_\theta^2-\modu{\alpha}^2\right]}{\sqrt{\pi}}\sum_{n,n^\prime=0}^{\infty}  \frac{\, \alpha^{n} \,{\alpha^\ast}^{n^\prime}\,H_n(X_\theta)\, H_{n^\prime}(X_\theta)\,e^{-i\,(n-n^\prime)\theta}}{n!\,n^\prime!\,2^{(n+n^\prime)/2}} \nonumber\\
&\times \exp\left[-\left(n-n^\prime\right)^2\kappa\tau\right]\,\delta_{\left[(n-h)/l\right],(n-h)/l}\,\delta_{\left[(n^\prime-h)/l\right],(n^\prime-h)/l}.
\label{ch2omega_Tau_final}
\end{align}

\begin{figure}[H]
\centering
\includegraphics[scale=0.55]{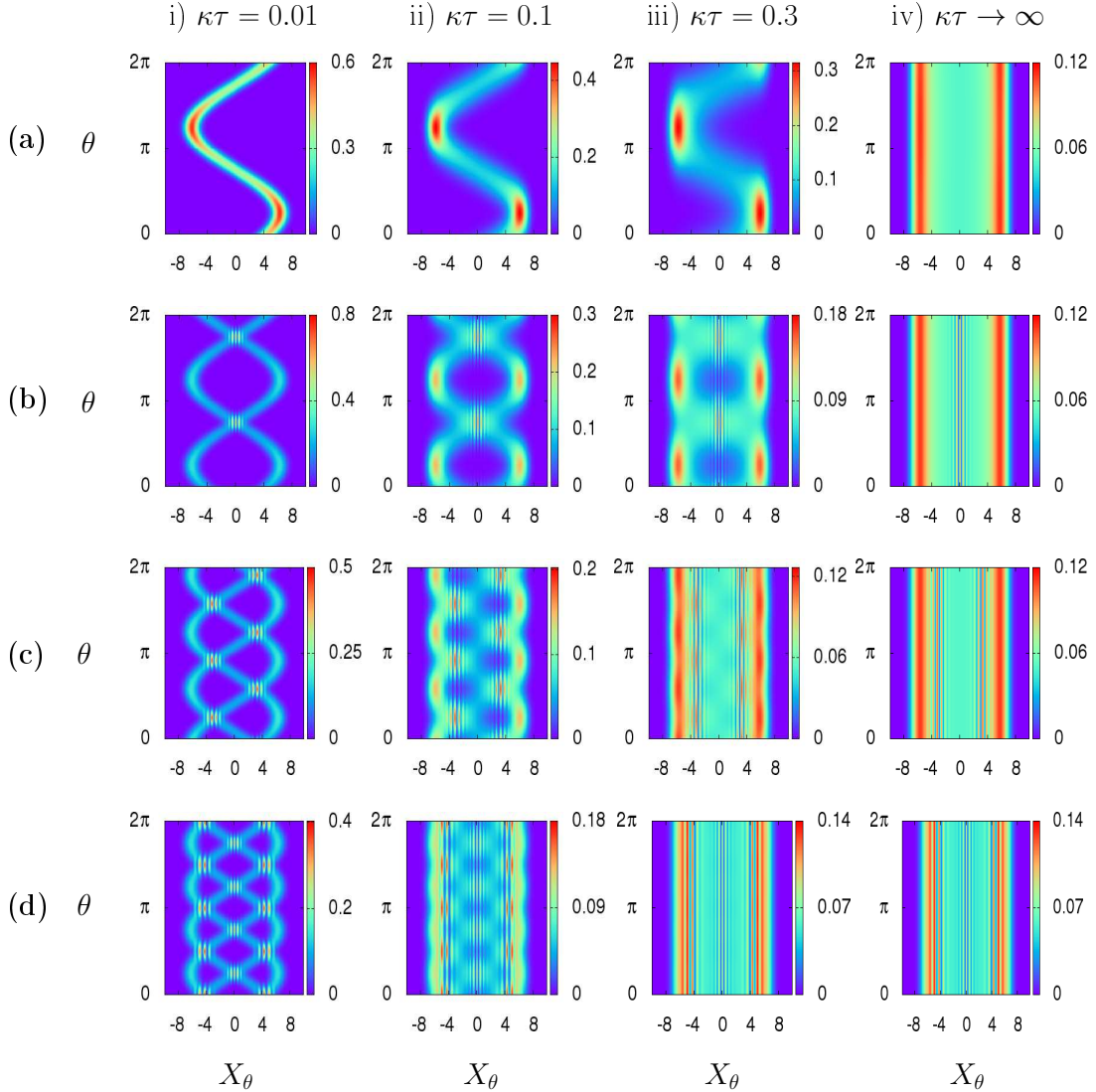}
\caption{Optical tomograms of the coherent state and superposed coherent states $\ket{\psi_{l,h}}$ in the presence of phase damping  for (a) $l=1$, (b) $l=2$, (c) $l=3$, and (d) $l=4$ at times (i) $\kappa \tau =0.01$, (ii) $\kappa \tau =0.1$, (iii) $\kappa \tau =0.3$, and (iv) $\kappa \tau\rightarrow\infty$. For all these plots $h=0$ and $\modu{\alpha}^2=20$.}
\label{ch2fig:PhaseDamp}
\end{figure}
For reference, we have plotted the effect of phase damping on the optical tomogram of a coherent state $\ket{\alpha}$ with field strength $\modu{\alpha}^2=20$ in Fig.~\ref{ch2fig:PhaseDamp}(a). In Figs.~\ref{ch2fig:PhaseDamp}(b)-\ref{ch2fig:PhaseDamp}(d) we show the optical tomogram of the state $\ket{\psi_{l,0}}$  in the presence of phase damping at different  times $\kappa\tau$ (scaled time) for $l=2,\,3$ and $4$, respectively. For all these plots, the value of the coupling constant $\kappa$ is taken to be $0.01$. Here we see that the sinusoidal strands in the optical tomogram of the state  retain their structure  only for a short  time $\kappa\tau$.  

In the long-time limit $\rho_t(\tau\rightarrow\infty)$, the optical tomograms  $\omega\left(X_\theta, \theta,\tau\rightarrow\infty\right)$ of the the initial superposed coherent states $\ket{\psi_{l,h}}$ is given by
 \begin{equation}
\omega_{l,h}\left(X_\theta, \theta,\tau\rightarrow\infty\right)=\frac{l^2\,N_{l,h}^{2}\,e^{{-X_\theta}^2-\modu{\alpha}^2}}{\sqrt{\pi}}\sum_{n=0}^{\infty} \frac{\modu{\alpha}^{2n}\,H_{n}^{2}(X_\theta)}{2^{n}\,\left(n!\right)^2}\,\delta_{\left[(n-h)/l\right],(n-h)/l}.
\end{equation}
The above optical tomogram is independent of the phase $\theta$; this is displayed in  the last column of Fig.~\ref{ch2fig:PhaseDamp}. We have repeated the above analysis for the state $\ket{\psi_{l,h}}$ with $l>4$ and found the similar results.  The optical tomogram of the states, in the long-time scales,  show a completely different structure for amplitude damping and phase damping models of the decoherence. This can be used to understand the type of interaction  the system is having with its environment.

\section{Conclusion}
We have studied the optical tomogram of a linear superposition of coherent states. A closed-form analytical expression for the optical tomogram of the superposed coherent states is derived. We have shown that the signatures of the macroscopic superposition states are captured in the optical tomogram. In general, the optical tomogram of a macroscopic superposition state composing $l$ coherent states shows a structure with $l$ sinusoidal strands. Interactions of the system with  its environment are inevitable in a real experimental setting, and  we found the manifestations of decoherence  directly  in the optical tomogram. Our analytical results may provide an opportunity for the experimentalist to compare and verify the experimentally obtained optical tomogram of superposed coherent states with that of the theoretical one.  Theoretical results on decoherence can be used to find out how much the decoherence models capture the effects of environmental interactions in an actual experimental setting. The results presented in this chapter will be useful for the characterization of the wave-packet fractional revival as it is associated with the generation of the macroscopic superposition states. Wave-packet propagation in a nonlinear medium provides us a good framework for illustrating this aspect.  Therefore, in the next chapter, we extend our investigation to the optical tomogram of an initial wave packet evolving in a nonlinear medium.


\chapter{VISUALIZING REVIVALS AND FRACTIONAL REVIVALS USING AN OPTICAL TOMOGRAM}\label{Ch_CSevolution}
\thispagestyle{plain}
\section{Introduction}
The aim of this chapter is twofold: first, to find the signatures of revivals and fractional revivals directly in the optical tomogram, which in turn can help experimentalists  avoid the errors that can accumulate during the reconstruction process, and second,  to study the effects of amplitude decay and phase damping models of decoherence on the optical tomogram of the states at the instants of fractional revivals.  The revival time and fractional revival times of the wave packet evolution are associated with, respectively, the first- and second-order terms in the Taylor series expansion of the energy spectrum, $E_n$, around the energy $E_{n_0}$ corresponding to the peak of the initial wave packet. Hence, it is sufficient to consider the wave packet evolution governed by an effective Hamiltonian whose energy eigenvalues are at most quadratic functions of $n$. We consider the dynamics of a single-mode field governed by a nonlinear Hamiltonian
\bea
H=\hbar \chi a^{\dag ^2}{a}^2=\hbar\, \chi\, \rm{\bf N}(\rm{\bf N}-1),
\label{ch3kerrhamiltonian}
\eea
where $a$ and $a^\dag$ are the usual photon annihilation and creation operators, respectively, ${\rm{\bf N}}=a^\dag a$, and $\chi$ is  a positive constant.  The eigenstates of the operator $\rm{\bf N}$ are the Fock states ${\ket{n}}$, where $n=0,\,1,\,\dots,\,\infty$. The numerical value of $\chi$ merely sets the time scale. Here, and in the rest of this thesis, $\hbar$ has been set equal to unity. The Hamiltonian given above is physically relevant in the context of the propagation of a single-mode field in a Kerr-like medium \citep{Milburn1986a,Kitagawa1986}. 
For ready reference,  in next section, we review some of the relevant results pertaining to revivals and fractional revivals of an initial coherent state propagating in the Kerr-like medium.

\section{Collapse and revival of an initial coherent state evolving in a Kerr-like medium}
\label{EvolutionCS}
Consider the evolution of an initial coherent state, $\ket{\psi(0)}=\ket{\alpha}$, in a Kerr-like medium governed by the Hamiltonian in Eq.~(\ref{ch3kerrhamiltonian}).  The  coherent state can be represented in the Fock state basis as
\begin{equation}
\ket{\psi(0)}=\ket{\alpha}=e^{-|\alpha|^2/2}\,\sum_{n=0}^{\infty}\,\frac{\alpha^{n}}{\sqrt{n!}}\ket{n}.
\label{ch3FockCS}
\end{equation}
The state of the field at any instant $t$ can be  written as
\begin{equation}
\ket{\psi(t)}=U(t)\ket{\psi(0)},\label{ch3SchrodingerEvolution}
\end{equation}
where the unitary time evolution operator corresponding to the Hamiltonian in Eq.~(\ref{ch3kerrhamiltonian}) is given by
\begin{equation}
U(t)=\exp\left[-i \chi t N(N-1)\right]. \label{ch3TimeAEvolution}
\end{equation}
Substituting Eq.~(\ref{ch3FockCS}) in Eq.~(\ref{ch3SchrodingerEvolution}), we get
\begin{equation}
		\ket{\psi(t)}=e^{-\modu{\alpha}^2/2}\sum_{n=0}^{\infty}\frac{\alpha^n e^{-i\chi t n(n-1)}}{\sqrt{n!}}\ket{n}.\label{ch3psi(t)initialCS}
	\end{equation}
Collapses and revivals of wave packets are observed during the evolution of wave packets in the medium.  It can be shown that the state $\ket{\psi(t)}$ given in Eq.~(\ref{ch3psi(t)initialCS}) revives periodically with revival time $\trev=\pi/\chi$. It also shows fractional revivals when the wave packet is split into a finite number of scaled copies of initial wave packet.    Between time $t=0$ and $t=T_{\rm rev}$, $\ket{\psi(t)}$ shows $k$-subpacket fractional revivals at times
\begin{equation}
t=j \pi /k \chi,
\end{equation}
where  $j=1,2,\dots,(k-1)$ for a given value of   $k(>1)$  with the condition that $j$ and $k$ are mutually prime integers. Here onwards we use the notation $(r, s)=1$ to denote the two mutually prime integers $r$ and $s$.

The time evolution operator $U(t)$ at fractional revival time $t=\pi/k\chi$ (where $k$ is an integer), given by
\begin{equation}
U\left(\pi/k\chi\right)=\exp\left[-\frac{i\pi}{k} {\rm{\bf N}}({\rm{\bf N}}-1)\right],
\end{equation}
possesses interesting periodicity properties. This follows from the fact that the eigenvalues of the operator ${\rm{\bf N}}$ (denoted by $N$) are integers. For odd
integer values of $k$,
\begin{equation}
\exp\left[-\frac{i\pi}{k}(N+k)(N+k-1)\right]=\exp\left[-\frac{i\pi}{k}N(N-1)\right].\label{ch3periodicity1}
\end{equation}
Similarly, for even integer values of $k$,
\begin{equation}
\exp\left[-\frac{i\pi}{k}(N+k)^2\right]=\exp\left[-\frac{i\pi}{k}N^2\right].\label{ch3periodicity2}
\end{equation}
As a result, $U\left(\pi/k\chi\right)$ can be expanded in a Fourier series with $e^{−i\,2\pi s/k}$ as basis functions, i.e,
\begin{align}
\exp\left[-\frac{i\pi}{k}N(N-1)\right]=\sum_{s=0}^{k-1} f_{s} \exp\left(-\frac{i\,2 \pi s}{k} N\right)
\end{align}
for odd integer values of $k$, and
\begin{align}
\exp\left[-\frac{i\pi}{k}N^2\right]=\sum_{s=0}^{k-1} g_{s} \exp\left(-\frac{i\,2 \pi s}{k}N\right)
\end{align}
for even integer values of $k$. The Fourier coefficients $f_s$ and $g_s$ in the above expansion are given by \citep{Tara1993}
\begin{eqnarray}
f_s&=&\frac{1}{\sqrt{k}}\exp\left[\frac{i\,\pi s(s+1)}{k}\right]\exp\left[\frac{-i\,\pi \,(k^2-1)}{4k}\right],\label{ch3Fourier_fs}\\
g_s&=&\frac{1}{\sqrt{k}}\exp\left[\frac{i\,\pi s^2}{k}\right]\exp\left[-\frac{i\,\pi}{4} \right].\label{ch3Fourier_gs}
\end{eqnarray}
Using the above equations and the equation
\begin{equation}
e^{i\chi\,{\rm{\bf N}}}\ket{\alpha}=\ket{\alpha\,e^{i\chi}},
\end{equation} 
we get the state at the instants of $k$-subpacket fractional revival time $t=\pi/k\chi$ as 
\begin{equation}
\ket{\psi(\pi/k \chi)}=\ket{\psi^{(k)}}=\begin{cases}
\sum_{s=0}^{k-1} f_{s}\ket{\alpha \,e^{-i\,2\pi s/k}} & \text{if $k$ is odd}\\
\sum_{s=0}^{k-1} g_{s}\ket{\alpha \,e^{i\,\pi/k}\,e^{-i\,2\pi s/k}} & \text{if $k$ is even.}
\end{cases} \label{ch3superpositionofcs}
\end{equation} 
Thus, at $k$-subpacket fractional revival times a discrete  superposition of $k$ coherent states are generated.  For example, at time $t= T_{\rm rev}/4$
\begin{eqnarray}
\ket{\psi^{(4)}}&=&\frac{1}{\sqrt{8}}\left[(1-i)\,\ket{\alpha\,e^{i\,\pi/4}}\,+\,\sqrt{2}\ket{\alpha\,e^{-i\,\pi/4}}\right.\nonumber\\
&&\left.-\,(1-i)\,\ket{\alpha\,e^{-i\,3\pi/4}}\,+\,\sqrt{2}\ket{\alpha\,e^{i\,3\pi/4}}\right],
\label{ch3super4}
\end{eqnarray}
which is a superposition of four coherent states. 

It has been shown that the distinctive signatures of $k$-subpacket fractional revivals are captured in the $k^{\rm th}$ moment of position and momentum operators, given by 
\begin{equation}  
\hat{x}=\frac{(a+a^{\dagger})}{\sqrt{2}}\quad {\rm and}\quad \hat{p}=\frac{(a-a^{\dagger})}{i\sqrt{2}},
\end{equation}
respectively, 
but not in lower moments \citep{Sudheesh2004}. The  $k$-subpacket fractional revivals are also captured in the higher ($>k$) moments of $\hat{x}$ or $\hat{p}$.
Revival and fractional revival phenomena have been examined as well by information entropic approach. Studies based on the R\'{e}nyi uncertainty relation for the fractional revivals of infinite square well potential and  quantum bouncer have been reported  in \citep{Romera2008, Santos2010}. 

\begin{figure}[H]
\centering
\includegraphics[scale=0.5]{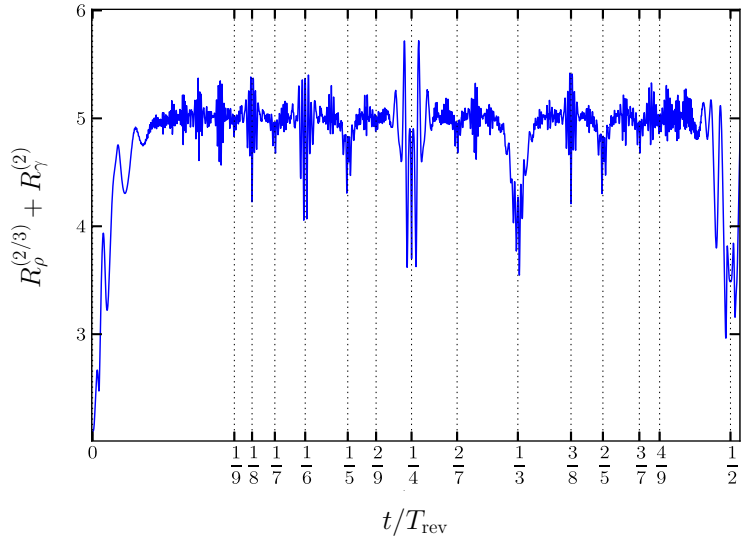} 
\caption{Time evolution of $R_{\rho}^{(2/3)}+R_{\gamma}^{(2)}$ for an initial coherent state with $\modu{\alpha}^2=35$. The main fractional revivals are indicated by vertical dotted lines.}
\label{ch3renyiCS}
\end{figure}

We calculate  the R\'{e}nyi entropy in position and momentum coordinate spaces to find the siganture of revivals and fractional revivals for an initial cohrent state evolving in the Kerr-like medium \citep{Rohith2014}. The R\'{e}nyi entropy is defined in terms of a generalized probability density $f(x)$ as \citep{Bialynicki2006}
\bea
R_{f}^{(\zeta)} \equiv \frac{1}{1-\zeta} \ln \int_{-\infty}^{\infty}[f(x)]^{\zeta} dx \qq \rm for \q 0<\zeta < \infty.\label{ch3RenyiEntropy}
\eea
In terms of the probability density in position and momentum spaces,
$\rho(x)=|\psi(x)|^2$ and $\gamma(p)=|\phi(p)|^2$, respectively, the R\'{e}nyi uncertainty relation is given by
\bea
R_{\rho}^{(\zeta)}+R_{\gamma}^{(\eta)} \geq -\frac{1}{2(1-\zeta)} \ln \frac{\zeta}{\pi}-\frac{1}{2(1-\eta)} \ln\frac{\eta}{\pi},
\label{ch3uncertainty}
\eea
with the condition $1/\zeta+1/\eta=2$. As $\zeta\rightarrow 1$ and $\eta\rightarrow 1$ the R\'{e}nyi uncertainty relation reduces to Shannon's, $S_\rho + S_\gamma \geq 1+\ln (\pi)$. For a Gaussian wave packet (For example, a coherent state), the sum of the R\'{e}nyi entropies $R_{\rho}^{(\zeta)}+R_{\gamma}^{(\eta)}$ reaches its lower bound. The entropy function takes local minima at fractional revival times   and thus the signatures of fractional revivals are given by the local minima of $R_{\rho}^{(\zeta)}(t)+R_{\gamma}^{(\eta)}(t)$ \citep{Romera2007}. We calculate the sum  $R_{\rho}^{(\zeta)}+R_{\gamma}^{(\eta)}$ for the time-evolved state $\ket{\psi(t)}$, given in Eq.~(\ref{ch3psi(t)initialCS}). The coordinate and momentum  space representations of the state $\ket{\psi(t)}$ can be written as
\begin{equation}
\psi(x,t)=e^{-\modu{\alpha}^2/2}\sum_{n=0}^{\infty}\frac{\alpha^n\,H_n(x)\,e^{-x^2/2}}{n! \sqrt{2^n \sqrt{\pi}}}e^{-i\chi t n(n-1)}\label{ch3CoordinateWaveFunCS}
\end{equation}
and 
\begin{equation}
\phi(p,t)=e^{-\modu{\alpha}^2/2}\sum_{n=0}^{\infty}\frac{(-i\,\alpha)^n\,H_n(p)\,e^{-p^2/2}}{n! \sqrt{2^n \sqrt{\pi}}}e^{-i\chi t n(n-1)},\label{ch3MomentumWaveFunCS}
\end{equation}
respectively. In order to evaluate the R\'{e}nyi entropies in conjugate spaces from Eqs.~(\ref{ch3CoordinateWaveFunCS}) and (\ref{ch3MomentumWaveFunCS}), the integral in Eq.~(\ref{ch3RenyiEntropy}) is evaluated numerically by means of trapezoidal rule. We choose $\zeta=2/3$ and $\eta=2$ for the calculation. Figure~\ref{ch3renyiCS} displays the time evolution of $R_{\rho}^{(2/3)}+R_{\gamma}^{(2)}$ for an initial coherent state in a Kerr-like medium.  In this figure we have plotted up to $T_{\rm rev}/2$ because it captures all the important fractional revivals.  The main fractional revivals are denoted by the vertical dotted lines in Fig.~\ref{ch3renyiCS}. 
We have seen that the signatures of revivals and fractional revivals are captured in the quantities, such as the Wigner function, expectation values of the quadrature variables, and R\'{e}nyi entropy. However, none of these quantities are directly measurable, and the estimation of these quantities require a faithful reconstruction of the density matrix of the state from the optical tomogram obtained from homodyne measurements. 

\section{Signatures of revivals and fractional revivals in an optical tomogram}
\label{tomographic}
\begin{figure}[h]
\centering
\includegraphics[scale=0.6]{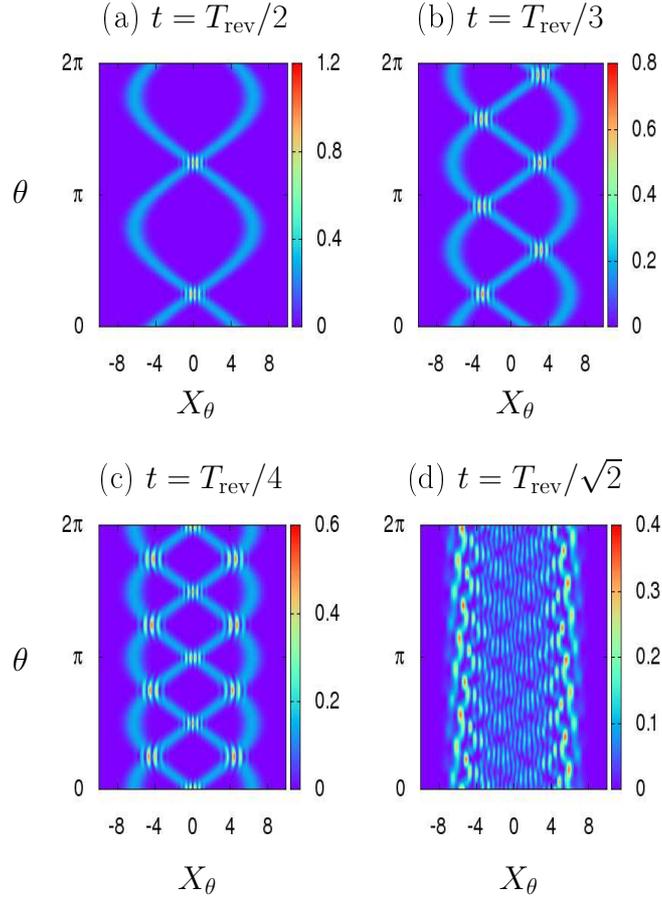} 
\caption{Time-evolved optical tomogram $\omega\left(X_{\theta},\theta,t\right)$ for an initial coherent state $\ket{\alpha}$ with field strength $\modu{\alpha}^2=20$ at times (a) $t=\trev/2$, (b) $t=\trev/3$, (c) $t=\trev/4$, and (d) $t=\trev/\sqrt{2}$. At a $k$-subpacket fractional revival time $t=\pi/k\chi$, the optical tomogram of the state shows structures with $k$ sinusoidal strands. The structures with sinusoidal strands are completely absent in the optical tomogram for the collapsed state at time $t=\trev/\sqrt{2}$.}
\label{ch3fig:optCS}
\end{figure}

In this section, we calculate the optical tomogram of the time-evolved state given in Eq.~(\ref{ch3psi(t)initialCS}) and look for the signatures of revivals and fractional revivals directly in the optical tomogram. We recall from Chapter~\ref{ch_OT_PSI_lh} that the optical tomogram of the coherent state $\ket{\alpha}$, given by Eq.~(\ref{ch2OT_CS}), is a structure with a single sinusoidal strand in the $X_{\theta}$-$\theta$ plane, and the maximum intensity of the optical tomogram along the $X_{\theta}$ axis occurs at $X_{\theta}=\sqrt{2\modu{\alpha}^2}\cos\delta$. Here, the quantity $\delta$ is the argument of the complex number $\alpha$. The optical tomogram of the state at any instant during the evolution of a coherent state $\ket{\alpha}$ is calculated by substituting Eq.~(\ref{ch3psi(t)initialCS}) in Eq.~(\ref{ch2opt_tomo_def_purestate}):
\begin{eqnarray}
\omega\left(X_{\theta},\theta,t\right)=\frac{\exp\left[-\modu{\alpha}^2-X_{\theta}^2\right]}{\sqrt{\pi}}
\modu{\sum_{n=0}^{\infty}\frac{\alpha^n\,e^{-i\,\chi
t n(n-1)}\,e^{-i\,n\theta}
H_n\left(X_{\theta}\right)}{n!\,2^{n/2}}}^2.
\label{ch3optCS}
\end{eqnarray}
In the following we analyze the optical tomogram $\omega\left(X_{\theta},\theta,t\right)$ at the instants of fractional revivals. At a $k$-subpacket fractional revival time $t=\pi/k\chi$, Eq.~(\ref{ch3optCS}) can be simplified to get the optical tomogram of the state $\ket{\psi^{(k)}}$ as
\begin{align}
\omega^{(k)}\left(X_{\theta},\theta\right)
=\frac{1}{\sqrt{\pi}} \modu{\sum_{s=0}^{k-1}\,f_{s,k} \exp{\left[-\frac{X_{\theta}^2}{2}-\frac{\modu{\alpha}^2}{2}-\frac{\alpha_{s}^2\,e^{-i2\theta}}{2}+\sqrt{2}\,\alpha_{s}\,X_{\theta}\,e^{-i\, \theta}\right]}}^2,
\label{ch3optCS_l}
\end{align}
where 
\begin{align}
f_{s,k}&=\begin{cases}
f_s & \text{if $k$ is odd}\\
g_s & \text{if $k$ is even},
\end{cases} \label{ch3FourierCoefficients}\\
\alpha_{s}&=\begin{cases}
\alpha\, e^{-i\,2 \pi s/k} & \text{if $k$ is odd}\\
\alpha\,e^{i\pi/k}\,e^{-i\,2 \pi s/k} & \text{if $k$ is even.}
\label{ch3alpha_s}
\end{cases}
\end{align}
Figures \ref{ch3fig:optCS}(a)-\ref{ch3fig:optCS}(c), show the optical tomograms of the state $\ket{\psi^{(k)}}$ for $k=2,\,3$, and $4$, corresponding to the two-, three-, and four-subpacket fractional revivals of the initial coherent state, respectively. 
 The value of the field strength $\modu{\alpha}^2$ used to plot the tomograms  is $20$.  The state $\ket{\psi^{(2)}}$ is a superposition of the coherent states $\ket{i\alpha}$ and $\ket{-i\alpha}$ with weights $(1-i)/2$ and $(1+i)/2$ (the Fourier expansion coefficients in Eq.~(\ref{ch3superpositionofcs})), respectively. The optical tomogram of this state is a structure with two sinusoidal strands. Thus, a structure with two sinusoidal strands  in the optical tomogram of the time-evolved state for an initial coherent state at $\trev/2$ is  a signature of two-subpacket fractional revival. Note that, the optical tomogram of the states $\ket{\psi^{(2)}}$, shown in Fig.~\ref{ch3fig:optCS}(a), is different from the optical tomogram of the even coherent state, shown in Fig.~\ref{ch2fig:opt_l2}(a). Along the $X_\theta$ axis, the maximum intensity of the optical tomogram of the state $\ket{\psi^{(2)}}$ occur at locations $X_{\theta}=\sqrt{2\modu{\alpha}^2}\,\cos(\delta+\pi/2)$ and $X_{\theta}=\sqrt{2\modu{\alpha}^2}\,\cos(\delta+3\pi/2)$, whereas, for an even coherent state, this occur at locations $X_{\theta}=\sqrt{2\modu{\alpha}^2}\,\cos(\delta)$ and $X_{\theta}=\sqrt{2\modu{\alpha}^2}\,\cos(\delta+\pi)$. In Fig.~\ref{ch3fig:optCS}(a), the maximum intensity of the optical tomogram along the $X_\theta$ axis occur at $X_{\theta}=\sqrt{40}\,\cos(3\pi/4)$ and $X_{\theta}=\sqrt{40}\,\cos(7\pi/4)$, corresponding to the sinusoidal strands of $\ket{i\alpha}$ and $\ket{-i\alpha}$, respectively. The quantum interference regions between the states $\ket{i\alpha}$ and $\ket{-i\alpha}$ are reflected in the optical tomogram of the state $\ket{\psi^{(2)}}$ at locations in the $X_{\theta}$-$\theta$ plane, where the two sinusoidal strands intersect, showing a large oscillation in the optical tomogram. 

The optical tomogram of the state $\ket{\psi^{(3)}}$, which is a state at the three-subpacket fractional revival,  plotted in Fig.~\ref{ch3fig:optCS}(b),  displays a structure with three sinusoidal strands.  Similarly, the optical tomogram of the state  $\ket{\psi^{(4)}}$, which is a state at the four-subpacket fractional revival,  plotted in Fig.~\ref{ch3fig:optCS}(c), shows a structure with four sinusoidal strands. The optical tomorams of the states $\ket{\psi^{(3)}}$ and $\ket{\psi^{(4)}}$ are different from the optical tomogram of the state $\ket{\psi_{3,0}}$ and $\ket{\psi_{4,0}}$ (see Figs.~\ref{ch2fig:opt_l34}(a) and \ref{ch2fig:opt_l34}(b)), respectively. We repeated the analysis for higher-order fractional revivals ($k>4$) and found the general result that the optical tomogram of the time-evolved state at $k$-subpacket fractional revival time shows a structure with $k$ sinusoidal strands.

During the evolution of the coherent state $\ket{\alpha}$, the wave packet may also show the collapse phenomenon at specific instants of time $t=T_{rev}/s$, where $s$ is any  irrational number \citep{Robinett2004}. The collapse phenomenon corresponds to the destruction of a wave packet during its evolution in a nonlinear medium due to the destructive interference of states comprising the wave packet \citep{Robinett2004,Rempe1987,Yeazell1990,Meacher1991,Greiner2002,Kirchmair2013}. The state at the instant of collapse is known as the collapsed state. At the instant of collapse  the state $\ket{\psi(t)}$ is not  a finite superposition of coherent states.  It has been shown that such collapsed states of the fields are of great importance because of their high nonclassical nature and can give a large amount of entanglement when these states are split on a beam splitter with vacuum in the second input port \citep{Rohith2016b}. To study the nature of the optical tomogram during the collapse of the wave packet, we plot the optical tomogram in Eq.~(\ref{ch3optCS}) at collapse time $t=\trev/\sqrt{2}$.  The optical tomogram at this instant is shown in Fig.~\ref{ch3fig:optCS}(d). The sinusoidal strands are not visible in the optical tomogram for the collapsed state, which implies that the optical tomogram of a collapsed state is qualitatively different from that of the state at the instants of fractional revivals. Fig.~\ref{ch3fig:optCS_revival} shows the revival of the initial state  at $t=\trev$. We can conclude that signatures of revivals and fractional revivals are captured in the optical tomogram of the time-evolved states. The optical tomogram at the instants of $k$-subpacket fractional revivals shows $k$  sinusoidal strands for an initial coherent state, which has one strand in its optical tomogram \citep{Rohith2015}.
\begin{figure}[H]
\centering
\includegraphics[height=6.5 cm, width= 5.5 cm]{revival.eps} 
\caption{Optical tomogram $\omega\left(X_{\theta},\theta,t\right)$ for an initial coherent state $\ket{\alpha}$ with field strength $\modu{\alpha}^2=20$ at revival time $t=\trev$.}
\label{ch3fig:optCS_revival}
\end{figure}

\section{Decoherence}
\label{Tomogramdecoherence}
In the previous sections, we have analyzed the optical tomograms of pure quantum states undergoing unitary evolution in the Kerr  medium. However, real optical nonlinearities are noisy and suffer various kinds of losses.  This leads to decoherence of the quantum states generated in the medium. 
In this section, we study the effect of environment-induced decoherence on the optical tomogram of the time-evolved states (the state $\ket{\psi^{(k)}}$ given in Eq.~(\ref{ch3superpositionofcs})) at the instants of $k$-subpacket fractional revival times. The  density matrix of the state $\ket{\psi^{(k)}}$ at time $\tau=0$ is given by  
\begin{align}
\rho^{(k)}(\tau=0)=\ket{\psi^{(k)}}\bra{\psi^{(k)}}.
\end{align}
We use the amplitude decay model and the phase damping models of decoherence described in Section~\ref{ch2Tomogramdecoherence}  to study the decoherence dynamics of the state $\rho^{(k)}$.
\subsection{Amplitude decay model}
\begin{figure}[H]
\centering
\includegraphics[scale=0.7]{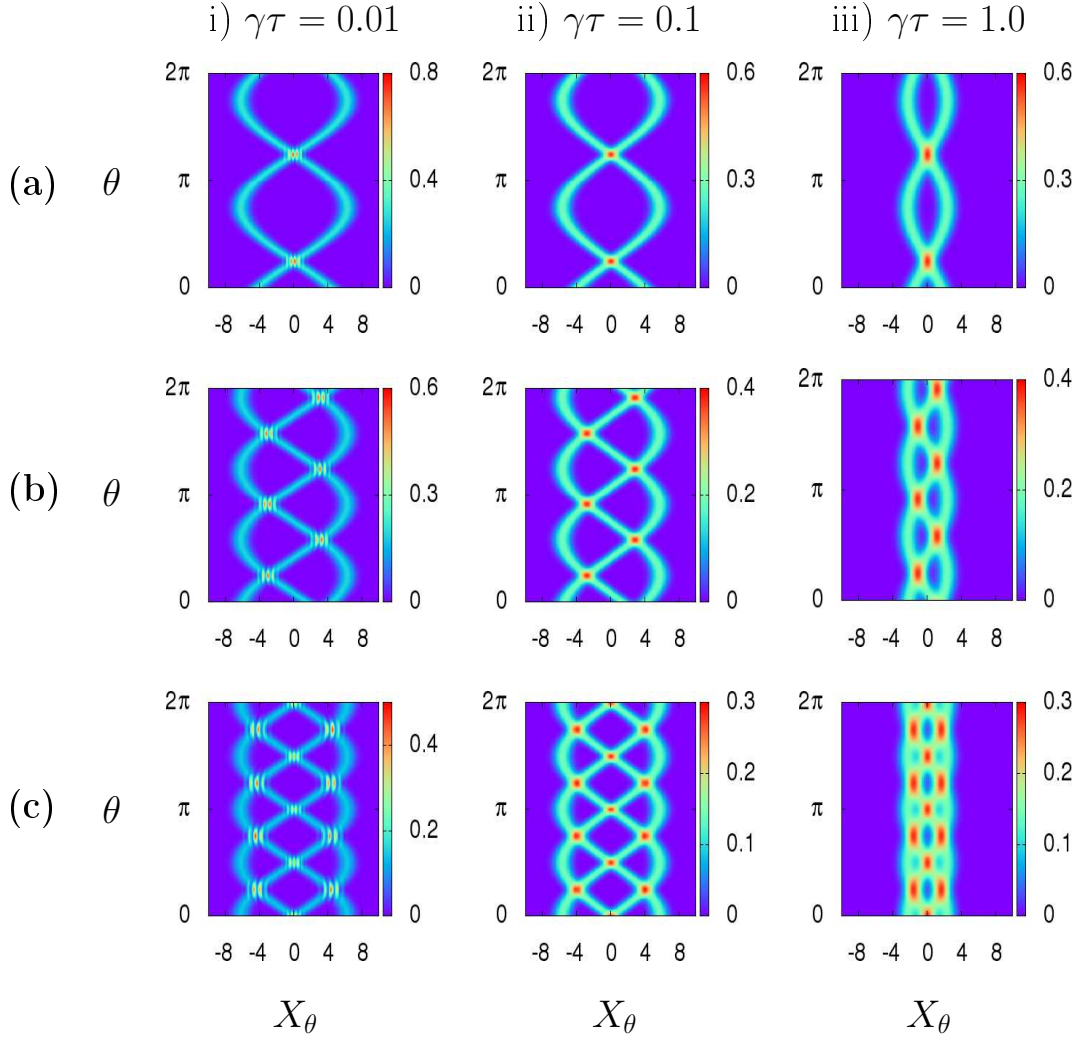}
\caption{Optical tomograms of the time-evolved states at (a) two-, (b) three-, and (c) four-subpacket fractional revival times for an initial coherent state with $\modu{\alpha}^2=20$, in the presence of amplitude damping  at  (i) $\gamma \tau =0.01$, (ii) $\gamma \tau =0.1$, and (iii) $\gamma \tau =1.0$.}
\label{ch3fig:AmplitudeDamp}
\end{figure}

Using Eq.~(\ref{ch2master}), the zero-temperature master equation for the density matrix $\rho^{(k)}$ can written as 
\begin{equation}
\frac{d \rho^{(k)} }{d \tau}=\gamma \left(2 a \rho^{(k)} a^\dag-a^\dag a \rho^{(k)} -\rho^{(k)} a^\dag a  \right), \label{ch3master}
\end{equation}
where $\gamma$ is the rate of decay. The  solution of Eq.~(\ref{ch3master}) is given by 
\begin{align}
\rho^{(k)}(\tau)=\sum_{s,s^\prime=0}^{k-1} f_{s,k}\,f_{s^\prime,k}^{\ast}\exp\left[-\left(\modu{\alpha}^2-\alpha_s\,\alpha_{s^\prime}^\ast\right)\left(1-e^{-2\gamma \tau}\right)\right]\, \ket{\alpha_s\,e^{-\gamma \tau}}\bra{\alpha_{s^\prime}\,e^{-\gamma \tau}}.\label{ch3master_solution}
\end{align}
The optical tomograms of the state $\rho^{(k)}(\tau)$ is calculated as 
\begin{align}
\omega^{(k)}\left(X_{\theta},\theta,\tau\right)=&\frac{1}{\sqrt{\pi}}\sum_{s,s^\prime=0}^{k-1} f_{s,k}\,f_{s^\prime,k}^{\ast}\exp\left[-\left(\modu{\alpha}^2-\alpha_s\,\alpha_{s^\prime}^\ast\right)\left(1-e^{-2\gamma \tau}\right)\right]\nonumber\\
&\times \zeta(X_{\theta},\theta,\alpha_s,\tau)\,\zeta^\ast (X_{\theta},\theta,\alpha_{s^\prime},\tau),
\end{align}
where the quantities $\alpha_s$ and $\zeta$ are defined in Eqs.~(\ref{ch3alpha_s}) and (\ref{ch2opt_amplitude_factor}), respectively.
In Fig.~\ref{ch3fig:AmplitudeDamp}, we plot the optical tomograms of the time-evolved state at the instants of two-, three-, and four-subpacket fractional revival times in the presence of amplitude damping. 

The structures with sinusoidal strands are not lost when  the interaction of the state with the environment is for a short duration of time. The sinusoidal strands get close together and get distorted  with an increase in   time $\gamma\tau$  and they merge for large $\gamma\tau$. Figures~\ref{ch3fig:AmplitudeDamp}(a)-\ref{ch3fig:AmplitudeDamp}(c) show the merging of two, three, and four sinusoidal strands corresponding to the optical tomograms of the states $\ket{\psi^{(2)}}$, $\ket{\psi^{(3)}}$, and $\ket{\psi^{(4)}}$, respectively.  It follows from Eq.~(\ref{ch3master_solution}) that, in the long-time limit, the state $\rho^{(k)}$ reduces to the vacuum state for which the optical tomogram is given by Eq.~(\ref{ch2Opt_vacuum}). This optical tomogram, shown in Fig.~\ref{ch2fig:longtime}, is a structure with single straight strand. We repeated the analysis described above for the states at the instants of higher-order fractional revivals and found  similar results.

\subsection{Phase damping model}
\begin{figure}[H]
\centering
\includegraphics[scale=0.7]{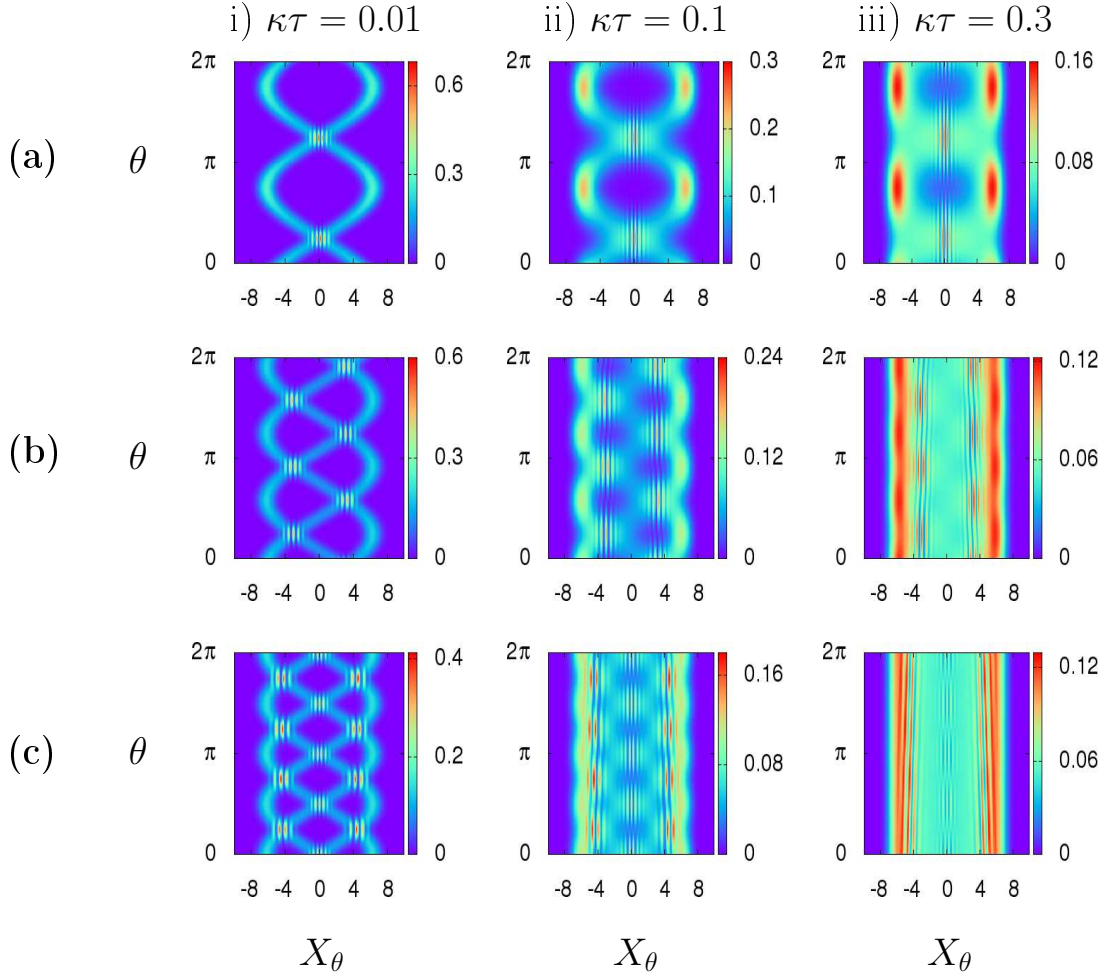}
\caption{Optical tomograms of the states at (a) two-, (b) three-, and (c) four-subpacket fractional revival times in the presence of phase damping. The plots are done for (i) $\kappa \tau =0.01$, (ii) $\kappa \tau =0.1$, and (iii) $\kappa \tau =0.3$ with $\modu{\alpha}^2=20$.}
\label{ch3fig:PhaseDamp}
\end{figure}

Using Eq.~(\ref{ch2master_phase}), we obtain the zero-temperature master equation for the state  $\rho^{(k)}$ in the phase damping model as
\begin{equation}
\frac{d \rho^{(k)}}{d \tau}=\kappa \left(2 A \rho^{(k)} A^\dag-A^\dag A \rho^{(k)} -\rho^{(k)} A^\dag A  \right), \label{ch3master_phase}
\end{equation}
where $\kappa$ is the rate of decoherence. Using Eq.~(\ref{ch2master_ph_solution}), the matrix elements of $\rho^{(k)}(\tau)$ for the initial state $\rho^{(k)}(\tau=0)$ is calculated as 
\begin{align}
\rho^{(k)}_{n,n^\prime}(\tau)=\frac{\exp\left[-\left(n-n^\prime\right)^2\kappa\tau-\modu{\alpha}^2\right]}{\sqrt{n!\,n^\prime!}}\sum_{s,s^\prime=0}^{k-1} f_{s,k}\,f_{s^\prime,k}^{\ast}\,\alpha_s^{n}
{\alpha_{s^\prime}^\ast}^{n^\prime}
\label{ch3PhDec_density_elements_CS}
\end{align}
It is clear from above equation that the diagonal elements of the matrix $\rho^{(k)}(\tau)$  do not decay due to phase damping. Substituting Eq.~(\ref{ch3PhDec_density_elements_CS}) in Eq.~(\ref{ch2omega_Tau}), we get the optical tomogram of the state $\ket{\psi^{(k)}}$ under phase damping as
\begin{align}
\omega^{(k)}\left(X_\theta, \theta,\tau\right)=&\frac{\exp\left[{-X_\theta}^2-\modu{\alpha}^2\right]}{\sqrt{\pi}}\sum_{n,n^\prime=0}^{\infty} \frac{H_n(X_\theta)\, H_{n^\prime}(X_\theta)e^{-i\,(n-n^\prime)\theta}}{2^{(n+n^\prime)/2}\,n!\,n^\prime!}\nonumber\\
&\times\sum_{s,s^\prime=0}^{k-1} f_{s,k}\,f_{s^\prime,k}^{\ast}\,\alpha_s^{n}
{\alpha_{s^\prime}^\ast}^{n^\prime}.
\label{ch3omega_Phasedamp}
\end{align}
In Fig.~\ref{ch3fig:PhaseDamp}, we show the optical tomograms of the states at two-, three-, and four-subpacket fractional revival times in the presence of phase damping. The sinusoidal strands in the optical tomogram of the state  retain their structure   only for a short  time $\kappa\tau$. The phase damping shows a faster decay of the sinusoidal strands in the optical tomogram of the states. The faster  decay of the state  is very noticeable in the case of states at higher-order fractional revivals (See Fig.~\ref{ch3fig:PhaseDamp}(c)). In the long-time limit, the state $\rho^{(k)}$ reduces to 
\begin{align}
\rho^{(k)}(\tau\rightarrow\infty)=e^{-\modu{\alpha}^2}\sum_{n=0}^{\infty} \frac{\modu{\alpha}^{2n}}{n!} \ket{n}\bra{n},
\end{align}
and the corresponding optical tomogram reads
 \begin{equation}
\omega^{(k)}\left(X_\theta, \theta,\tau\rightarrow\infty\right)=\frac{e^{{-X_\theta}^2-\modu{\alpha}^2}}{\sqrt{\pi}}\sum_{n=0}^{\infty} \frac{\modu{\alpha}^{2n}\,H_{n}^{2}(X_\theta)}{2^{n}\,\left(n!\right)^2}.
\end{equation}
This optical tomogram is shown in the last column of the Fig.~\ref{ch2fig:PhaseDamp}(a).  We have repeated the above analysis for the states at higher-order fractional revival times and found similar results. 
 
\section{Conclusion}
\label{Summary}
We have studied the optical tomograms of  the states obtained by the evolution of a coherent state in a Kerr-like medium.  We have shown that the signatures of revivals and fractional revivals are captured directly in the optical tomograms of the quantum states. The optical tomogram of the time-evolved state at the instants of fractional revivals shows structures with sinusoidal strands.  In general, the optical tomogram of the time-evolved state at $k$-subpacket fractional revivals is  a structure with $k$ sinusoidal strands in the $X_\theta$-$\theta$ plane. There are no sinusoidal strands present when the initial state collapses during the evolution. Our results will be helpful
for the study of revivals and fractional revivals directly from the optical tomogram of
the states generated by homodyne measurements. Since our methods avoids the reconstruction of the density matrix or the quasiprobability distribution, more comprehensive is the information about the state measured, and the fractional revival phenonmenon can be studied with high accuracy.
The analysis described in this chapter can be repeated for different initial states. A problem of considerable interest is to study in detail the revivals and fractional revivals of initial superposed wave packets. In the next chapter, we study the fractional revivals of superposed coherent states.
\chapter{FRACTIONAL REVIVALS OF SUPERPOSED COHERENT STATES}\label{Ch_SCSevolution}
\thispagestyle{plain}
\section{Introduction}
The universal scenario of revivals and fractional revivals described in \citep{Averbukh1989} applies to an arbitrary initial states,  including a superposition of several wave packets. However, the generic analytical expressions of the revival and fractional revival phenomena, discussed in a wide class of  systems \citep{Robinett2004}, are mainly dealing with an initial single wave packet. A problem of considerable  interest is to study in detail the revivals and fractional revivals of an initial  superposed wave packets. In this chapter, we study the fractional revivals of an initial macroscopic superposition state as it propagates through a nonlinear medium. Our aim is to investigate if there is any change in the fractional revival time depending upon the number of subpackets composing the initial superposition state. We also study how the selective identification of the fractional revivals, using the moments of quadrature variables and the optical tomogram of the time-evolved state, depend on the number of subpackets composing the initial superposition state. For this purposes, we consider the propagation of an initial superposed coherent state $\ket{\psi_{l,h}}$, given in Eq.~(\ref{ch2psi_lh}), through the Kerr medium, governed by the Hamiltonian given in Eq.~(\ref{ch3kerrhamiltonian}). 
The number state representation of the state $\ket{\psi_{l,h}}$, which we will use extensively, is given by
 \begin{eqnarray}
\ket{\psi_{l,h}}=l\,N_{l,h}\,\,e^{-|\alpha|^2/2}\,\sum_{n=0}^{\infty} \frac{\alpha^{ln+h}}{\sqrt{(l\,n+h)!}}\,\ket{l\,n+h}.
 \label{ch4generalsupernumber}
 \end{eqnarray}
It consists of an arithmetic infinite progression of Fock states with suitable amplitudes, having the state $\ket{h}$ as the initial term and a common difference equal to $l$ between successive terms.  Such an initial state shows revivals at an integer multiple of the time $\trev=\pi/\chi$. The dynamics of the initial state $\ket{\psi_{l,h}}$ can be analyzed using the methods described in the previous chapter. 
In the subsequent sections, we discuss the dynamics of the initial even coherent state of order $l$, which is the state $\ket{\psi_{l,h}}$ with $h=0$.

\section{Evolution of the even coherent state of order $2$}
\label{Evolution2CS}

Consider the evolution of an initial even coherent state of order $2$, obtained by setting $l=2$ and $h=0$ in Eq.~(\ref{ch2psi_lh}):
\begin{equation}
\ket{\psi_{2,0}}=N_{\rm 2,0}\left[\ket{\alpha}+\ket{-\alpha}\right].
\end{equation}
The corresponding Fock state representation of the even coherent state of order $2$ is given by
 \begin{align}
\ket{\psi_{2,0}}=2N_{2,0}\,\,e^{-|\alpha|^2/2}\,\sum_{n=0}^{\infty} \frac{\alpha^{2n}}{\sqrt{(2\,n)!}}\,\ket{2\,n}.
 \label{ch4general2n}
 \end{align}
The Fock state representation of the even coherent state of order $2$ contains only the even photon excitations.

\begin{figure}[H]
\centering
\includegraphics[scale=0.65]{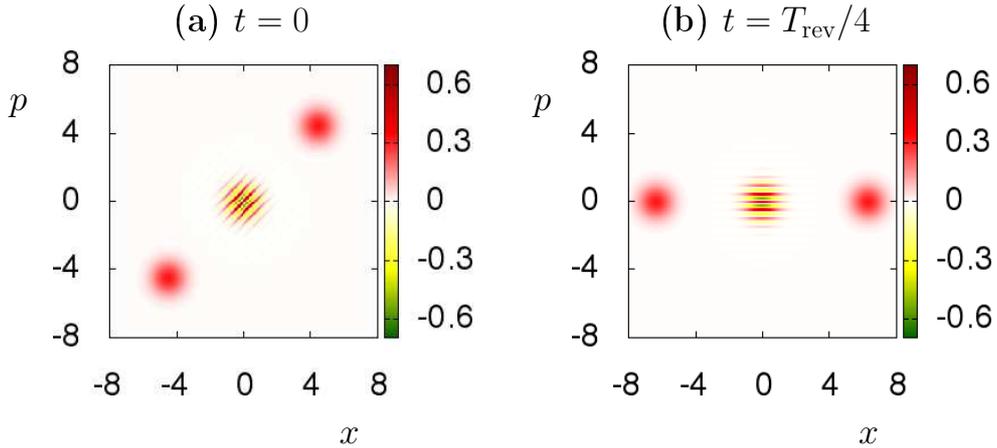}  
\caption{Wigner   function at  times (a) $t=0$ and (b) $t=T_{\rm rev}/4$, for an initial even coherent state $\ket{\psi_{2,0}}$ with $|\alpha|^2=20$. The unitary time evolution operator at  $t=T_{\rm rev}/4$  rotates the initial even coherent state through an angle of $45$ degrees in the clockwise direction in phase space.}
\label{ch4Wigner2CST0}
\end{figure}

The state at time $t$ for an initial even coherent state $\ket{\psi_{2,0}}$ is given by
\begin{equation}
\ket{\psi_{2,0}(t)}=2\,N_{2,0}\,e^{-\modu{\alpha}^2/2}\sum_{n=0}^{\infty}\frac{\alpha^{2n}}{\sqrt{(2n)!} }e^{-i\chi t\, 2n(2n-1)}\ket{2n}\label{ch4StateAttimet2N}
\end{equation}
At time $t=\trev/k=\pi/k\chi$, the state $\ket{\psi_{2,0}(t)}$ can be written as
\begin{equation}
\ket{\psi_{2,0}(t=\trev/k)}=\ket{\psi^{(k)}_{2,0}}=\sum_{s=0}^{k-1}\sum_{r=0}^{1} f_{s,k}\ket{\alpha_{r,s}}, \label{ch4psi_trevbyk}
\end{equation} 
where $f_{s,k}$ is defined in Eq.~(\ref{ch3FourierCoefficients}) and 
\begin{align}
\alpha_{r,s}&=\begin{cases}
\alpha\, e^{i\, \left(\pi r - 2\pi s/k\right)} & \text{if $k$ is odd}\\
\alpha\,e^{i\, \left(\pi r - 2\pi s/k\right)}\,e^{i\pi/k} & \text{if $k$ is even.}
\end{cases}
\end{align}
See Eq.~(\ref{APDX1psi_trevbyk}) in Appendix \ref{Appendix_superposedCS} for more details. Between $t=0$ and $t=T_{\rm rev}$, at $t=jT_{\rm rev}/4$, where $j=1,\,2,\,{\rm and}\, 3$,  the state is again an even coherent state of order $2$ but rotated in phase space:
\begin{align}
\ket{\psi(T_{\rm rev}/4)}=N_{2,0}\,\Big[\ket{\alpha\,e^{-i\pi/4}}+\ket{-\alpha\,e^{-i\pi/4}}\Big],
\end{align}
\begin{align}
\ket{\psi(T_{\rm rev}/2)}=N_{2,0}\,\Big[\ket{\alpha\,e^{i\pi/2}}+\ket{-\alpha\,e^{i\pi/2}}\Big],
\end{align}
\begin{align}
\ket{\psi(3T_{\rm rev}/4)}=N_{2,0}\,\Big[\ket{\alpha e^{i\pi/4}}+\ket{-\alpha e^{i\pi/4}}\Big].
\end{align}
The $k$ subpackets composing the superposition state $\ket{\psi^{(k)}}$ can be visualized in phase space using the Wigner function of the state. The Wigner function of a state $\ket{\psi}$ is defined as \citep{Wigner1932,Agarwal1970}
\begin{align}
W(\beta)=\frac{2\,e^{2\modu{\beta}^2}}{\pi^2}\int d^2 z \langle -z\left|\psi\rangle\langle \psi \right| z\rangle\, e^{2\left(z^\ast \beta-z\beta^\ast\right)}, \label{ch3Def:WignerFunction}
\end{align}
where $\ket{z}$ is a coherent state. In terms of the position $x$ and momentum $p$, the complex variable $\beta=(x+i\,p)/\sqrt{2}$. 
The Wigner function of the state $\ket{\psi^{(k)}_{2,0}}$ is calculated as
\begin{align}
W_{2,0}^{(k)}(\beta)=\frac{2\,e^{2\modu{\beta}^2-\modu{\alpha}^2}}{\pi} \sum_{s,s^\prime=0}^{k-1}\sum_{r,r^\prime=0}^{1} f_{s,k}\,f_{s^\prime,k}^\ast\,e^{-\left(2 \beta-\alpha_{r,s}\right)\left(2\beta^\ast-\alpha_{r^\prime,s^\prime}\right)}. \label{ch4WignerFunfor_psi20}
\end{align}
This equation can be obtained from the Eq.~(\ref{APDX1WignerFunfor_psilh_evolution}) in Appendix \ref{Appendix_superposedCS} for the case $l=2$ and $h=0$. Figure~\ref{ch4Wigner2CST0} shows the Wigner function of the initial state $\ket{\psi_{2,0}}$ and the state $\ket{\psi^{(4)}_{2,0}}$ with $\modu{\alpha}^2=20$. The value of the argument of $\alpha$ is set to be $\delta=\pi/4$ throughout this chapter. The unitary time evolution operator at  $t=T_{\rm rev}/4$  rotates the initial even coherent state of order $2$ through an angle of $45$ degrees in the clockwise direction in phase space.   

\begin{figure}[H]
\centering
\includegraphics[scale=0.85]{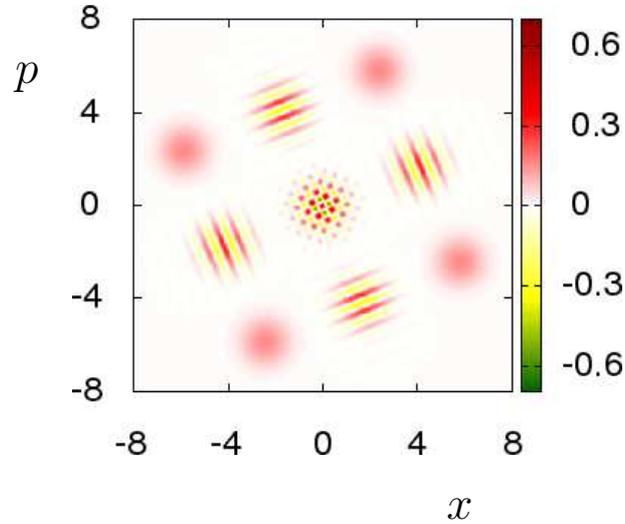} 
\caption{Wigner function at two-subpacket fractional revival time $t=T_{\rm rev}/8$ for an initial even coherent state of order $2$ with $\modu{\alpha}^2=20$.}
\label{ch4Wigner2CST8}
\end{figure}

Here, we find that the $k$-subpacket fractional revival occurs at time $t=j\,\trev/4k$ where $j=1,\, 2,\, \dots,\, (4k-1)$ for a given value of   $k (>1)$ with $(j, 4k)=1$. At $k$-subpacket fractional revival time, the initial wave packet splits into $k$ phase rotated even coherent states of order $2$ \citep{Rohith2014}.  In contrast, we have seen in Chapter \ref{Ch_CSevolution} that for an initial coherent state $k$-subpacket fractional revival occurs  at  $t=j\trev/k$, where   $j=1,\, 2,\, \dots,\, (k-1)$ for a given value of   $k(>1)$  with $(j, k)=1$. For example, the two-subpacket fractional revival for an initial even coherent state of order $2$ occurs at time $t=T_{\rm rev}/8$ and the state at this time is a superposition of two even coherent state of order $2$,
\begin{align}
\ket{\psi^{(8)}_{2,0}}=&C_1\,N_{2}\Big[\ket{\alpha\,e^{i\pi/8}}+\ket{-\alpha\,e^{i\pi/8}}\Big]\nonumber\\
&+C_2\,N_{2}\Big[\ket{\alpha\,e^{-i3\pi/8}}+\ket{-\alpha\,e^{-3i\pi/8}}\Big],
\label{ch4stwoecs}
\end{align}
where $C_1=(1-i)/2$ and $C_2=(1+i)/2$.
Figure~\ref{ch4Wigner2CST8} clearly shows a superposition of two even coherent state of order $2$ at time $t=\trev/8$.

\begin{figure}[H]
\centering
\includegraphics[scale=0.5]{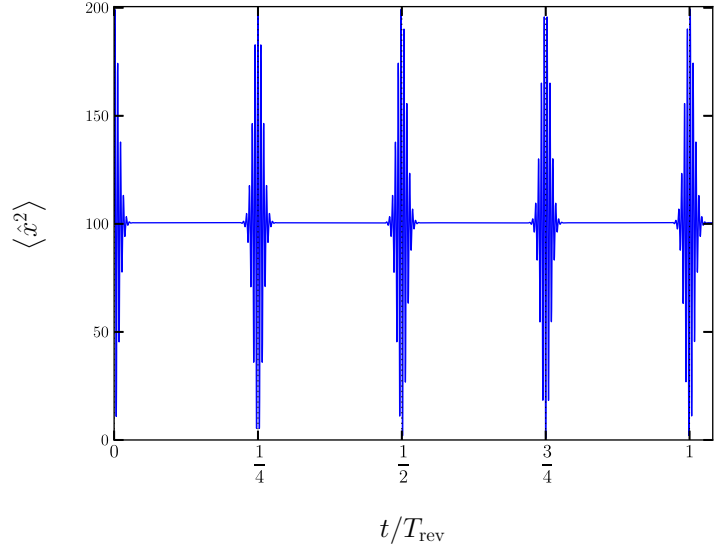} 
\caption{$\aver{\hat{x}^2(t)}$ as a function of $t/T_{\rm rev}$ for an initial even coherent state $\ket{\psi_{2,0}}$ with $|\alpha|^2=100$. Between time $t=0$ and $T_{\rm rev}$,  the second moment of the position operator is a constant most of the time except at times   $T_{\rm rev}/4$, $\trev/2$, and  $3\trev/4$. At these instants, the second moment shows a rapid variation, which are the signatures of wave-packet rotations.}
\label{ch4x2vstevencs}
\end{figure}

All odd moments of the operators $\hat{x}$ and $\hat{p}$ vanish at all times for the initial even coherent state of order $2$. The expectation value of $\hat{x}^2$  at any time can be obtained as explicit functions of $t$ in the form
\begin{align}
\aver{\hat{x}^2(t)}=&2\,N_{2,0}^{2}\,\modu{\alpha}^2\, \Big[e^{-\modu{\alpha}^2 \left( 1-\cos  4\chi t  \right)}\cos \left( 2\chi t+\modu{\alpha}^2 \sin  4\chi t -\frac{\pi }{4} \right)\nonumber\\
&+ e^{-\modu{\alpha}^2 \left( 1+\cos \left( 4\chi t \right) \right)}\cos \left( 2\chi t-\modu{\alpha}^2 \sin \left( 4\chi t \right)-\frac{\pi }{4} \right)\Big]+\modu{\alpha}^2 +\frac{1}{2}.
\end{align}
Between $t=0$ and $\trev$, the above expression for $\aver{\hat{x}^2}$ is static most of the time except at  $t=\trev/4$, $\trev/2$ and $3\,\trev/4$, for sufficiently large value of $\modu{\alpha}^2$. Thus, the second moment of the position operator $\hat{x}^2$ captures the signatures of wave-packet rotations in phase space at times $\trev/4$,  $\trev/2$ and $3\,\trev/4$. Figure~\ref{ch4x2vstevencs} shows the variation of the expectation value $\aver{\hat{x}^2}$ versus time for the initial even coherent state of order $2$. 

The expressions for the $2k^{\rm th}$ moments of $\hat{x}$ and $\hat{p}$ can be deduced readily from the general result
\begin{align}
\aver{a^{2k}}=&2\,N^2_{2,0}\, {\alpha}^{2k}\exp\left[ -i\,2k(2k-1)\chi t\right]\left\{\exp\left[-\modu{\alpha}^2(1-\cos 4k\chi t)-i\,\modu{\alpha}^2 \sin 4k\chi t\right]  \right.\nonumber\\
&\left.+ \exp\left[-\modu{\alpha}^2(1+\cos 4k\chi t)+i\,\modu{\alpha}^2 \sin 4k\chi t\right]\right\},
\end{align} 
where $k$ is a positive integer. The time dependence of the $2k^{\rm th}$ moments of $\hat{x}$ are
strongly controlled by the factors $\exp[-\modu{\alpha}^2(1\pm\cos\,4k\chi t)]$, $k=1,\, 2,\, \dots$, that modulates the oscillatory term. \begin{figure}[H]
\centering
\includegraphics[scale=0.7]{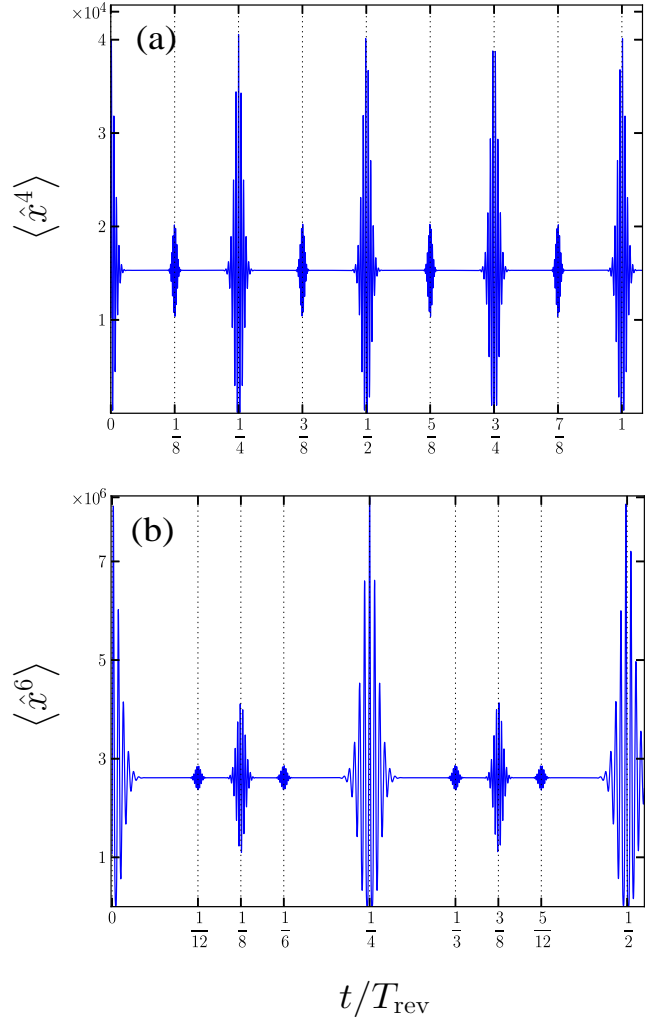}
\caption{Temporal evolution of higher moments of $\hat{x}$ for an initial even coherent state $\ket{\psi_{2,0}}$  with $\modu{\alpha}^2=100$. (a) Between $t=0$ and $t=T_{\rm rev}$,  $\aver{\hat{x}^4(t)}$ is a constant most of the time except at   $t=j\,T_{\rm rev}/8$, where $j=1,\, 2,\, \dots,\, 7$. At these instants, the fourth moment of $\hat{x}$ show a rapid variation, which are the signatures of  two-subpacket fractional revival and wave-packet rotations.  (b) In this figure we have plotted between $t=0$ and  $\trev/2$ for a  better view. $\aver{\hat{x}^6(t)}$ is a constant most of the time except at   $t=j\,T_{\rm rev}/12$, where $j=1,\, 2,\, \dots,\, 6$. At these instants, the sixth moment of $\hat{x}$ show a rapid variation, which are the signatures of  three- and two-subpacket fractional revivals, and wave-packet rotations.}
\label{ch4x2x6forevencs}
\end{figure}
Between $t=0$ and $t=T_{\rm rev}$, these factors act as a strong damping factor for large values of $\modu{\alpha}^2$, except at  fractional revival times $t=j\trev/4k$. It can be concluded that the $k$-subpacket fractional revivals are captured in the $2k^{\rm th}$ moment of $\hat{x}$ or $\hat{p}$, but not in lower moments, in contrast with the case of an initial coherent state. These results are illustrated in Fig.~\ref{ch4x2x6forevencs}. Figure~\ref{ch4x2x6forevencs}(a) shows the temporal evolution of the expectation value $\aver{\hat{x}^4}$. It  shows rapid oscillations around $t/\trev=j/8$, where $j=1,\, 2,\, \dots,\, 7$, between $t=0$ and $\trev$. Therefore the fourth moment of $\hat{x}$ versus time captures the signature of the two-subpacket fractional revivals at $t/\trev=j/8$ where $j=1,\, 2,\, \dots, 7$ with $(j, 8)=1$ and wave-packet rotations at $t/\trev=j/4$ where $j=1,\, 2,\, 3$. Figure~\ref{ch4x2x6forevencs}(b) is a plot of $\aver{\hat{x}^6}$ versus time, which shows the signatures of the three-subpacket fractional revival.


\begin{figure}[h]
\centering
\includegraphics[scale=0.6]{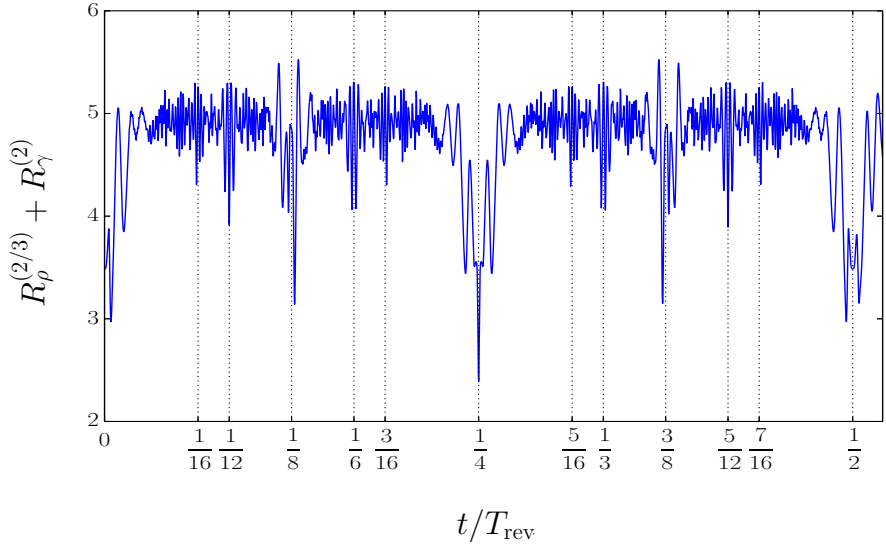} 
\caption{Time evolution of $R_{\rho}^{(2/3)}+R_{\gamma}^{(2)}$ for an initial even coherent state of order $2$ with $\modu{\alpha}^2=30$. The main fractional revivals are indicated by vertical dotted lines.}
\label{ch4entropyevencs}
\end{figure} 
The signatures of fractional revivals are identified by tracking the time evolution of the sum of R\'{e}nyi entropies in conjugate spaces. We study the time evolution of the sum $R_{\rho}^{(2/3)}+R_{\gamma}^{(2)}$, where $R_{\rho}^{(2/3)}$ and $R_{\gamma}^{(2)}$ are the R\'{e}nyi entropies in position and momentum spaces, respectively. The R\'{e}nyi entropies $R_{\rho}^{(2/3)}$ and $R_{\gamma}^{(2)}$ are calculated by plugging the probability densities of the state $\ket{\psi_{2,0}(t)}$ in  position and momentum spaces, that is $\modu{\psi_{2,0}(x,t)}^2$ and $\modu{\phi_{2,0}(p,t)}^2$, respectively, in Eq.~(\ref{ch3RenyiEntropy}). The integrations are performed numerically by using the trapezoidal rule. Figure~\ref{ch4entropyevencs} shows the  sum of the R\'{e}nyi entropies in position and momentum spaces as a function of time for an initial even coherent state $\ket{\psi_{2,0}}$ with $\modu{\alpha}^2=30$. The signatures of fractional revivals are indicated by the local minima of  the sum $R_{\rho}^{(2/3)}+R_{\gamma}^{(2)}$.

%
Next, we study the optical tomogram of the time-evolved state $\ket{\psi_{2,0}(t)}$ given in Eq.~(\ref{ch4StateAttimet2N}). The time-evolved optical tomogram for initial even coherent state $\ket{\psi_{2,0}}$ is calculated as
\begin{align}
\omega_{2,0}\left(X_{\theta},\theta,t\right)=\frac{4\,N_{2,0}^2\,e^{-\modu{\alpha}^2}\,e^{-X_{\theta}^2}}{\sqrt{\pi}}\modu{\sum_{n=0}^{\infty}\frac{\alpha^{2n}\, H_{2n}\left(X_{\theta}\right)\,e^{-i\, 2n\theta}\,e^{-i\,\chi t\, 2n(2n-1)}}{(2n)!\,2^{n}}}^2.
\label{ch4optEven}
\end{align}
At time $t=0$, the Eq.~(\ref{ch4optEven}) gives the optical tomogram of the even coherent state $\ket{\psi_{2,0}}$, which is  a structure with two sinusoidal strands in the $X_{\theta}$-$\theta$ plane (see Fig.~\ref{ch2fig:opt_l2}(a)). 
\begin{figure}[H]
\centering
\includegraphics[scale=0.6]{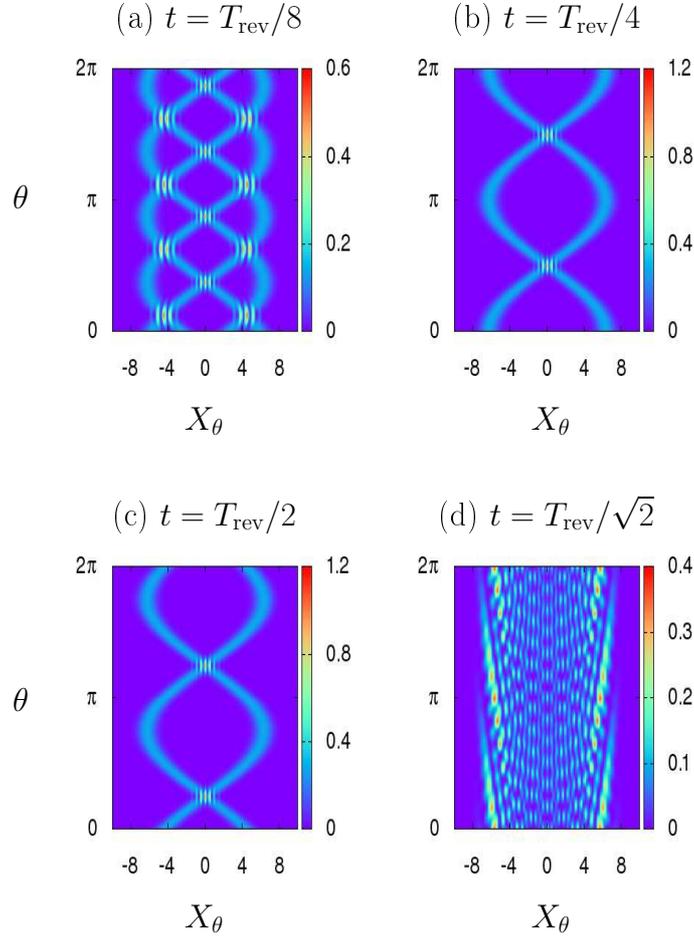}
\caption{Time-evolved optical tomogram $\omega_{2,0}\left(X_{\theta},\theta\right)$  of the initial even coherent state $\ket{\psi_{2,0}}$ with $\modu{\alpha}^2=20$ at  (a) $t=0$, (b) $t=\trev/8$, (c) $t=\trev/4$, (d) $t=\trev/2$, (e) $t=\trev/\sqrt{2}$, and (f) $\trev$, respectively. At the instants of  $k$-subpacket fractional revivals,  the optical tomogram of the time-evolved state of the  initial even coherent state of order $2$ displays a structure with $2k$ sinusoidal strands.}
\label{ch4fig:optEven}
\end{figure}

In Figs.~\ref{ch4fig:optEven}(a)-\ref{ch4fig:optEven}(d), we plot the optical tomogram given in Eq.~(\ref{ch4optEven}) at different instants during the evolution of the  initial even coherent state $\ket{\psi_{2,0}}$ in the medium. Figure~\ref{ch4fig:optEven}(a) shows the optical tomogram of the time-evolved state  at  $\trev/8$, which corresponds   to two-subpacket fractional revival. It displays  a structure with four  sinusoidal strands, which is a signature of two-subpacket fractional revival for the initial even coherent state $\ket{\psi_{2,0}}$. 
The time-evolved optical tomogram for initial even coherent state $\ket{\psi_{2,0}}$ is also analyzed at higher-order fractional revival times,  and we found that, at the instants of $k$-subpacket fractional revivals, the optical tomogram of the time-evolved state for the initial even and odd coherent state displays a structure with $2k$ sinusoidal strands \citep{Rohith2015}. 

\begin{figure}[H]
\centering
\includegraphics[scale=0.6]{OTECSrevival.eps}
\caption{Time-evolved optical tomogram $\omega_{2,0}\left(X_{\theta},\theta\right)$  of the initial even coherent state $\ket{\psi_{2,0}}$ with $\modu{\alpha}^2=20$ at  revival time $t=\trev$}
\label{ch4fig:opt2CSrevival}
\end{figure}
At the instants of rotated wave packets, the state is again a superposition of two coherent states. Figures~\ref{ch4fig:optEven}(b) and \ref{ch4fig:optEven}(c) show the optical tomogram  of rotated wave packets at time  $t=\trev/4$ and $\trev/2$. The optical tomogram shows a structure with two sinusoidal strands, as expected.  These tomograms  are qualitatively different from the optical tomogram shown in Fig.~\ref{ch2fig:opt_l2}(a). The locations of the sinusoidal strands, where the maximum intensity of the optical tomogram along the $X_{\theta}$ axis occurs, in these optical tomograms are shifted due to the phase-space rotation of the quantum states during the evolution of the initial even coherent state $\ket{\psi_{2,0}}$ in the medium.  Figure~\ref{ch4fig:optEven}(d) shows the optical tomogram of a collapsed state at time $t=\trev/\sqrt{2}$,  which again confirms our result that  sinusoidal strands are absent in the optical tomogram of the collapsed state. The optical tomogram of the time-evolved state at the revival time is shown in Fig.~\ref{ch4fig:opt2CSrevival}. Our analysis  shows a  clear distinction between the time evolution of an initial coherent state, presented in the previous chapter, and the initial even coherent state of order $2$. In the next section we study the dynamics of the initial even coherent state of order $3$.

\section{Evolution of the even coherent state of order $3$}
\label{Evolution3CS}
\begin{figure}[h]
\centering
\includegraphics[scale=0.75]{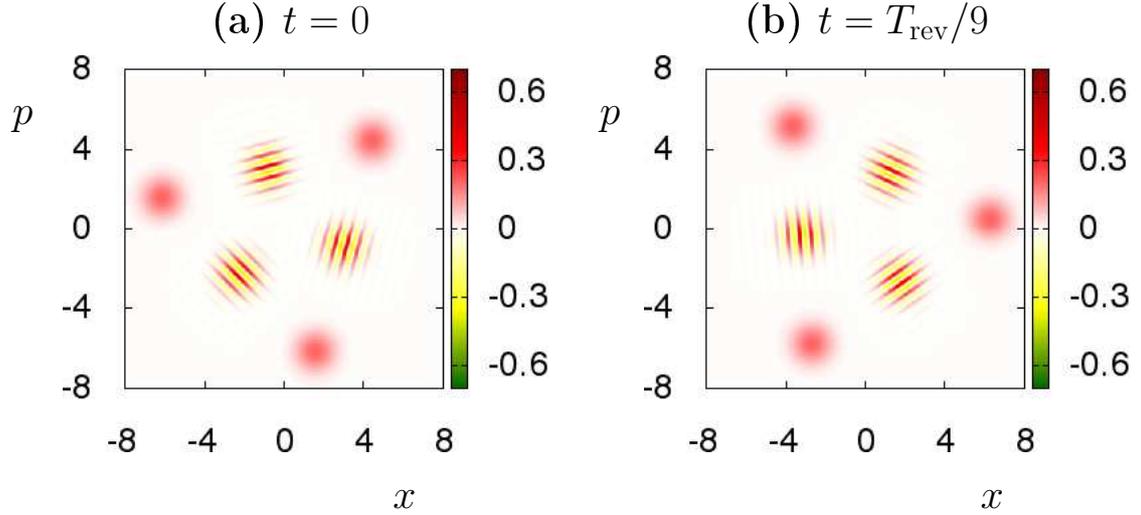} 
\caption{Wigner function at  times (a) $t=0$ and (b) $t=T_{\rm rev}/9$ for the  initial  state $\ket{\psi_{3,0}}$  with $|\alpha|^2=20$.  Both the figures show a superposition of three coherent states. The unitary time evolution operator at $t=T_{\rm rev}/9$ rotates the initial state through an angle of $40$ degrees ($2\pi/9$ radians) in the clockwise direction in phase space.}
\label{ch4Wigner3nt0and9}
\end{figure}

Setting $l=3$ and $h=0$ in Eq.~(\ref{ch2psi_lh}), we obtain the even coherent state of order $3$ \citep{Peng1990,Napoli1999} as
\begin{eqnarray}
\ket{\psi_{3,0}}=N_{3,0}\left[\ket{\alpha}+\ket{\alpha\,e^{i2\pi/3}}+\ket{\alpha\,e^{-i2\pi/3}}\right].
\label{ch4ketpsi3}
\end{eqnarray}
Its Fock state representation is given by
\begin{eqnarray}
\ket{\psi_{3,0}}=3N_{3,0}\,\,e^{-|\alpha|^2/2}\,\sum_{n=0}^{\infty}\,\frac{\alpha^{3n}}{\sqrt{3n!}}\ket{3n}.
\end{eqnarray}
The state at time $t$ for an initial state $\ket{\psi_{3,0}}$ is given by
\begin{equation}
\ket{\psi_{3,0}(t)}=3\,N_{3,0}\,e^{-\modu{\alpha}^2/2}\sum_{n=0}^{\infty}\frac{\alpha^{3n}}{\sqrt{(3n)!} }e^{-i\chi t\, 3n(3n-1)}\ket{3n}.\label{ch4StateAttimet3N}
\end{equation}
At time $t=\trev/k=\pi/k\chi$, the state $\ket{\psi_{3,0}(t)}$ can be written as
\begin{equation}
\ket{\psi_{3,0}(t=\trev/k)}=\ket{\psi^{(k)}_{3,0}}=\sum_{s=0}^{k-1}\sum_{r=0}^{2} f_{s,k}\ket{\alpha_{r,s}}, \label{ch4psi_trevbykfor3N}
\end{equation} 
where 
\begin{align}
\alpha_{r,s}&=\begin{cases}
\alpha\, e^{i\,2\pi \left( r/3 - s/k\right)} & \text{if $k$ is odd}\\
\alpha\,e^{i\,2\pi \left(r/3 - s/k\right)}\,e^{i\pi/k} & \text{if $k$ is even.}
\end{cases}
\end{align}
See Eq.~(\ref{APDX1psi_trevbyk}) in Appendix \ref{Appendix_superposedCS} for more details. The time evolution of the initial state $\ket{\psi_{3,0}}$ shows fractional revivals and rotations at different instants when compared to the initial coherent state and the initial even coherent state of order $2$. For the initial state $\ket{\psi_{3,0}}$, the rotations in phase space occur at $t=jT_{\rm rev}/9$, where $j=1,\, 2,\, \dots,\, 8$, between $t=0$ and $t=T_{\rm rev}$. It is necessary to bear in mind that, for initial coherent state there  is no rotation  and for an initial even coherent state of order $2$ the rotations occur at $t=jT_{\rm rev}/4$, where $j=1$, $2$, and  $3$.   For example, at time $t=T_{\rm rev}/9$ the initial state $\ket{\psi_{3,0}}$ evolves to
\begin{equation}
\ket{\psi^{(9)}_{3,0}}=N_{3,0}\left[\ket{\alpha\,e^{-i2\pi/9}}+\ket{\alpha\,e^{i4\pi/9}}+\ket{\alpha\,e^{-i8\pi/9}}\right].
\end{equation}
The Wigner function of the state $\ket{\psi^{(k)}_{3,0}}$ is calculated as
\begin{align}
W_{3,0}^{(k)}(\beta)=\frac{2\,e^{2\modu{\beta}^2-\modu{\alpha}^2}}{\pi} \sum_{s,s^\prime=0}^{k-1}\sum_{r,r^\prime=0}^{2} f_{s,k}\,f_{s^\prime,k}^\ast\,e^{-\left(2 \beta-\alpha_{r,s}\right)\left(2\beta^\ast-\alpha_{r^\prime,s^\prime}\right)}. \label{ch4WignerFunfor_psi30}
\end{align}
Figure~\ref{ch4Wigner3nt0and9} shows the plots of the Wigner function for  the states $\ket{\psi_{3,0}}$ and $\ket{\psi^{(9)}_{3,0}}$. The unitary time evolution operator at $t=T_{\rm rev}/9$ rotates the initial state through an angle of $40$ degrees ($2\pi/9$ radians) in the clockwise direction in phase space. 

\begin{figure}[h]
\centering
\includegraphics[scale=0.75]{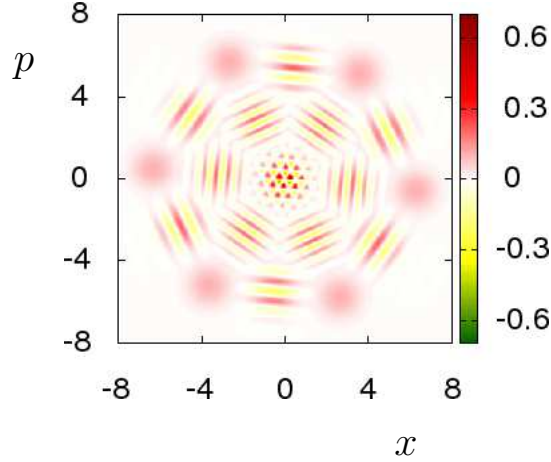} 
\caption{Wigner function at  two-subpacket fractional revival time $t=T_{\rm rev}/18$ for the initial  state  $\ket{\psi_{3,0}}$ with $\modu{\alpha}^2=20$. It shows a superposition of two initial states of the form $\ket{\psi_{3,0}}$.  }
\label{ch4Wigner3Nat18}
\end{figure}

In this case, the $k$-subpacket fractional revival occurs at time $t=j\,\trev/9k$, where $j=1,\, 2,\, \dots,\, (9k-1)$, for a given value of   $k (>1)$ with $(j, 9k)=1$.   For example, two-subpacket fractional revival for an initial  state  occurs at $t=T_{\rm rev}/18$, and the state at this time is a superposition of two  states of the form $\ket{\psi_{3,0}}$:
\begin{align}
\ket{\psi(\trev/18)}=&C_1\,N_{3}\left[\ket{\alpha\,e^{-i11\pi/18}}+\ket{\alpha\,e^{i\pi/18}}+\ket{\alpha\,e^{i13\pi/18}}\right]\nonumber\\
&+C_2\,N_{3}\left[\ket{\alpha\,e^{-i17\pi/18}}+\ket{\alpha\,e^{-i5\pi/8}}+\ket{\alpha\,e^{i7\pi/18}}\right],
\label{ch4psi3at2}
\end{align}
where $C_1=(1-i)/2$ and $C_2=(1+i)/2$. Figure~\ref{ch4Wigner3Nat18} clearly shows a superposition of two initial states of the form given in Eq.~(\ref{ch4ketpsi3}) with different $\alpha$ values as given in Eq.~(\ref{ch4psi3at2}) at $t=\trev/18$.

\begin{figure}[h]
\centering
\includegraphics[scale=0.5]{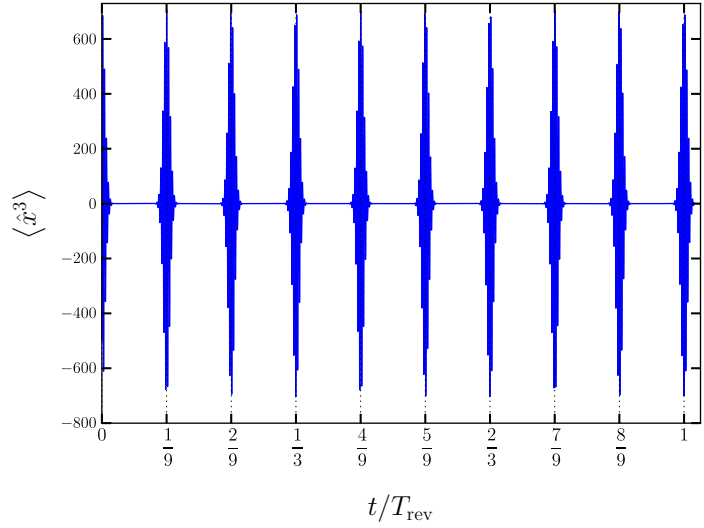} 
\caption{$\aver{\hat{x}^3(t)}$ as a function of $t/T_{\rm rev}$ for the initial state  $\ket{\psi_{3,0}}$ with $\modu{\alpha}^2=100$. Between $t=0$ and $T_{\rm rev}$,  the third moment of $\hat{x}$ is a constant most of the time except at  $t=jT_{\rm rev}/9$  where  $j=1$, $2$, $\dots$, $8$. At these instants of time, the time-evolved state is a rotated  initial wave packet.}
\label{ch4x3vst3N}
\end{figure}

Only the $3k^{\rm th}$ (where $k=1, 2, \dots$) moment of $\hat{x}$ and $\hat{p}$ gives non-zero value and all other moments are identically equal to  zero at all times for the initial state $\ket{\psi_{3,0}}$. The expectation value of $\hat{x}^3$  at any time for an initial state $\ket{\psi_{3,0}}$ is
\begin{align}
\aver{\hat{x}^3(t)}=&3\,N_{3,0}^{2}\,\modu{\alpha}^{3}\, \left[e^{-\modu{\alpha}^2 \left( 1-\cos  6\chi t  \right)}\cos \left( 6\chi t+\modu{\alpha}^2 \sin  6\chi t -{3\pi }/{4} \right)\right.\nonumber\\
&+\left. e^{-\modu{\alpha}^2 \left( 1-\sin \left( 6\chi t -\pi/6\right) \right)}\cos \left( 6\chi t+\modu{\alpha}^2 \cos \left( 6\chi t+\pi/6 \right)-{3\pi }/{4} \right)\right.\nonumber\\
&+\left. e^{-\modu{\alpha}^2 \left( 1+\sin \left( 6\chi t +\pi/6\right) \right)}\cos \left( 6\chi t-\modu{\alpha}^2 \cos \left( 6\chi t-\pi/6 \right)-{3\pi }/{4} \right)\right].
\end{align}
 Between $t=0$ and $\trev$, the above expression for $\aver{\hat{x}^3}$ is zero most of the times except at  $t=jT_{\rm rev}/9$, where $j=1, 2, \dots, 8$  for  sufficiently large value of $\modu{\alpha}^2$. These instants correspond to wave-packet rotations in phase space.  Figure~\ref{ch4x3vst3N} shows the variation of the expectation value $\aver{\hat{x}^3}$ versus time for the initial state $\ket{\psi_{3,0}}$. It shows that the  wave-packet rotations in phase space are captured in the third moment of $\hat{x}$.

\begin{figure}[H]
\centering
\includegraphics[scale=0.7]{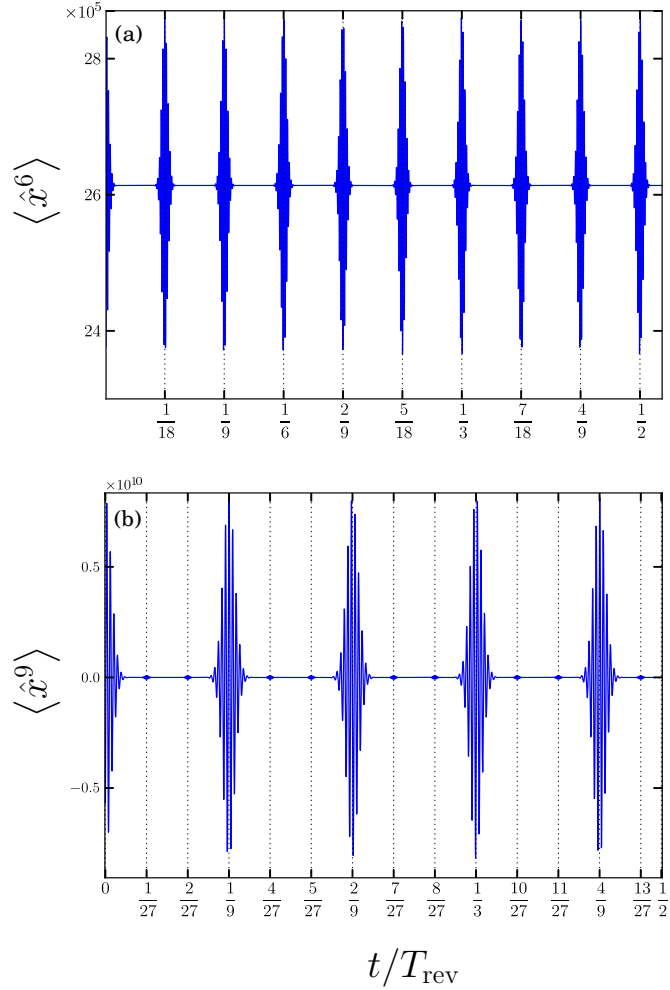}
\caption{Temporal evolution of higher moments of $\hat{x}$ for an initial state $\ket{\psi_{3,0}}$  with $\modu{\alpha}^2=100$ between time $t=0$ and $\trev/2$.   (a) $\aver{\hat{x}^6(t)}$ is a constant most of the time except at   $t=j\,T_{\rm rev}/18$, where $j=1$, $2$, $\dots$, $9$. At these instants, the sixth moment of $\hat{x}$ show a rapid variation, which are the signatures two-subpacket fractional revival and wave-packet rotations. (b) $\aver{\hat{x}^9(t)}$ is a constant most of the time except at   $t=j\,T_{\rm rev}/27$, where $j=1$, $2$, $\dots$, $13$. At these instants, the ninth moment of $\hat{x}$ show a rapid variation, which are the signatures of three-, two-subpacket fractional revivals  and rotations.}
\label{ch4x6x9for3N}
\end{figure}

The expressions for the higher moments of $\hat{x}$ and $\hat{p}$ can be deduced readily from the general result
 \begin{align}
\aver{a^{3k}}=&3\,N^2_{3,0} {\alpha}^{3k}\exp\left[-i\,3k(3k-1)\chi t\right]\left\{\exp\left[-\modu{\alpha}^2(1-\cos 6k\chi t)-i\,\modu{\alpha}^2 \sin 6k\chi t\right] \right.\nonumber\\
&\left.+ \exp\left[-\modu{\alpha}^2\left(1-\sin (6k\chi t-\pi/6)\right)-i\,\modu{\alpha}^2 \cos (6k\chi t+\pi/6)\right]\right.\nonumber\\
&\left. + \exp\left[-\modu{\alpha}^2\left(1+\sin (6k\chi t+\pi/6)\right)+i\,\modu{\alpha}^2 \cos (6k\chi t-\pi/6)\right]\right\}.
\end{align}
The time dependence of $3k^{\rm th}$ moments of the operator $\hat{x}$ is
strongly controlled by the factors $\exp[-\modu{\alpha}^2(1-\cos 6k\chi t)]$ and  $\exp\left[-\modu{\alpha}^2\left(1\pm\sin (6k\chi t-\pi/6)\right)\right]$, where $k$ is a positive integer, that modulates the oscillatory term. Between $t=0$ and $t=T_{\rm rev}$, these factors act as a strong damping factor for large values of $\modu{\alpha}^2$, except at fractional revival times $t=j\trev/9k$. It can be concluded that the $k$-subpacket fractional revivals are captured in the $3k^{\rm th}$ moment of $\hat{x}$ or $\hat{p}$. These results are illustrated in Fig.~\ref{ch4x6x9for3N}. Figure~\ref{ch4x6x9for3N}(a) shows the temporal evolution of the expectation 
value $\aver{\hat{x}^6(t)}$. We have plotted the graph in between $t=0$ and $\trev/2$ for a better view. It confirms that  the sixth moment of $\hat{x}$  captures the signatures of  two-subpacket fractional revival and rotations.  Figure~\ref{ch4x6x9for3N}(b) is a plot of $\aver{\hat{x}^9(t)}$ versus time which shows the signatures of three- and two-subpacket fractional revivals and rotations.  

\begin{figure}[h]
\centering
\includegraphics[scale=0.6]{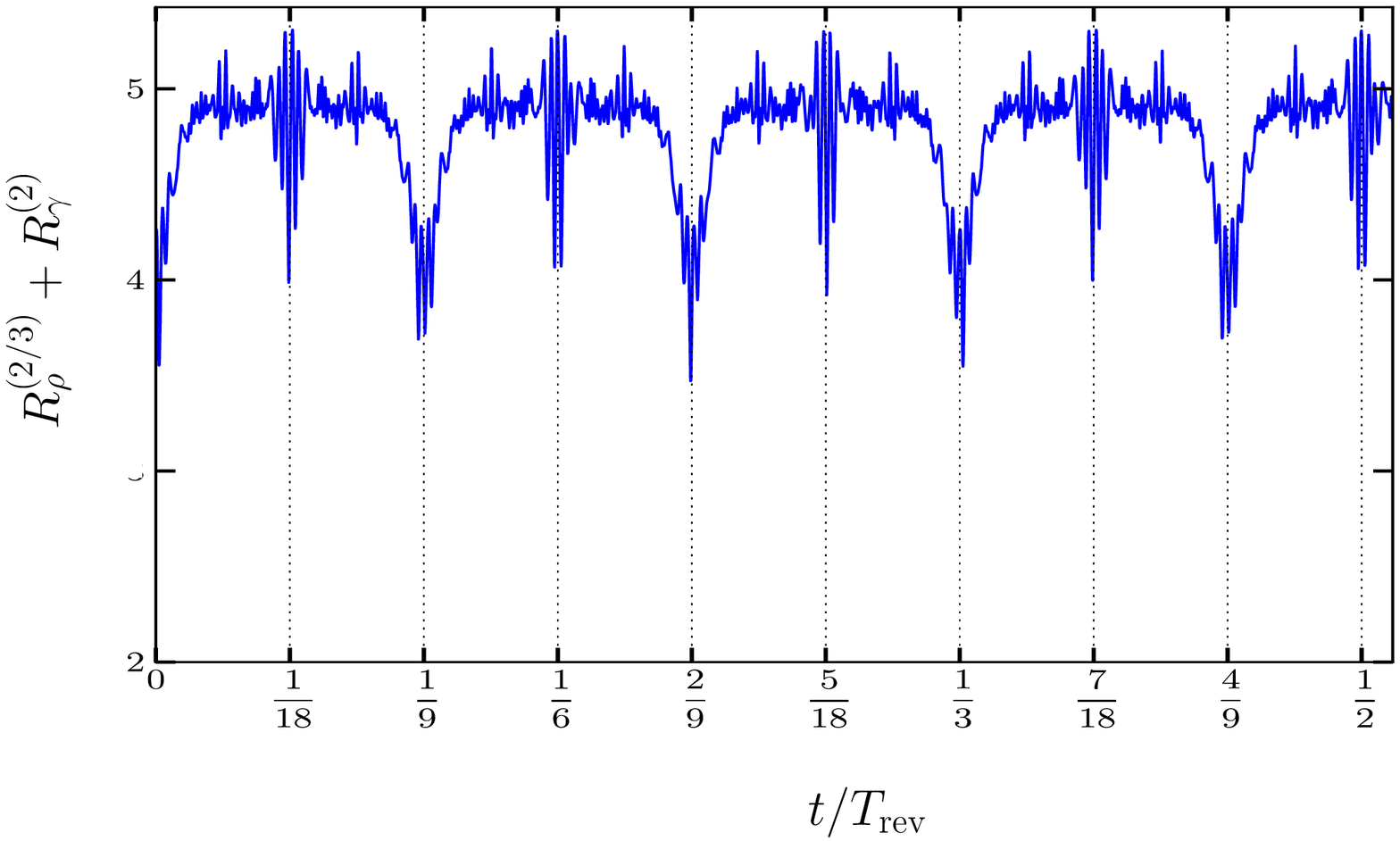} 
\caption{Time evolution of $R_{\rho}^{(2/3)}+R_{\gamma}^{(2)}$ for the initial state $\ket{\psi_{3,0}}$  with $\modu{\alpha}^2=30$. The main fractional revivals are indicated by vertical dotted lines.}
\label{ch4entropy3N}
\end{figure}

Figure~\ref{ch4entropy3N} shows the  sum of the R\'{e}nyi entropies in conjugate spaces, $R_{\rho}^{(2/3)}+R_{\gamma}^{(2)}$, as a function of time for the initial state $\ket{\psi_{3,0}}$ with $\modu{\alpha}^2=30$. It confirms our analysis based on the expectation values. The main fractional revivals are indicated by vertical dotted lines in the figure. 

\begin{figure}[H]
\centering
\includegraphics[scale=0.7]{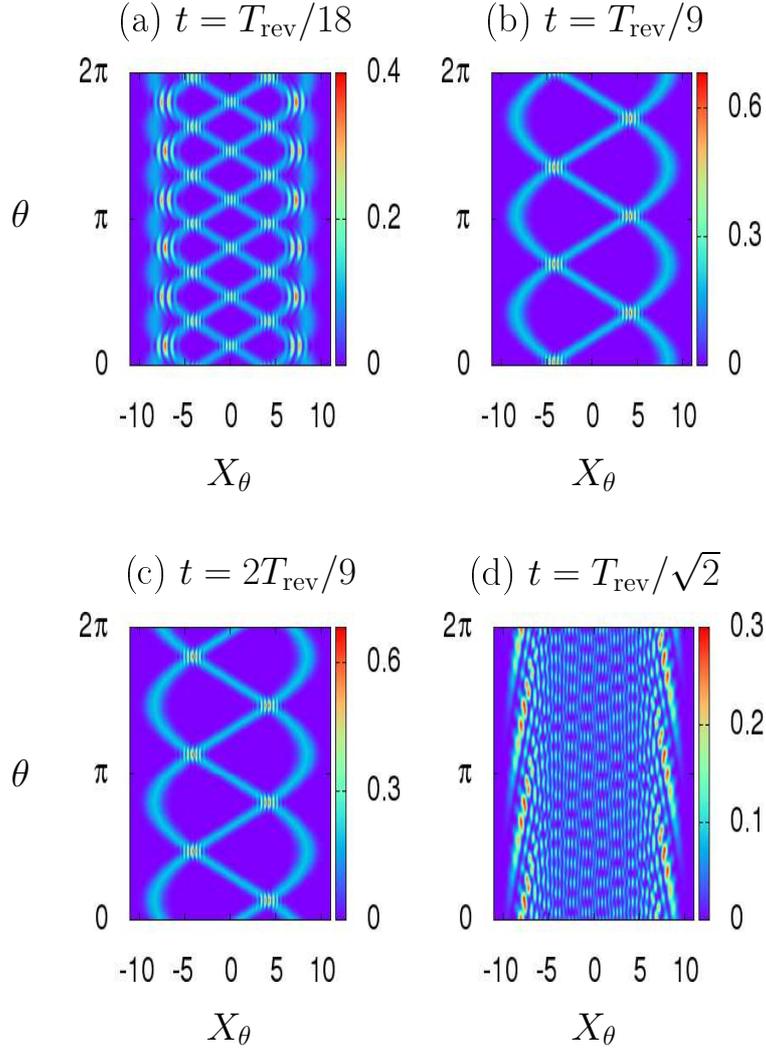}
\caption{Time-evolved optical tomogram $\omega_{3,0}\left(X_{\theta},\theta,t\right)$  for the initial state $\ket{\psi_{3,0}}$ with $\modu{\alpha}^2=35$ at  times (a) $t=\trev/18$, (b) $t=\trev/9$, (c) $t=2\trev/9$, and (d) $t=\trev/\sqrt{2}$, respectively. At the instants of  $k$-subpacket fractional revivals,  the optical tomogram of the time-evolved state of the  initial even coherent state displays a structure with $3k$ sinusoidal strands.}
\label{ch4fig:opt3CS}
\end{figure}

The time evolution of the optical tomogram for the initial state $\ket{\psi_{3,0}}$ is calculated as
\begin{align}
\omega_{3,0}\left(X_{\theta},\theta,t\right)=&\frac{9\,N_{3,0}^2\,\exp\left[-\modu{\alpha}^2-X_{\theta}^2\right]}{\sqrt{\pi}}\modu{\sum_{n=0}^{\infty}\frac{\alpha^{3n}\,H_{3n}\left(X_{\theta}\right)\,e^{-i\chi t\, 3n(3n-1)}\,e^{-i 3n\theta} }{(3n)!\,2^{3n/2}}}^2.
\label{ch4opt3CS}
\end{align}
At time $t=0$, the Eq.~(\ref{ch4opt3CS}) gives the optical tomogram of the state $\ket{\psi_{3,0}}$, which is a structure with three sinusoidal strands in the $X_\theta$-$\theta$ plane (see Fig.~\ref{ch2fig:opt_l34}(a)). In Figs.\ref{ch4fig:opt3CS}(a)-\ref{ch4fig:opt3CS}(d), we plot the optical tomogram given in Eq.~(\ref{ch4opt3CS}) at different instants during the evolution of the state $\ket{\psi_{3,0}}$ in Kerr-like medium. For better resolution of the sinusoidal strands in the optical tomogram, we increase the value of $\modu{\alpha}^2$ to $35$. Figure~\ref{ch4fig:opt3CS}(a) shows the optical tomogram of the time-evolved state  at  $\trev/18$, which corresponds   to two-subpacket fractional revival for the initial state $\ket{\psi_{3,0}}$. It displays  a structure with six  sinusoidal strands, which is a signature of the two-subpacket fractional revival for the initial state $\ket{\psi_{3,0}}$.  The optical tomogram of the time-evolved state for  initial state $\ket{\psi_{3,0}}$ has also been analyzed at higher-order fractional revival times  and we found that, at the instants of $k$-subpacket fractional revivals, the optical tomogram of the time-evolved state for the initial state $\ket{\psi_{3,0}}$ displays a structure with $3k$ sinusoidal strands.
 
Figures~\ref{ch4fig:opt3CS}(b) and \ref{ch4fig:opt3CS}(c) show the optical tomograms of the rotated initial wave packet at times $t=\trev/9$ and $t=2\trev/9$, respectively. The optical tomograms at these instants display a structure with three sinusoidal strands. These optical tomograms are qualitatively different from the optical tomogram of the initial state $\ket{\psi_{3,0}}$.  The optical tomogram of a collapsed state at time $t=\trev/\sqrt{2}$ for initial  state $\ket{\psi_{3,0}}$ is shown in Fig.\ref{ch4fig:opt3CS}(d). The sinusoidal strands are absent in the optical tomogram of the collapsed state. The optical tomogram of the time-evolved state at revival time is shown in Fig.~\ref{ch4fig:opt3CSrevival}.
\begin{figure}[h]
\centering
\includegraphics[scale=0.7]{OT3CSrevival.eps}
\caption{Time-evolved optical tomogram $\omega_{3,0}\left(X_{\theta},\theta,t\right)$  for the initial state $\ket{\psi_{3,0}}$ with $\modu{\alpha}^2=35$ at revival time $t=\trev$.}
\label{ch4fig:opt3CSrevival}
\end{figure}

So far, we have investigated the dynamics of the initial even coherent states of order $2$ and $3$. In the next section we summarize our results for the initial state $\ket{\psi_{l,h}}$ for a general $l$ and $h$.

\section{Conclusion}
\label{conclusion}
We have extended the foregoing analysis for the initial state  $\ket{\psi_{l,h}}$ with a general $l$  and $h$ values and found the following results: 
\begin{enumerate}[i)]
\item For a given $l$ value, the fractional revival time is independent of the $h$ value. 
\item The time-evolved state at $t=j\trev/l^2$, where $j=1$, $2$, $\dots$, $(l^2-1)$, is a rotated initial wave packet.
\item The $k$-subpacket fractional revival occur at  $t=j\,\trev/l^2k$ where $j=1$, $2$, $\dots$, $(l^2k-1)$ for a given value of   $k\, (>1)$ with  $(j,l^2k)=1$.
\item The distinctive signatures of $k$-subpacket fractional revivals are captured in $(lk)^{\rm th}$ moments of the operators $\hat{x}$ and $\hat{p}$.
\item The optical tomogram of the time-evolved at $k$-subpacket fractional revival time shows a structure with ($lk$) sinusoidal strands in the $X_{\theta}$-$\theta$ plane.
\end{enumerate}

\begin{figure}[H]
\centering
\includegraphics[scale=0.7]{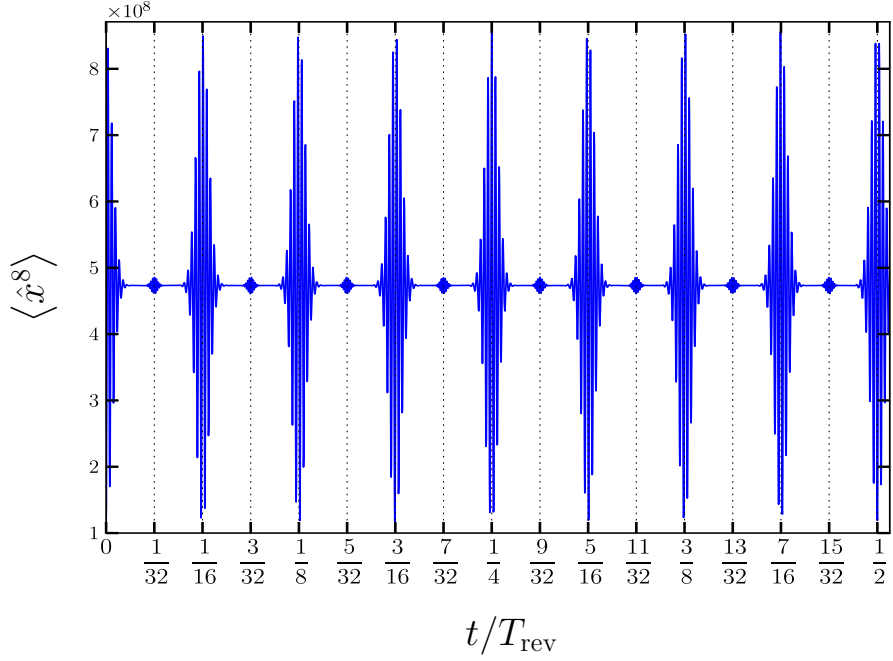} 
\caption{$\aver{\hat{x}^8}$ as a function of $t/T_{\rm rev}$ for the initial state  $\ket{\psi_{4,0}}$ with $\modu{\alpha}^2=100$.  Between $t=0$ and $t=T_{\rm rev}/2$,  $\aver{x^8(t)}$ is a constant most of the time except at   $t=j\,T_{\rm rev}/32$, where $j=1$, $2$, $\dots$, $16$. In this range, the two-subpacket fractional revivals occur at $t/\trev=1/32$, $3/32$, $5/32$, $7/32$, $9/32$, $11/32$, $13/32$, and $15/32$ and its signatures are captured in the eighth moment of $\hat{x}$. In the given range, the eighth moment of $\hat{x}$ also captures the wave packet rotations at $t/\trev=j/16$, where $j=1$, $2$, $\dots$, $8$.}
\label{ch4x8for4N}
\end{figure}
We do not write it down the analysis for higher values of $l$ because it is  repetitive, but for completeness we discuss the dynamics of an initial state  $\ket{\psi_{4,0}}$. According to the result (iii) quoted above, the two-subpacket  fractional revivals of the initial state $\ket{\psi_{4,0}}$ occur at  $t=j\,\trev/32$. This initial state can be written as a superposition of four coherent states  (see Eq.~(\ref{ch2psi_lh})). Indeed, the  initial state $\ket{\psi_{4,0}}$  shows two-subpacket fractional revival at time $t=\trev/32$:
\begin{align*}
 \ket{\psi(\trev/32)}=&N_{4,0} C_1\Big[\ket{\alpha e^{-i 31\pi/32}}+\ket{\alpha e^{-i15\pi/32}}+\ket{\alpha e^{i \pi/32}}+\ket{\alpha e^{i 17\pi/32}}\Big] \nonumber \\
 &+N_{4,0} C_2\Big[\ket{\alpha e^{-i 23\pi/32}}+\ket{\alpha e^{-i7\pi/32}}+\ket{\alpha e^{i 9\pi/32}}+\ket{\alpha e^{i 25\pi/32}}\Big],
 \end{align*}
where $C_1=(1-i)/2$ and $C_2=(1+i)/2$. Figure~\ref{ch4x8for4N}  shows the temporal evolution of the expectation value $\aver{\hat{x}^8}$ for the initial state $\ket{\psi_{4,0}}$. We have plotted till $t=\trev/2$ for a better view, i.e., $j$  runs only up to $16$ instead of $31$ in the result (iii).  It  captures the signature of two-subpacket fractional revival at $t=j\trev/32$ where $j=1$, $2$, $\dots$,  $16$ with $(j, 32)=1$  and rotations at $t=j\trev/16$, where $j=1$, $2$, $\dots$, $8$, which confirms our general result (ii) quoted above. 

The experimental manifestations of our results are  possible using the continuous-variable optical homodyne tomography.   The moments of the operators $\hat{x}$ and $\hat{p}$ can be experimentally measured using the homodyne correlation techniques with a weak local oscillator \citep{Shchukin2005}. It may be possible to measure the R\'{e}nyi entropy using the techniques described in \citep{Daley2012,Abanin2012}. 
\chapter{SIGNATURES OF ENTANGLEMENT IN AN OPTICAL TOMOGRAM} \label{Ch_EntangledOT}
\thispagestyle{plain}
\section{Introduction}\label{ch5introduction}
In the preceding chapters, we have analyzed the optical tomogram of a single-mode electromagnetic field. We have shown that the signatures of a macroscopic superposition state are captured in the optical tomogram of the state, enabling the selective identification of the macroscopic superposition states directly from its optical tomogram. Since the fractional revival phenomenon of a wave packet is associated with the generation of a macroscopic superposition state, these signatures help in the characterization of revivals and fractional revivals of an initial wave packet evolving in a nonlinear medium. For an initial single wave packet, the optical tomogram of the time-evolved state shows a structure with $k$ sinusoidal strands at $k$-subpacket fractional revival time \citep{Rohith2015}. Further, we have shown that the fractional revival time depends on the number of subpackets composing the superposition state \citep{Rohith2014}. For an initial superposed coherent state $\ket{\psi_{l,h}}$,
the optical tomogram of the state shows a structure with $lk$ sinusoidal strands at $k$-subpacket fractional revival time.

In this chapter, we extend our investigations to the optical tomogram of the two-mode states of the electromagnetic field. To be specific, we study the optical tomogram of the two-mode states generated at the output of a beam splitter. Depending upon the nature of the input fields, a beam splitter can generate both separable and entangled two-mode state at the output. A beam splitter generates a separable two-mode state if both the input fields are classical, and it generates an entangled state if one of the input fields is nonclassical \citep{Kim2002}. 
As mentioned in Chapter \ref{Ch_Introduction}, the characterization of the entangled states generated by this process can be done by optical homodyne tomography. Two homodyne detection arrangements, one for each mode, are used to characterize a two-mode entangled state of light. Various types of entangled states have been characterized recently \citep{DAuria2009,Yao2012,Lvovsky2013,Morin2014}.  A conditional measurement on one of the modes of entangled states may change the state in the other mode due to entanglement, and such changes may show up in the optical tomogram of the state.  The main goal of this chapter is to find the signatures of entanglement in the optical tomogram of the state, without reconstructing the density matrix of the state.  For this purpose, we investigate the optical tomograms of maximally entangled coherent states created in a beam splitter. Such an investigation will avoid the computational complexity of finding the two-mode density matrix of the state from its optical tomogram, in order to determine whether the state is entangled or not. In the next section, we discuss the generation of separable as well as entangled two-mode states of light using a beam splitter.


\section{Beam splitting action}
\label{beamsplitting}
\begin{figure}[h]
	\centering
	\includegraphics[scale=0.6]{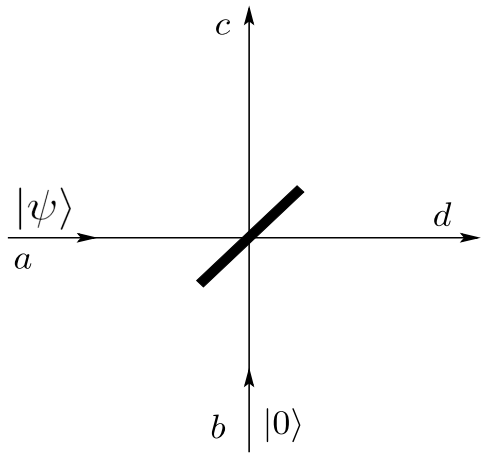} 
   \caption{A $50/50$ beam splitter with $\ket{\psi}$ in the horizontal port and $\ket{0}$ in the vertical port. Here $a$ and $b$ ($c$ and $d$) are the input (output) field modes. }
  \label{ch5beamsplitter}
   \end{figure}

Consider a $50/50$ beam splitter with zero phase difference between reflected and transmitted port. The unitary operator for the beam splitter reads 
\bea
U_{BS}=\exp\left[\frac{\pi}{4}(a^\dag b-a b^\dag)\right],\label{ch5BSoperator}
\eea
where $a$ and $b$ are the bosonic operators for the input field modes. The output field modes of the beam splitter are designated by $c$ and $d$. A schematic diagram of the beam splitter is given in Fig.~\ref{ch5beamsplitter}. 
We consider both classical and nonclassical states  in the horizontal input port (mode $a$) of the beam splitter and study the optical tomogram of the output states. In both of these cases, we take vacuum state $\ket{0}$ in the vertical input port (mode $b$) of the beam splitter. The states considered for classical and nonclassical states are  coherent state, and  even and odd coherent states, respectively.
Beam splitting action on the coherent state $\ket{\psi}=\ket{\alpha}$, where $\alpha$ ($=\modu{\alpha}e^{i\delta}$) is a complex number, with vacuum $\ket{0}$ generates the separable state 
\bea
\ket{\Phi}_{ss}=\ket{\beta}_c\otimes\ket{\beta}_d,
\eea
where $\beta=\alpha/\sqrt{2}$. Next, we consider beam splitting of even and odd coherent states $\ket{\psi_{2,h}}$ with $h=0$ and $h=1$ (obtained by setting $l=2$ in Eq.~(\ref{ch2psi_lh})), respectively. In this case, we get entangled states at the output modes of the beam splitter. The state $\ket{\Phi}_{h}$ of the beam splitter output modes is calculated using the unitary operator given in Eq.~(\ref{ch5BSoperator}):
\bea
\ket{\Phi}_{h}=N_{2,h}\left[\ket{\beta}_c\ket{\beta }_d+e^{i\pi h}\ket{-\beta}_c\ket{-\beta }_d\right].\label{ch5entstate}
\eea
The entanglement of the state $\ket{\Phi}_{h}$ can easily be calculated using  the von Neumann entropy
\begin{equation}
E=-Tr\left[\rho_k \log \rho_k\right],\label{ch5VNE}
\end{equation}
where $\rho_k$ ($k$ = mode $c$ or $d$) is the reduced density matrix of either of the subsystems $c$ or $d$. For the state $\ket{\Phi}_{h}$, the reduced density matrix $\rho_k$ is obtained as
\begin{align}
\rho_k=N_{2,h}^2\sum_{r,r^\prime=0}^{1} \exp\left\{-i\,\pi (r-r^\prime) h-\modu{\beta}^2\left[1-e^{i\pi (r-r^\prime)}\right]\right\}{\ket{\beta_r}_{k}}\, _{k}\bra{\beta_{r^\prime}},
\label{ch5reduceddensitymatrix}
\end{align}
where $\beta_r=\beta\,e^{i\pi r}$ and $\beta_{r^\prime}=\beta\,e^{i\pi r^\prime}$. Using Eqs.~(\ref{ch5VNE}) and (\ref{ch5reduceddensitymatrix}), we have numerically evaluated the entanglement of the state $\ket{\Phi}_{h}$ as a function of $\modu{\alpha}^2$.

\begin{figure}[h]
   \centering
   \includegraphics[scale=0.45]{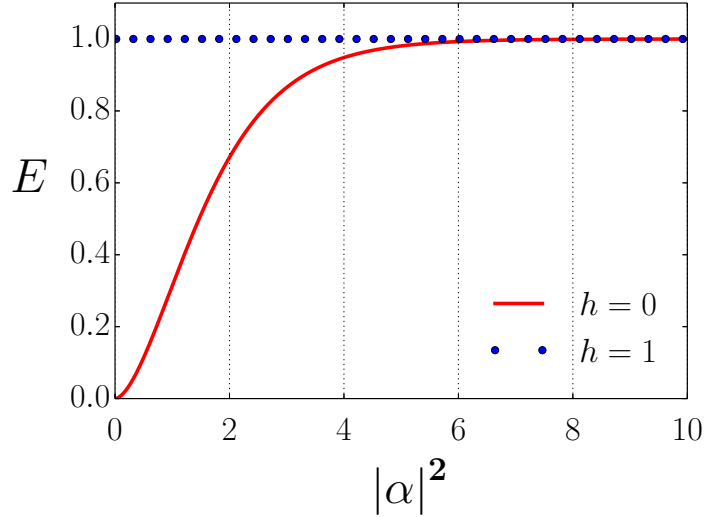}
   \caption{Variation of entanglement of the state $\ket{\Phi}_{h}$ as a function of $\modu{\alpha}^2$.} 
	\label{ch5fig:entanglement}
   \end{figure} 
Figure \ref{ch5fig:entanglement} shows the variation of entanglement of the state $\ket{\Phi}_{h}$ as a function of $\modu{\alpha}^2$. With increase in $\modu{\alpha}^2$ value, the entanglement of the state $\ket{\Phi}_{0}$ increases from zero and saturates (say from $\modu{\alpha}^2=6$ onwards) to unity for large $\modu{\alpha}^2$, whereas the entanglement of the state $\ket{\Phi}_{1}$ is always unity for any  value of $\modu{\alpha}^2>0$. In the limit of large $\modu{\alpha}^2$, the coherent states $\ket{\alpha}$ and $\ket{-\alpha}$ composing the superposition state $\ket{\psi_{2,h}}$ form an orthogonal basis and thus the entangled state, given in Eq.~(\ref{ch5entstate}), is already in the Schmidt decomposition form \citep{vanEnk2003}. Since the two Schmidt coefficients have the equal magnitude, $1/\sqrt{2}$, the entanglement of the state $\ket{\Phi}_{h}$ can be found to be $E=\log_2 (2)=1$ ebits, which is the maximum entanglement possible in two dimensions (2D). Hence, for large $\modu{\alpha}^2$, the state $\ket{\Phi}_{h}$ is a maximally entangled state in 2D. It should be noticed that the entanglement of the output state, given in Eq.~(\ref{ch5entstate}), is  a measure of the nonclassicality of the input state $\ket{\psi_{2,h}}$ \citep{Asboth2005,Miranowicz2015}.

\section{Optical tomogram of the two-mode state}
\label{OpticalTomography}
In this section, we  calculate  the optical tomograms of the separable state $\ket{\Phi}_{ss}$ and the maximally entangled state $\ket{\Phi}_{h}$. For a two-mode state with density matrix $\rho$, the optical tomogram is given by \citep{Amosov2012}
\begin{align}
\omega\left(X_{\theta_1},\theta _1; X_{\theta_2},\theta _2\right)=\bra{X_{\theta_1},\theta_1}\bra{X_{\theta_2},\theta_2}\rho\ket{X_{\theta_2},\theta_2}\ket{X_{\theta_1},\theta_1},\label{ch52modeOTdef}
\end{align}
where $\ket{X_{\theta_i},{\theta}_i}$, with $i=1$ and $2$, is the eigenvector of the Hermitian operator $\hat{X}_{\theta_i}$ (defined in Eq.~(\ref{ch2QuadratureOperator})) with eigenvalue $X_{\theta_i}$. The quantities $X_{\theta_1}$ and $\theta_1$ ($X_{\theta_2}$  and $\theta_2$) are the quadrature and the phase of local oscillator in homodyne detection setup for mode $c$ (mode $d$), respectively. The phase of the local oscillators varies in the domain $0\leq \theta_1,\theta_2 \leq 2\pi$. For a pure two-mode state $\ket{\psi}$, the Eq.~(\ref{ch52modeOTdef}) can be rewritten as
\begin{align}
\omega\left(X_{\theta_1},\theta _1; X_{\theta_2},\theta _2\right)=\modu{\bra{\psi}\ket{X_{\theta_1},\theta_1}\ket{X_{\theta_2},\theta_2}}^2.\label{ch52modeOTdefAlternate}
\end{align}
The optical tomograms of the two-mode coherent states of charged particle moving in a varying magnetic field have been investigated \citep{Manko2012b}. 

Substituting $\ket{\Phi}_{ss}$ in Eq.~(\ref{ch52modeOTdefAlternate}), we get the optical tomogram of the two-mode separable state as 
\begin{align}
\omega_{ss}(X_{\theta_1},\theta _1;X_{\theta_2},\theta _2)=\omega_1(X_{\theta_1},\theta _1)\times\omega_2(X_{\theta_2},\theta _2),
\end{align}
where $\omega_1(X_{\theta_1},\theta _1)$, $\omega_2(X_{\theta_2},\theta _2)$ are the optical tomogram of the coherent state $\ket{\beta}$ in mode $c$ and mode $d$, respectively. These optical tomograms can be written as (using Eq.~(\ref{ch2OT_CS}))
\begin{align}
\omega_i(X_{\theta_i},{\theta_i})=\frac{1}{\sqrt{\pi}}\exp\left[-\left(X_{\theta_i}-\sqrt{2}\modu{\beta}\cos(\delta-{\theta_i})\right)^2\right].\label{ch5subsystemtomo}
\end{align}
Therefore, the optical tomogram of a separable state can be written as the product of optical tomograms of the subsystems \citep{Ibort2009}. We recall from Chapter~\ref{ch_OT_PSI_lh} that, the optical tomogram of a coherent state is a structure with single sinusoidal strand. 
Hence, the optical tomogram in mode $c$ (mode $d$) will always be a structure with a single sinusoidal strand for any values of parameters in mode $d$  (mode $c$) and $\delta$. 

Again, by substituting Eq.~(\ref{ch5entstate}) in Eq.~(\ref{ch52modeOTdefAlternate}), the two-mode optical tomogram for the entangled state $\ket{\Phi}_{h}$ is obtained as
\begin{align}
\omega_{h}(X_{\theta_1},\theta _1;X_{\theta_2},\theta _2)=\frac{N^2_{2,h}}{\pi}\modu{\sum_{r=0}^{1}e^{-i\pi r h} ~\eta(X_{\theta_1},\theta_1,\beta_r)\eta(X_{\theta_2},\theta_2,\beta_r)}^2, \label{ch5tomogram}
\end{align}
where
\begin{align}
\eta(X_{\theta_i},\theta_i,\beta_r)=\exp\left(-\frac{\modu{\beta_r}^2}{2}-\frac{{X_{\theta_i}}^2}{2}+\sqrt{2}\beta_r X_{\theta_i} e^{-i{\theta_i}}- \frac{\beta^2_r e^{-i{2\theta_i}}}{2}\right).\label{ch5eta}
\end{align}
It is clear from the Eq.~(\ref{ch5tomogram}) that, the optical tomogram of the entangled state $\ket{\Phi}_h$ cannot be written in the form of the product of subsystem tomograms, that is 
\begin{align}
\omega_{h}(X_{\theta_1},\theta _1;X_{\theta_2},\theta _2)\neq \omega_1(X_{\theta_1},\theta _1)\times\omega_2(X_{\theta_2},\theta _2).
\end{align}

\section{Signature of entanglement}
\begin{figure}[H]
   \centering
	\includegraphics[height=6 cm,width=\textwidth]{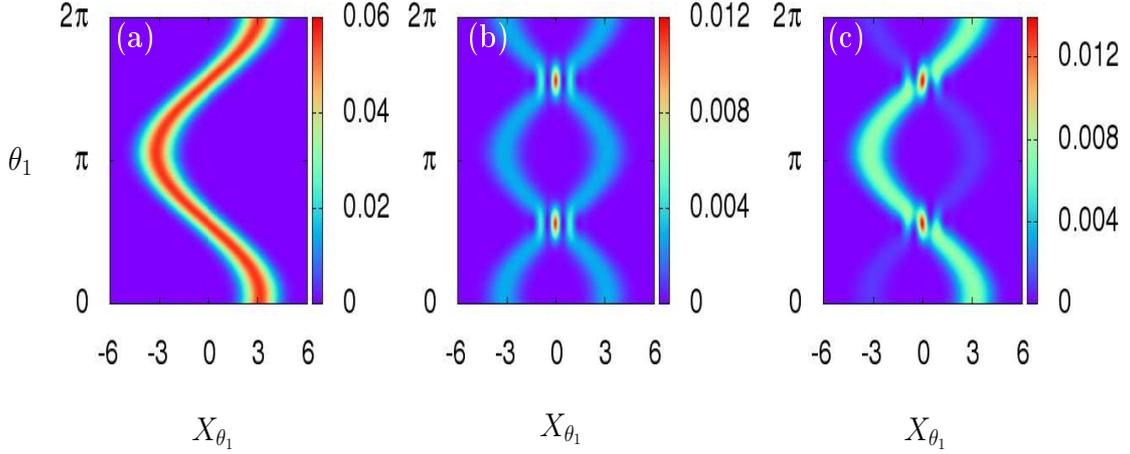}
   \caption {Optical tomograms $\omega_{0}(X_{\theta_1},\theta _1;X_{\theta_2},\theta _2)$ in mode $c$ for the entangled state $\ket{\Phi}_0$ with $\modu{\alpha}^2=10$, $\delta=0.2$, and $X_{\theta_2}=2.0$, for different relative phases $\modu{\delta-\theta_2}$ of the quadrature measurement in mode $d$: (a) $0.3$, (b) $\pi/2$  and (c)  $\pi/2-\pi/40$. The optical tomogram  shows  a sinusoidal single-stranded structure for $\modu{\delta-\theta_2}=0.3$. The optical tomogram shows sinusoidal double-stranded structure for $\modu{\delta-\theta_2}=\pi/2$. The optical tomogram for $\modu{\delta-\theta_2}=3\pi/2$ is exactly the  same as in the case of $\modu{\delta-\theta_2}=\pi/2$.  The plot in the last column is an optical tomogram of an intermediate case which shows the transition of the single-stranded structure to the double-stranded structure.} 
\label{ch5fig:l2tomo}
   \end{figure}

In this section, we analyze the optical tomogram of the entangled state  $\ket{\Phi}_{h}$  in detail, and look for the signatures of entanglement in the optical tomogram of the state. In the following, we present our analysis for initial even coherent state $\ket{\psi_{2,0}}$. For initial even coherent state, the entangled output state $\ket{\Phi}_{0}=N_{0}\left[ \ket{\beta }\ket{\beta}+\ket{-\beta }\ket{-\beta}\right]$. A measurement of field quadrature in any of the modes will collapse the entanglement between the modes. The quantum state in one mode is correlated with the quadrature measurement in the other mode. For example, a measurement of quadrature $\hat{X}_{\theta_2}$ in mode $d$ will project the state $\ket{\Phi}_{0}$ to the state $\ket{\phi}_{0,c}$ in mode $c$:
\bea
\ket{\phi}_{0,c}=\tilde{N}_{0}\left[\psi_{\beta}(X_{\theta_2},\theta_2)\ket{\beta}+\psi_{-\beta}(X_{\theta_2},\theta_2)\ket{-\beta}\right],\label{ch5phi_c}
\eea
where $\psi_{\pm \beta}(X_{\theta_2},\theta_2)=\langle X_{\theta_2},\theta_2 \ket{\pm \beta}$ is the quadrature representation of the coherent state $\ket{\pm \beta}$, and 
\begin{align}
\tilde{N}_{0}=&\left\{\modu{\psi_{\beta}(X_{\theta_2},\theta_2)}^2+\modu{\psi_{-\beta}(X_{\theta_2},\theta_2)}^2\right.\nonumber\\
&\left.+e^{-2\modu{\beta}^2}\left[\psi_{\beta}^{\ast}(X_{\theta_2},\theta_2)\psi_{-\beta}(X_{\theta_2},\theta_2)+\psi_{\beta}(X_{\theta_2},\theta_2)\psi_{-\beta}^{\ast}(X_{\theta_2},\theta_2)\right]\right\}^{-1/2}.
\end{align}
Based on the relative strength of the coefficients $\psi_{\beta}(X_{\theta_2},\theta_2)$ and $\psi_{-\beta}(X_{\theta_2},\theta_2)$,  the state $\ket{\phi}_{0,c}$ can be one of following: $\ket{\beta}$,  $\ket{-\beta}$ and  a superposition of $\ket{\beta}$ and  $\ket{-\beta}$. The probability for occurring the state $\ket{\pm \beta}$ is proportional to 
\begin{align}
\modu{\psi_{\pm \beta}(X_{\theta_2},\theta_2)}^2 =&\frac{1}{\sqrt{\pi}}\exp\left[-2\modu{\beta}^2\cos^2 (\delta-\theta_2)-X^2_{\theta_2} \pm 2\sqrt{2}X_{\theta_2}\modu{\beta}\cos (\delta-\theta_2) \right].
\label{ch5psi_beta}
\end{align}
For $X_{\theta_2}\neq 0$, relative strength of the probabilities crucially depend on the last term in Eq.~(\ref{ch5psi_beta}). In the range $0\leq \modu{\delta-\theta_2}< \pi/2$ and $3\pi/2< \modu{\delta-\theta_2}\leq 2\pi$, the state $\ket{\phi}_{0,c}$ can be approximated to the coherent state $\ket{\beta}$ because  $\modu{\psi_{\beta}(X_{\theta_2},\theta_2)}^2 \gg \modu{\psi_{-\beta}(X_{\theta_2},\theta_2)}^2$, which gives a structure with a single strand for the optical tomogram in mode $c$ (Note that the optical tomogram for a coherent state $\ket{\beta}$ in $X_{\theta_i}$-$\theta_i$ plane is a structure with single strand in Eq.~(\ref{ch5subsystemtomo})). Also, in the range $\pi/2< \modu{\delta-\theta_2} < 3\pi/2$, the state $\ket{\phi}_{0,c}$ can be approximated to the coherent state $\ket{-\beta}$ because $\modu{\psi_{-\beta}(X_{\theta_2},\theta_2)}^2 \gg \modu{\psi_{\beta}(X_{\theta_2},\theta_2)}^2$.  Thus, the optical tomogram in mode $c$ will be a structure with a single strand corresponds to the coherent state $\ket{-\beta}$. Figure \ref{ch5fig:l2tomo}(a) shows single-stranded structure in the optical tomogram $\omega_{0}(X_{\theta_1},\theta _1;X_{\theta_2},\theta _2)$ for  $\modu{\alpha}^2=10$, $\delta=0.2$,  $X_{\theta_2}=2.0$ and $\modu{\delta-\theta_2}=0.3$.

 It can be shown that $\modu{\psi_\beta(X_{\theta_2},\theta_2)}^2=\modu{\psi_{-\beta}(X_{\theta_2},\theta_2)}^2$ for $\modu{\delta-\theta_2}=\pi/2$ or $3\pi/2$ (within the periodicity of optical tomogram). For these two values of $\modu{\delta-\theta_2}$, the probability for occurring $\ket{\beta}$ and $\ket{-\beta}$ in mode $c$ is $50:50$, and hence the state $\ket{\phi}_{0,c}$ reduces to even coherent state of the form $\left[\ket{\beta}+\ket{-\beta}\right]$. The optical tomogram in mode $c$ will display a double-stranded structure, in which, one strand corresponds to $\ket{\beta}$ and the other corresponds to $\ket{-\beta}$. The optical tomograms for $\modu{\delta-\theta_2}=\pi/2$ and  $3\pi/2$ are exactly the same because the state $\ket{\phi}_{0,c}$ reduces to same even coherent state  for both of these values.   The optical tomogram in mode $c$ for $\modu{\delta-\theta_2}=\pi/2$ with $X_{\theta_2}=2.0$ is shown in Fig.~\ref{ch5fig:l2tomo}(b). Quantum interference between the state $\ket{\beta}$ and $\ket{-\beta}$ are reflected in the optical tomogram at regions in $X_{\theta_1}$-$\theta_1$ plane, where the two strands intersect.   The state $\ket{\phi}_{0,c}$ is neither a coherent state nor an even coherent state in the vicinity of $\modu{\delta-\theta_2}=\pi/2$ or $3\pi/2$.  The optical tomogram of such intermediate state, corresponding to $\modu{\delta-\theta_2}=\pi/2-\pi/40$, is shown in Fig.~\ref{ch5fig:l2tomo}(c). In the next section, we analyze the Mandel's $Q$ parameter of the state $\ket{\phi}_{0,c}$ to understand this transition.
When $X_{\theta_2}=0$, the state $\ket{\phi}_{0,c}$ is always an even coherent state, without any condition on  $\modu{\delta-\theta_2}$. This displays a structure with two sinusoidal strands in the optical tomogram.

\section{Mandel's $Q$ parameter}
\label{mandel}
\begin{figure}[H]
      \centering
	\includegraphics[height=6 cm,width=8 cm]{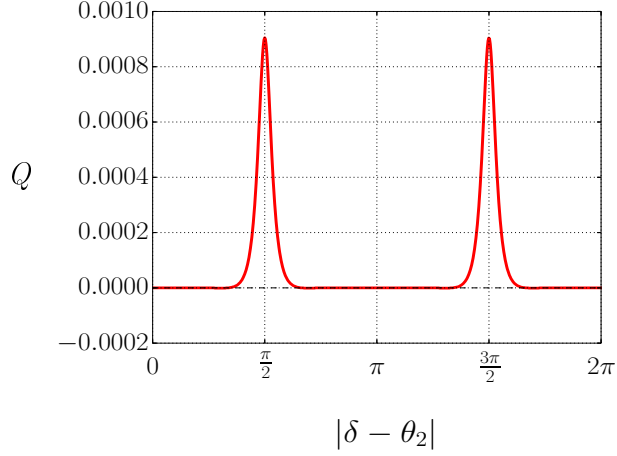}   
   \caption{Mandel's $Q$ parameter for the state $\ket{\phi}_{0,c}$ as a function of the relative phase $\modu{\delta-\theta_2}$ of the measurement in mode $d$ with $\modu{\alpha}^2=10$ and $X_{\theta_2}=2$. The positive value of $Q$ at $\modu{\delta-\theta_2}=\pi/2$ and $3\pi/2$ (and its vicinity) indicates the super-Poissonian statistics of the state $\ket{\phi}_{0,c}$ and for all other values (except for $\pi/2$ or $3\pi/2$ and its vicinity) of $\modu{\delta-\theta_2}$, the state $\ket{\phi}_{0,c}$ exhibit Poissonian statistics ($Q=0$).} 
\label{ch5fig:mandel}
   \end{figure}

We can also show the above features using the statistics of photon number in the state $\ket{\phi}_{0,c}$, specifically, in terms of the Mandel's $Q$ parameter, defined as \citep{Mandel1979}
\bea
Q=\frac{\aver{\hat{n}^2}-\aver{\hat{n}}^2}{\aver{\hat{n}}}-1,
\eea
where $\hat{n}$ is the photon number operator. A positive value of $Q$ indicates the super-Poissonian statistics of the field, and $Q=0$ indicates the Poissonian statistics exhibited by a coherent field. The $Q$ parameter of the state $\ket{\phi}_{0,c}$  is calculated as
\begin{equation}
Q=\frac{2\modu{\alpha}^2\,e^{-\modu{\alpha}^2}\,\cosh\left(2\modu{\alpha}X_{\theta_2}\cos(\delta-\theta_2)\right)\,\cos\left(2\modu{\alpha}X_{\theta_2}\sin(\delta-\theta_2)\right)}{\cosh^2\left(2\modu{\alpha}X_{\theta_2}\cos(\delta-\theta_2)\right)-4\,e^{-2\modu{\alpha}^2}\,\cos^2\left(2\modu{\alpha}X_{\theta_2}\sin(\delta-\theta_2)\right)}.
\end{equation}
Mandel's $Q$ parameter of the state $\ket{\phi}_{0,c}$ in mode $c$ as a function of relative phase $\modu{\delta-\theta_2}$ is plotted in Fig.~\ref{ch5fig:mandel}. It shows that at $\modu{\delta-\theta_2}=\pi/2$, $3\pi/2$ and its vicinity, the state exhibits super-Poissonian statistics and for all other $\modu{\delta-\theta_2}$ values, the state $\ket{\phi}_{0,c}$ shows Poissonian statistics corresponding to a coherent field, which is either $\ket{\beta}$ or $\ket{-\beta}$. 

The range of $\modu{\delta-\theta_2}$ values for which the intermediate states exist can be found using the full width at half maximum (FWHM) of the two peaks in the plot of $Q$ versus $\modu{\delta-\theta_2}$. The  FWHM of the peaks depends on the values of  $X_{\theta_2}$ and $\modu{\beta}^2$ and it is  $0.21$ radians for each  peak in the Fig.~\ref{ch5fig:mandel}.  The single-mode $Q$ parameter described here is sufficient  to verify our findings on the  optical tomograms, but it should be noted that two-mode normally ordered variances can also be studied to understand the super- or sub-Poisson photon-number correlations \citep{Miranowicz2010}. The different statistics of photon number exhibited by the state in mode $c$ upon changing the parameters in mode $d$, has been experimentally observed in the case of micro-macro entanglement of light \citep{Lvovsky2013} using the reconstructed density matrix. We have theoretically shown that, without reconstructing the density matrix of the system, the signature of entanglement can be directly observed in  the optical tomogram of the state.

\begin{figure}[H]
	\includegraphics[height=6 cm,width=\textwidth]{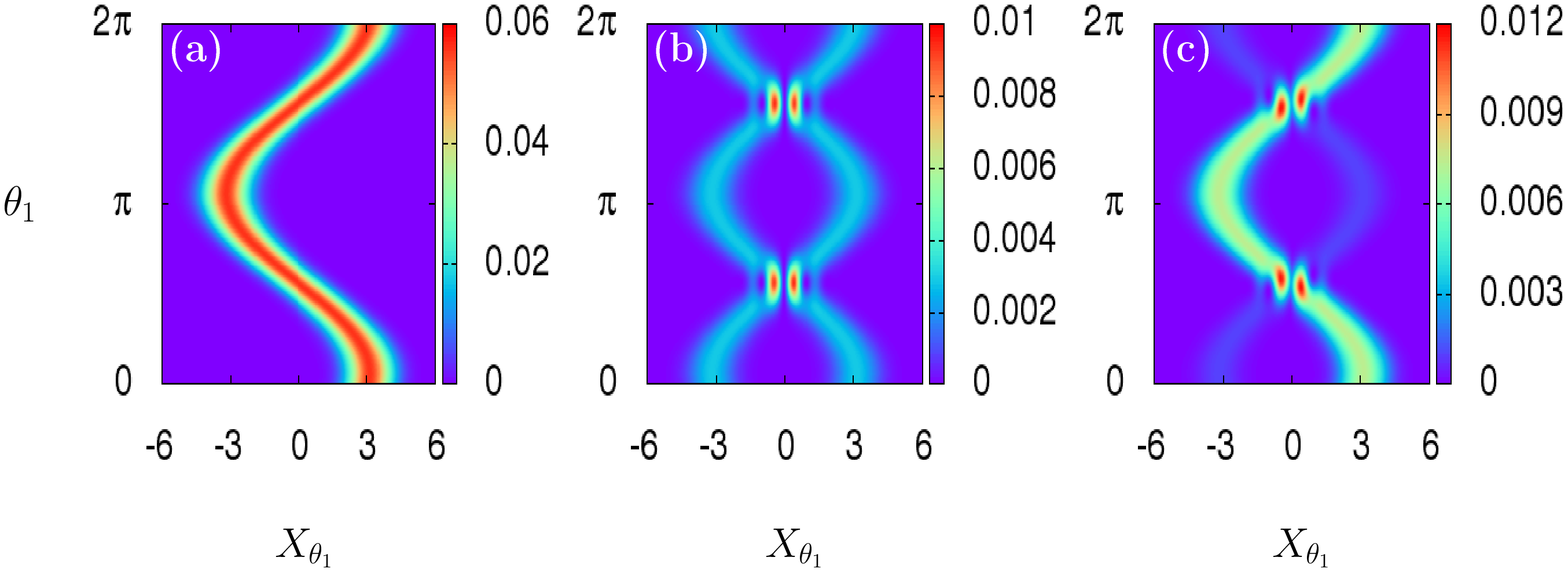}
   \caption{Same as in Fig.~\ref{ch5fig:l2tomo} but for $\ket{\Phi}_{1}$.} 
	\label{ch5fig:l2h1tomo}
   \end{figure}

When the initial state is an  odd coherent state (i.e., $h=1$), we get the entangled state $\ket{\Phi}_{1}$ given in Eq.~(\ref{ch5entstate}) at the  output modes of the beam splitter.  We have repeated the forgoing analysis for the entangled state $\ket{\Phi}_{1}$ and verified the results obtained earlier. 
The Figs.~\ref{ch5fig:l2h1tomo}(a)-\ref{ch5fig:l2h1tomo}(c) show the optical tomograms   $\omega_{1}(X_{\theta_1},\theta _1;X_{\theta_2},\theta _2)$  in mode $c$ for the entangled state $\ket{\Phi}_{1}$ with same set of parameters used in the case of the entangled state $\ket{\Phi}_{0}$. The forgoing analysis uses only a single-mode optical tomogram to study the signatures of entanglement in a two-mode system. This procedure reduces the number of homodyne measurements to be performed to determine whether the state $\ket{\Phi}_h$ is  entangled or not.
  
  \section{Robustness of the entangled states}
\label{decoherene}
In this section,  we study the effect of environment-induced decoherence on the optical tomogram of the entangled state $\ket{\Phi}_h$ generated at the output of the beam splitter.  We consider the two-mode extension of the amplitude decay and phase damping models of decoherence described in Section \ref{Tomogramdecoherence}. We assume that both of the output modes of the beam splitter interact independently with the external environment consisting of an infinite number of harmonic oscillators that are initially in the vacuum state. Such decoherence process can be effectively described by the quantum mechanical master equation for the two-mode density matrix of the system.

Let $\rho_{cd}$ be the density  matrix of bipartite field modes at the output of the beam splitter. The  interaction with the external environment leaves the system in a mixed state, and we use the  logarithmic negativity \citep{Vidal2002} to quantify the entanglement. The logarithmic negativity is defined as
\begin{equation}
E_N=\log_2 \parallel \rho_{cd}^{T_k} \parallel,
\end{equation}
where $\parallel \cdot \parallel$ denotes the trace norm operation, which  is equal to the sum of the absolute values of eigenvalues  for a Hermitian operator, and $\rho^{T_k}$ with $k=1$ (or $2)$ represents the partial transpose of $\rho_{cd}$ with respect to mode $c$  (or $d)$. In the subsequent sections, we discuss the effect of amplitude decay and phase damping models of decoherence on the optical tomogram of the entangled state $\ket{\Phi}_h$.

\subsection{Amplitude decay model}
In the  rotating wave approximation, the amplitude decay of the state $\ket{\Phi}_h$ due to photon absorption can be modelled by the interaction Hamiltonian 
\begin{align}
H_{\rm amp}^{(2)}=\sum_{j=0}^{\infty}\sum_{s=1}^{2}\gamma_s\left(a_s^\dag e_j+e_j^\dag a_s\right),
\end{align}
where  $\gamma_s$ is the coupling strength of the mode $a_s$ ($a_1$ corresponds to mode $c$ and $a_2$ corresponds to mode $d$)  with the environment modes $e_j$. The Markovian dynamics of the state $\rho_{cd}$ obeys the zero-temperature master equation \citep{Gardiner1991}
	\begin{equation}
		\pdv{\rho_{cd}}{\tau}=\sum_{s=1}^{2}\gamma_s \left(2\, a_s \,\rho_{cd}\, a_s^\dag-a_s^\dag\, a_s\, \rho_{cd} -\rho_{cd}\, a_s^\dag\, a_s  \right), \label{ch5master}
	\end{equation}
where $\tau$ is the time. Without loss of generality, we choose the coupling constants $\gamma_1=\gamma_2=\gamma=0.01$. Using the procedure described in \citep{Chaturvedi1991}, the solution  $\rho_{cd}(\tau)$ of the Eq.~(\ref{ch5master})  can be calculated  as 
\begin{align}
\rho_{cd}(\tau)=&N^2_{2,h}\sum_{r,r^\prime=0}^{1} e^{-i\pi h(r-r^\prime)}\exp\left[-2\modu{\beta}^2\left(1-e^{i\pi (r-r^\prime)}\right)\left(1-e^{-2\gamma \tau}\right)\right]\nonumber\\
&\times \ket{\beta_r\, e^{-\gamma \tau}}_{c} {\ket{\beta_{r^\prime}\, e^{-\gamma \tau}}_{d}}\,\,{_{c}\bra{\beta_r\, e^{-\gamma \tau}} _{d}\bra{\beta_{r^\prime}\, e^{-\gamma \tau}}}\label{ch5rho_cd(t)}.
\end{align}
The state $\rho_{cd}(\tau)$ is a mixed state for all the time $\tau >0$, and we numerically evaluate the logarithmic negativity $E_N$ of the state as a function of time $\tau$. The important steps for calculating the logarithmic negativity $E_N$ of the state $\rho_{cd}(\tau)$ are described in Appendix \ref{Appendix_signatureEntanglement}. Figure \ref{ch5fig:EntanglementDecay} shows the variation of entanglement $E_N$ of the state $\ket{\Phi}_h$ as a function of scaled time $\gamma \tau$ for $\modu{\alpha}^2=10$. It is interesting to see that the entanglement of both the states $\ket{\Phi}_0$ and $\ket{\Phi}_1$ decay in the same manner. The entanglement of the state $\ket{\Phi}_h$ vanishes to zero for large $\gamma \tau$.
\begin{figure}[H]
\centering
\includegraphics[height=7 cm,width=8 cm]{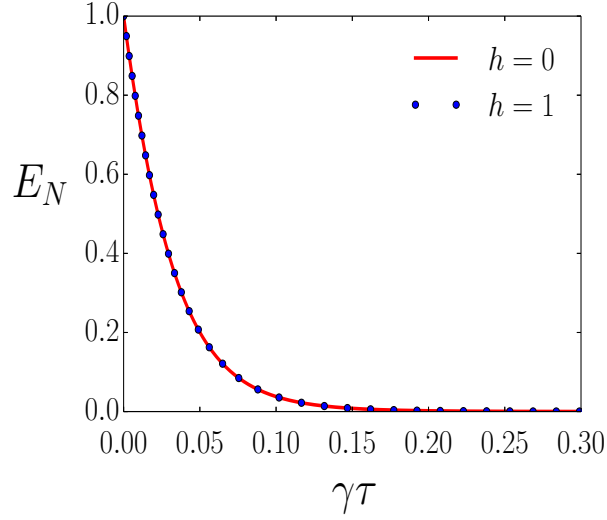}
\caption{Decay of entanglement of the state $\ket{\Phi}_h$ as a function of scaled time $\gamma \tau$ for $\modu{\alpha}^2=10$.}
\label{ch5fig:EntanglementDecay}
\end{figure}

The optical tomogram of the decohered state $\rho_{cd}(\tau)$, given in Eq.~(\ref{ch5rho_cd(t)}),  is obtained as 
\begin{align}
\omega_{h}\left(X_{\theta_1},\theta _1; X_{\theta_2},\theta _2;\tau\right)=&N^2_{2,h}\sum_{r,r^\prime=0}^{1} e^{i\pi h(r-r^\prime)}\exp\left[-2\modu{\beta}^2\left(1-e^{i\pi (r-r^\prime)}\right)\left(1-e^{-2\gamma \tau}\right)\right]\nonumber\\
&\times \zeta(X_{\theta_1},\theta_1,\beta_r,\tau)\zeta^\ast (X_{\theta_1},
\theta_1,\beta_{r^\prime},\tau)\nonumber\\
&\times\zeta(X_{\theta_2},\theta_2,\beta_r,\tau)\zeta^\ast (X_{\theta_2},\theta_2,\beta_{r^\prime},\tau),
\end{align}
where the quantity $\zeta$ is defined in Eq.~(\ref{ch2opt_amplitude_factor}).
In the following, we analyze the optical tomograms of the states in mode $c$ for different quadrature measurements in mode $d$ in the presence of amplitude damping. Figure \ref{ch5fig:l2h0decay} displays the optical tomograms $\omega_{0}(X_{\theta_1},\theta _1;X_{\theta_2},\theta _2;\tau)$ in mode $c$ for the entangled state $\ket{\Phi}_0$ at different times $\gamma \tau$ for the relative phases of measurements (a) $\modu{\delta-\theta_2}=0.3$, (b) $\modu{\delta-\theta_2}=\pi/2$, and (c) $\modu{\delta-\theta_2}=\pi/2-\pi/40$.

Recall that, the state in mode $c$ is the coherent state $\ket{\beta}$ for the quadrature measurement with relative phase $\modu{\delta-\theta_2}=0.3$. 
During the decoherence process, the coherent state remains as a coherent state with exponentially decreasing amplitude. The reduction in the amplitude of the coherent state $\ket{\beta}$ is clearly observed in Fig.~{\ref{ch5fig:l2h0decay}(a), where the sinusoidal strand shrinks in the horizontal direction (along the $X_{\theta_1}$ axis), and finally becomes a single straight strand for very large $\gamma \tau$. 
\begin{figure}[H]
   \centering
	\includegraphics[height=12 cm,width=\textwidth]{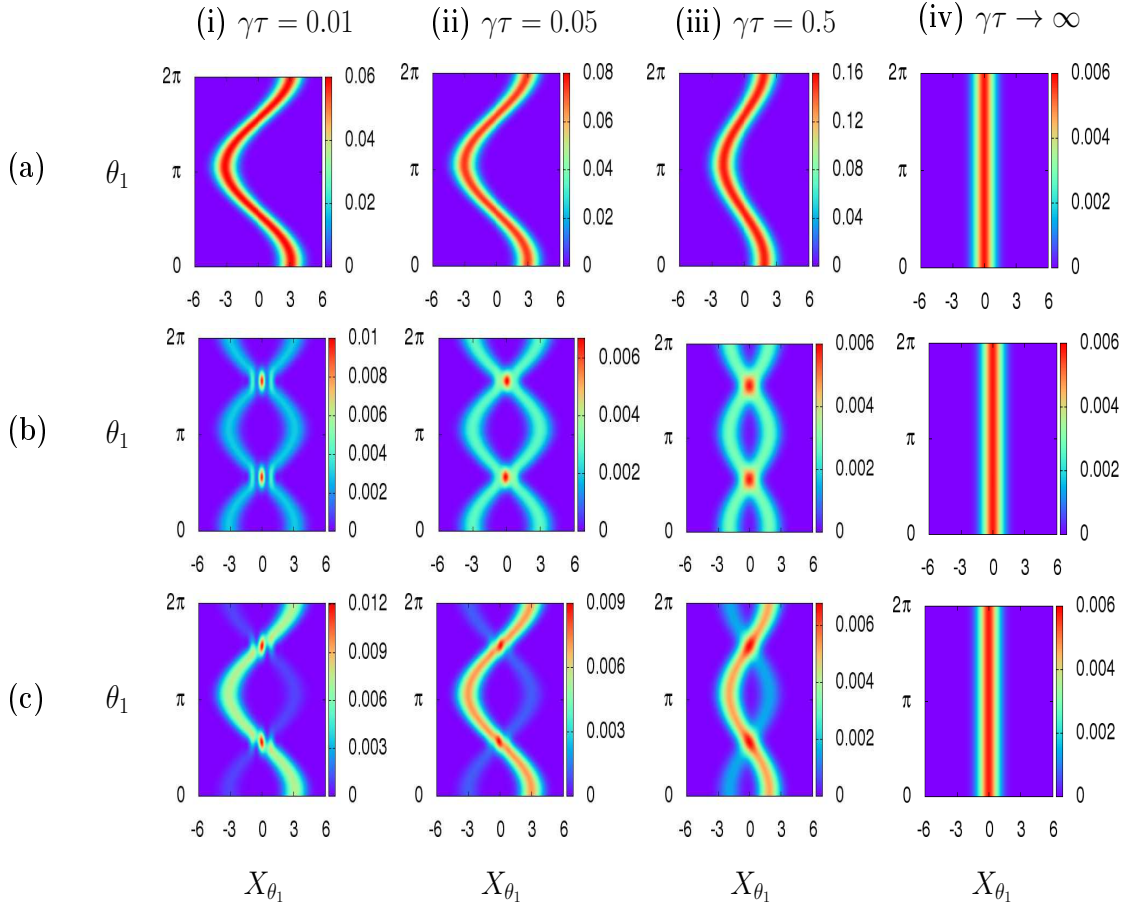}
   \caption {Optical tomograms $\omega_{0}(X_{\theta_1},\theta _1;X_{\theta_2},\theta _2;\tau)$ in mode $c$ for the entangled state $\ket{\Phi}_0$ in the presence of amplitude damping at times (i) $\gamma \tau=0.01$, (ii) $\gamma \tau=0.05$, (iii) $\gamma \tau=0.5$,  and (iv)  $\gamma \tau\rightarrow\infty$, for the relative phases of measurements (a)$\modu{\delta-\theta_2}=0.3$, (b) $\modu{\delta-\theta_2}=\pi/2$, and (c) $\modu{\delta-\theta_2}=\pi/2-\pi/40$. Here,  $\modu{\alpha}^2=10$, $\delta=0.2$ and $X_{\theta_2}=2.0$.} 
 \label{ch5fig:l2h0decay}  
   \end{figure}

Figure \ref{ch5fig:l2h0decay}(b) shows that the two sinusoidal strands of the even coherent state in the optical tomogram get close together with an increase in time and they merge for large $\gamma \tau$ \citep{Rohith2016a}. The merging of the sinusoidal strands with increase in  $\gamma \tau$ is due to the exponential decrease in the amplitude of the coherent states contributing to the superposition state $\left[\ket{\beta}+\ket{-\beta}\right]$ in mode $c$ for  $\modu{\delta-\theta_2}=\pi/2$. In Fig.~\ref{ch5fig:l2h0decay}(c), we plot the optical tomogram of the intermediate state for $\modu{\delta-\theta_2}=\pi/2-\pi/40$ at different times $\gamma \tau$. In this case, the two strands in the optical tomogram, one with high intensity and other with low intensity, merge with an increase  in the time $\gamma \tau$. For large $\gamma \tau$, the state $\ket{\Phi}_h$ reduces to the two-mode vacuum state $\ket{0}_c\ket{0}_d$, and the corresponding optical tomogram is given by
\begin{equation}
\omega_{h}(X_{\theta_1},\theta _1;X_{\theta_2},\theta _2;\tau\rightarrow\infty)=\frac{1}{\pi} e^{-X^2_{\theta_1}-X^2_{\theta_2}}.
\end{equation}
This optical tomogram is independent of the values of $\theta_1$ and $\theta_2$. Therefore, in the long-time limit, the optical tomogram in mode $c$ is a structure with single straight strand in the $X_{\theta_1}$-$\theta_1$ plane irrespective of the value of $\modu{\delta-\theta_2}$, which can be seen in the last column of the Fig.~\ref{ch5fig:l2h0decay}. We have repeated the above analysis for the entangled state $\ket{\Phi}_1$ and found similar results.

\subsection{Phase damping model}
In this model, the interaction of the  output modes $c$ and $d$ of the beam splitter with the environment modes $e_j$  can be described by the Hamiltonian 
\begin{align}
H_{\rm ph}^{(2)}=\sum_{j=0}^{\infty}\sum_{s=1}^{2}\kappa_s\left(A_s^\dag e_j+e_j^\dag A_s\right),
\end{align}
where  $A_s=a_s^\dag a_s$ and $\kappa_s$ is the coupling strength of the mode $a_s$ (mode $c$ or $d$)  with the environment. The Markovian dynamics of the two-mode state $\rho_{cd}$ is described by the  zero-temperature master equation
\begin{equation}
		\pdv{\rho_{cd}}{\tau}=\sum_{s=1}^{2}\kappa_s \left(2 A_s \rho_{cd} A_s^\dag-A_s^\dag A_s \rho_{cd} -\rho_{cd} A_s^\dag A_s  \right).\label{ch5TwomodePhaseDampingMasterEq}
	\end{equation}
The solution of Eq.~(\ref{ch5TwomodePhaseDampingMasterEq}) can be written in the Fock basis as 
\begin{align}
\rho_{cd}(\tau)=\sum_{m_1=0}^{\infty}\sum_{m_2=0}^{\infty}\sum_{n_1=0}^{\infty}\sum_{n_2=0}^{\infty} {\left[\rho_{cd}(\tau)\right]_{m_1 m_2;n_1 n_2}}\ket{m_1}_{c} {\ket{m_2}_{d}}\,\,{_{c}\bra{n_1}}\,{ _{d}\bra{n_2}}, \label{ch5PDsolution}
\end{align}
where the matrix elements are calculated as 
\begin{align}
{\left[\rho_{cd}(\tau)\right]_{m_1 m_2;n_1 n_2}}=&N^2_{2,h} e^{-2\modu{\beta}^2}\exp\left[-\sum_{s=1}^{2}\kappa_s\tau\left(n_s-m_s \right)^2\right]\nonumber\\
&\times\sum_{r,r^\prime=0}^{1} \frac{e^{-i\pi h(r-r^\prime)}\,{\beta}_{r}^{m_1}\,{\beta}_{r^\prime}^{m_2}\,{\beta^{\ast}_r}^{n_1}\,{\beta^{\ast}_{r^\prime}}^{n_2}}{\sqrt{m_1!\,m_2!\,n_1!\,n_2!}}.
\label{ch5PDmatrixelements}
\end{align}
A derivation of the Eq.~(\ref{ch5PDsolution}) is given in Appendix~\ref{Appendix_signatureEntanglementII}. As in the previous case, the state $\rho_{cd}(\tau)$ is a mixed state for $\tau>0$, and the amount of entanglement in the state $\rho_{cd}(\tau)$ is calculated numerically in terms of the logarithmic negativity $E_N$. We set the phase damping coupling constants $\kappa_1=\kappa_2=\kappa=0.01$, for the calculation. The variation of entanglement of the state $\ket{\Phi}_h$ as a function of the scaled time $\kappa\tau$ is shown in Fig.~\ref{ch5fig:PDEntanglementDecay}. A comparison of Fig.~\ref{ch5fig:PDEntanglementDecay} with  Fig.~\ref{ch5fig:EntanglementDecay} reveals that, the decay of entanglement of the state $\ket{\Phi}_h$ due to phase damping is much slower compared to that due to amplitude damping of the state. In the phase damping model, the entanglement of the state $\ket{\Phi}_h$ goes to zero for  times greater than $\kappa\tau=3.0$,  whereas in the amplitude damping model this happens from time $\gamma\tau=0.2$ itself.
  \begin{figure}[H]
\centering
\includegraphics[height=7 cm,width=8 cm]{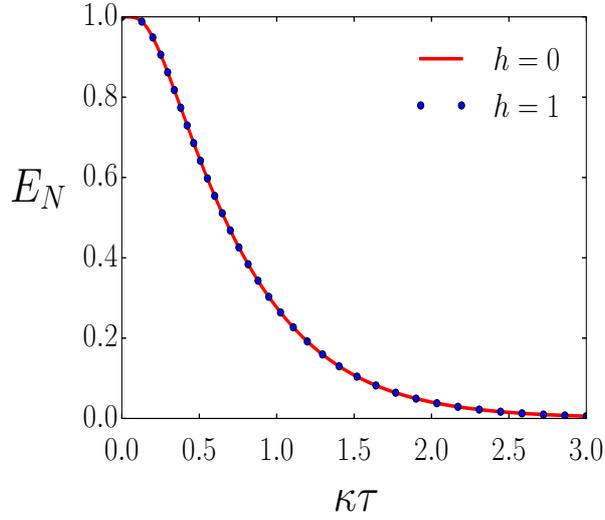}
\caption{Decay of entanglement of the state $\ket{\Phi}_h$ as a function of scaled time $\kappa \tau$ for $\modu{\alpha}^2=10$.}
\label{ch5fig:PDEntanglementDecay}
\end{figure}

The optical tomogram of the state $\ket{\Phi}_h$ in the presence phase damping is obtained by substituting Eq.~(\ref{ch5PDsolution}) in Eq.~(\ref{ch52modeOTdef}):
\begin{align}
\omega_{h}\left(X_{\theta_1},\theta _1; X_{\theta_2},\theta _2;\tau\right)=&\frac{N^2_{2,h} \exp\left[-2\modu{\beta}^2-X_{\theta_1}^2-X_{\theta_2}^2\right]}{\pi}\nonumber\\
&\times\sum_{m_1=0}^{\infty}\sum_{m_2=0}^{\infty}\sum_{n_1=0}^{\infty}\sum_{n_2=0}^{\infty} \frac{H_{m_1}(X_{\theta_1})H_{m_2}(X_{\theta_2})H_{n_1}(X_{\theta_1})H_{n_2}(X_{\theta_2})}{2^{(m_1+m_2)/2}\,2^{(m_2+n_2)/2}}\nonumber\\
&\times\exp\left[-\kappa\tau\left(n_1-m_1 \right)^2-\kappa\tau\left(n_2-m_2 \right)^2\right]\nonumber\\
&\times\exp\left[i\theta_1\left(n_1-m_1 \right)+i\theta_2\left(n_2-m_2 \right)\right]\nonumber\\
&\times\sum_{r,r^\prime=0}^{1} \frac{e^{-i\pi h(r-r^\prime)}\,{\beta}_{r}^{m_1}\,{\beta}_{r^\prime}^{m_2}\,{\beta^{\ast}_r}^{n_1}\,{\beta^{\ast}_{r^\prime}}^{n_2}}{m_1!\,m_2!\,n_1!\,n_2!}.\label{ch5PDOT}
\end{align}
Using Eq.~(\ref{ch5PDOT}), we analyze the optical tomogram of the states in mode $c$ for different quadrature measurements in mode $d$ in the presence of phase damping. Figure \ref{ch5fig:PDl2h0decay} displays the optical tomograms $\omega_{0}(X_{\theta_1},\theta _1;X_{\theta_2},\theta _2;\tau)$ in mode $c$ for the entangled state $\ket{\Phi}_0$ at different times $\kappa \tau$ for the relative phases of measurements (a) $\modu{\delta-\theta_2}=0.3$, (b) $\modu{\delta-\theta_2}=\pi/2$, and (c) $\modu{\delta-\theta_2}=\pi/2-\pi/40$. The structures with sinusoidal strands are not lost when the interaction of the state with environment is for a short time. The sinusoidal strands in the optical tomogram retain their structure only for a short time $\kappa\tau$. With an increase in the interaction time $\kappa\tau$, the sinusoidal strands in the optical tomogram get distorted and the signatures of the entanglement are lost.
\begin{figure}[H]
   \centering
	\includegraphics[height=12.8 cm,width=\textwidth]{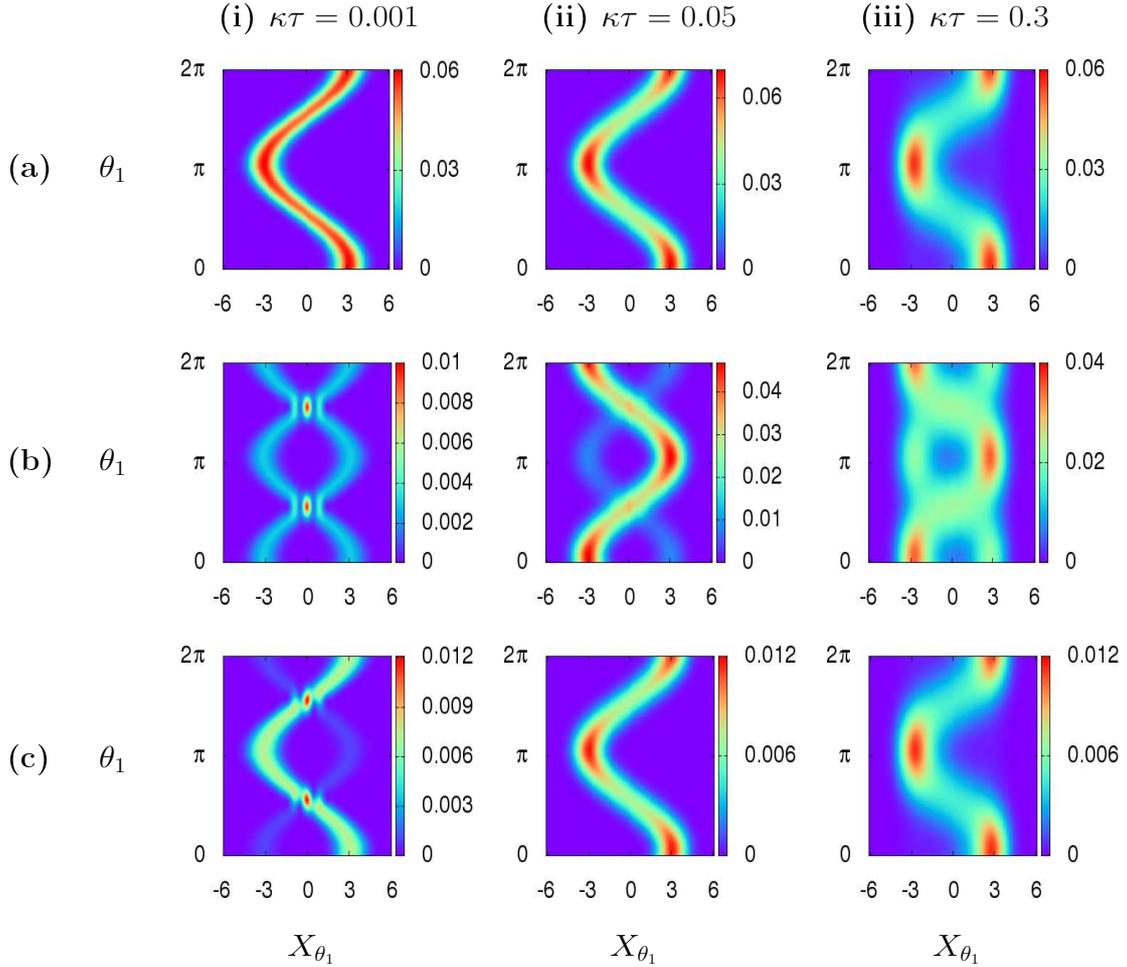}
   \caption {Optical tomograms $\omega_{0}(X_{\theta_1},\theta _1;X_{\theta_2},\theta _2;\tau)$ in mode $c$ for the entangled state $\ket{\Phi}_0$ in the presence of phase damping at times (i) $\kappa \tau=0.001$, (ii) $\kappa \tau=0.05$, and (iii) $\kappa \tau=0.3$, for the relative phases of measurements (a)$\modu{\delta-\theta_2}=0.3$, (b) $\modu{\delta-\theta_2}=\pi/2$, and (c) $\modu{\delta-\theta_2}=\pi/2-\pi/40$. Here,  $\modu{\alpha}^2=10$, $\delta=0.2$ and $X_{\theta_2}=2.0$.} 
 \label{ch5fig:PDl2h0decay}  
   \end{figure}
   
In the long time limit $\kappa \tau\rightarrow\infty$, the optical tomogram given in Eq.~(\ref{ch5PDOT}) reduces to 
\begin{align}
\omega_{h}\left(X_{\theta_1},\theta _1; X_{\theta_2},\theta _2;\tau\rightarrow\infty\right)=&\frac{2\,N^2_{2,h}\left[1+(-1)^h\right]}{\pi} \exp\left[-2\modu{\beta}^2-X_{\theta_1}^2-X_{\theta_2}^2\right]\nonumber\\
&\times\sum_{m=0}^{\infty}\sum_{n=0}^{\infty} \frac{\modu{\beta}^{2m}\,\modu{\beta}^{2n}\,H_{m}^2(X_{\theta_1})H_{n}^2(X_{\theta_2})}{2^{(m+n)}\,\left(m!\,n!\right)^2}.
\end{align}
This optical tomogram is independent of the values of $\theta_1$ and $\theta_2$; this is displayed in Fig.~\ref{ch5fig:PDlongtime}. We have repeated the analysis of phase damping for the entangled state $\ket{\Phi}_1$ and found similar results. 
\begin{figure}[h]
   \centering
	\includegraphics[height=6 cm,width=7cm]{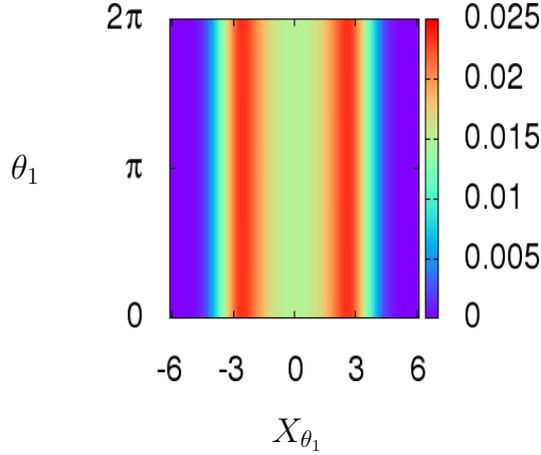}
   \caption {Optical tomograms $\omega_{0}(X_{\theta_1},\theta _1;X_{\theta_2},\theta _2;\tau)$ in mode $c$ for the entangled state $\ket{\Phi}_0$ in the presence of phase damping at long times $\gamma \tau\rightarrow\infty$.} 
 \label{ch5fig:PDlongtime}  
   \end{figure}

\section{Conclusion}
\label{conclusion}
We have obtained a closed-form  analytical expression for the optical tomogram of the maximally entangled coherent state generated at the output of the beam splitter. For separable two-mode states, the optical tomogram of the system can be written as the product of the optical tomograms of the subsystems. Whereas, for the entangled two-mode states, the optical tomogram in the mode $c$ shows different features when we change  the parameters  $X_{\theta_2}$ and $\theta_2$ in the mode $d$. Similarly, the optical tomogram in the mode $d$ will be affected by the parameters in mode $c$. Specifically, for the entangled state $\ket{\Phi}_{h}=N_{h}\left[\ket{\beta}\ket{\beta}+e^{i\pi h} \ket{-\beta}\ket{-\beta}\right]$, with $X_{\theta_2}\neq 0$, the optical tomogram in mode $c$ shows double-stranded structure if $\modu{\delta-\theta_2}=\pi/2$ or $3\pi/2$ and a single-stranded structure for all other values (except for $\pi/2$ and $3\pi/2$ and their vicinity) of $\modu{\delta-\theta_2}$. Our calculations not only avoids the computational complexity of finding the two-mode density matrix or the quasiprobability distribution of the state but it also reduces the number of homodyne measurements to be performed to determine whether the state $\ket{\Phi}_h$ is  entangled or not. The above results hold even during the short time  interaction of the system  with its external environment.
\chapter{ENTANGLEMENT DYNAMICS OF QUANTUM STATES IN A BEAM SPLITTER}\label{Ch_EntanglementDynamics}
\thispagestyle{plain}
\section{Introduction}  
It has been shown that a standard nonlinear optics interaction arising from  a Kerr nonlinearity, followed by a simple interaction with a beam splitter, produces a large amount of entanglement in an arbitrarily short time \citep{vanEnk2003}.  Here, the initial state considered was a coherent state, and the input state for the beam splitter are taken at specific instants (at fractional revival times)  during the time  evolution of the coherent state in the Kerr medium. 
In this chapter, we study, to a greater extent, the continuous dynamics of entanglement using the state at  any instants instead of  at specific instants during the  evolution of coherent state in the Kerr medium using the set-up in \citep{vanEnk2003}. We also investigate the optical tomogram of the entangled states generated in a beam splitter with a Kerr-like medium placed on one of its input arms and look for the signatures of entanglement in the optical tomogram of the entangled states generated at the instants of fractional revival times.   
In the next section, we describe the entanglement dynamics of the quantum states generated in a beam splitter using an initial coherent state evolving in a Kerr medium.

\section{Entanglement dynamics of an initial coherent state} 
\label{Sec_initialCS}
Consider the dynamics of an initial coherent state $\ket{\psi(0)}=\ket{\alpha}$, where $\alpha=\modu{\alpha}\,e^{i\delta}$, governed by the nonlinear Hamiltonian given in Eq.~(\ref{ch3kerrhamiltonian}). The time-evolved state $\ket{\psi(t)}$ is given by Eq.~(\ref{ch3psi(t)initialCS}). We write it again for ready reference:
\begin{equation}
		\ket{\psi(t)}=e^{-\modu{\alpha}^2/2}\sum_{n=0}^{\infty}\frac{\alpha^n e^{-i\chi t n(n-1)}}{\sqrt{n!}}\ket{n}.\label{ch6psi(t)initialCS}
	\end{equation}
We recall from Chapter \ref{Ch_CSevolution} that, the state $\ket{\psi(t)}$ revives periodically with the revival time $\trev=\pi/\chi$. Also, 
between time  $t=0$ and $t=\trev$, $\ket{\psi(t)}$  shows $k$-subpacket fractional revivals at times $t=j\pi/k\chi$, where $j=1$, $2$,...,$(k-1)$, for a given value of $k$ ($>1$) with the condition that $(j,k)=1$. 
Here, the interesting thing is that even if the initial wave packet is a classical one, $\ket{\psi(t)}$  becomes nonclassical during the evolution in the Kerr medium \citep{Yurke1986,Tara1993,Sudheesh2004}.

\begin{figure}[h]
\centering
\includegraphics[height=6 cm,width=10 cm]{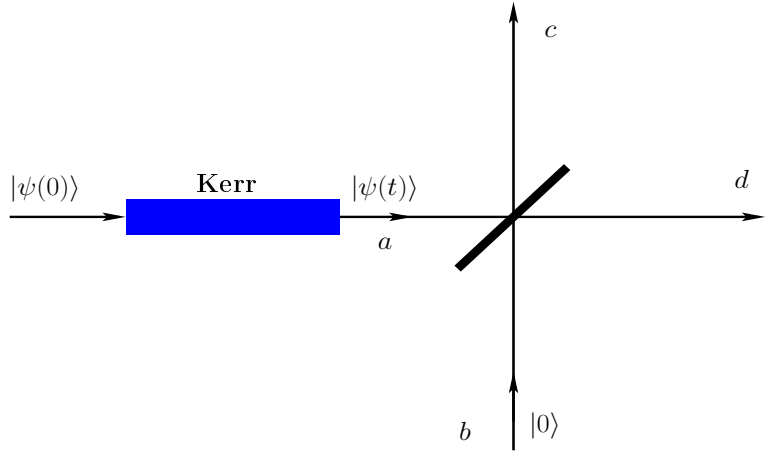}
\caption{A $50/50$ beam splitter with a Kerr medium in the horizontal input port.  The time-evolved state $\ket{\psi(t)}$, obtained by the evolution of $\ket{\psi(0)}$, is injected to the horizontal input port and the vacuum state $\ket{0}$ to the vertical input port of the beam splitter. Here, $a$ and $b$ ($c$ and $d$) are the input (output) field modes of the beam splitter.}
\label{ch6beamsplitter}
\end{figure}
Subsequently, we split the time-evolved state $\ket{\psi(t)}$ in a  beam splitter with the vacuum state $\ket{0}$ to generate entangled states. As in Chapter \ref{Ch_EntangledOT}, we consider a $50/50$ lossless beam splitter with zero phase difference between reflected and transmitted beam. A schematic representation of the  set-up used for our investigation is given in Fig.~\ref{ch6beamsplitter}. We recall that $a$ and $b$ ($c$ and $d$) are the input (output) field modes of the beam splitter. The state at a particular instant of time $t$, $\ket{\psi(t)}$, is achieved by adjusting the interaction length (time) of the medium. We inject the  field $\ket{\psi(t)}$  through the horizontal port (mode $a$) of the beam splitter and a vacuum state through the vertical port (mode $b$).   Thus, the  input state  to the beam splitter is   $\ket{\psi(t)}_a\ket{0}_b$.  
The state $\ket{\Phi(t)}$ of the output modes can be obtained using the unitary operator  $U_{BS}$ of the beam splitter given Eq.~(\ref{ch5BSoperator}):
\begin{equation}
\ket{\Phi(t)}=U_{BS}\,\big(\ket{\psi(t)}_a\ket{0}_b\big).\label{ch6phit}
\end{equation}
Substituting Eq.~(\ref{ch6psi(t)initialCS}) in Eq.~(\ref{ch6phit}), we get
\begin{align}
		\ket{\Phi(t)}=e^{-\modu{\alpha}^2/2}
\sum_{n=0}^\infty \frac{\alpha^n}{\sqrt{n!}}\frac{\exp\left[-i\chi t n(n-1)\right]}{\quad 2^{n/2}}\sum_{p=0}^{n}\dbinom{n}{p}^{1/2}\,\ket{p}_c\ket{n-p}_d.\label{ch6PhiinitialCS}
	\end{align}
The state $\ket{\Phi(t)}$ is a two-mode pure state of the field. 
The total density matrix for the state $\ket{\Phi(t)}$ is 
\begin{align}
\rho_{cd}(t)=&e^{-\modu{\alpha}^2}\sum_{n=0}^{\infty} \sum_{n^\prime=0}^{\infty}\frac{\alpha^n\,{\alpha^\ast}^{n^\prime}e^{-i\chi t\left[n(n-1)-n^\prime(n^\prime-1)\right]}}{\sqrt{n!\,n^\prime!}\,2^{(n+n^\prime)/2}}\sum_{p=0}^{n} \sum_{p^\prime=0}^{n^\prime}\dbinom{n}{p}^{1/2}\dbinom{n^\prime}{p^\prime}^{1/2}\nonumber\\
&\times\ket{p}_c\ket{n-p}_d {_c\bra{p^\prime}} {_d\bra{n^\prime-p^\prime}}\label{ch6DensityinitialCS}
\end{align}
We have numerically calculated the entanglement $E$ of the state $\ket{\Phi(t)}$ in terms of von Neumann entropy, using Eq.~(\ref{ch5VNE}),  and plotted it  in Fig.~\ref{ch6m0nu510} for various values of $\modu{\alpha}^2$ between the time  $t=0$ and $T_{\rm rev}$.

\begin{figure}[H]
\centering
\includegraphics[scale=0.6]{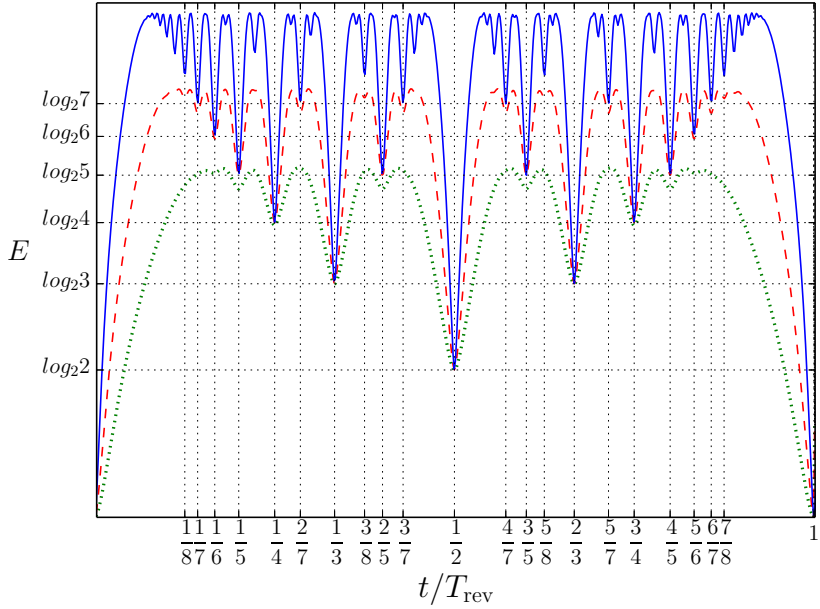}
\caption{Entanglement as a function of time $t/\trev$ from time $t=0$ to $t=\trev$ for an initial coherent state $\ket{\alpha}$ with $\modu{\alpha}^2=5$ (dotted), $10$ (dashed) and $20$ (solid). Local minima of the curves indicates the instants of fractional revivals.}
\label{ch6m0nu510}
\end{figure}
 It can be seen that at time $t=0$ the entanglement is zero because at this instant both the input states feeding into the beam splitter are classical states. The input state to the beam splitter is   $\ket{\alpha}_a \ket{0}_b$ and  the output state given in Eq.~(\ref{ch6PhiinitialCS}) takes the form $\ket{\Phi(0)}=\ket{\frac{\alpha}{\sqrt{2}}}_c\ket{\frac{\alpha}{\sqrt{2}}}_d$. This output state is a separable state and the  entanglement between the two output modes  $c$ and $d$ is  zero.  During the time evolution in the medium, the state  $\ket{\psi(t)}$  exhibits nonclassical behaviour and the output state $\ket{\Phi(t)}$ shows non-zero entanglement. 
The entropic entanglement potential (EEP), defined in \citep{Asboth2005} using relative entropy, is a measure of nonclassicality of the state in one of the input arms of the beam splitter with vacuum in the second input arm. For pure states, the EEP reduces to the von Neumann entropy in one of the  output arms.  It means that a nonzero value of  the  von Neumann entropy $E$ reveals the  nonclassicality   of the  pure state $\ket{\psi(t)}$. In other words, the state $\ket{\Phi(t)}$ will be  an entangled state whenever the state $\ket{\psi(t)}$ is a nonclassical state.

At $k$-subpacket fractional revival times, the time-evolved state $\ket{\psi(t)}$ is a superposition of $k$ phase-rotated coherent states, as given in Eq.~(\ref{ch3superpositionofcs}). Therefore, the output states given in Eq.~(\ref{ch6PhiinitialCS}) reduces to
\begin{equation}
\ket{\Phi^{(k)}}=\left\{\begin{array}{ll}
\sum_{s=0}^{k-1} f_{s}\ket{\beta_s}_c\ket{\beta_s}_d,\quad&
k \quad{\rm odd;} \\
\sum_{s=0}^{k-1} g_{s}\ket{\beta_s}_c\ket{\beta_s}_d,\quad&
k \quad{\rm even,}
 \end{array} \right.
\label{ch6entangledcs}
\end{equation}
where $f_s$ and $g_s$ are the Fourier coefficients (given in Eqs.~(\ref{ch3Fourier_fs}) and (\ref{ch3Fourier_gs}), respectively), $\beta_s=\alpha_s/\sqrt{2}$ and $\alpha_s$ is given in Eq.~(\ref{ch3alpha_s}). 
For large values of $\modu{\alpha}^2$,  the coherent states appearing in the superposition given in Eq.~(\ref{ch3superpositionofcs}),  form an orthogonal basis in $k$ dimension.  Thus, the  state $\ket{\Phi^{(k)}}$, is already written in the Schmidt decomposition \citep{Ekert1995}. It implies that the state $\ket{\Phi^{(k)}}$  is a maximally entangled state  in $k$ dimension with the Schmidt rank $k$ and  von Neumann entropy $E= \log_2\,k$ in one of the output arms of the beam splitter \citep{vanEnk2003}.  
 It is evident from the figure that for $\modu{\alpha}^2=5$, the states at two-, three-, and four-subpacket fractional revivals are maximally entangled states  with $E = \log_2 2$, $\log_2 3$, and $\log_2 4$, respectively. 
 The entanglement using von Neumann entropy  plot given in Fig.~\ref{ch6m0nu510} shows clear signatures of fractional revivals.  We have shown that, at the instants of fractional revivals the entanglement  takes  a local minimum in the von Neumann entropy  plot. 
 It can be verified from the figure  that the states at higher-order fractional revival times  are maximally entangled states for larger values of  $\modu{\alpha}^2$. The fractional revival times are marked in the figure with  vertical dotted lines. It should be noted that the  Phase entropy, Wehrl entropy, and   R\'{e}nyi entropy  take local minima at the instants of fractional revivals of wave packets  in a single-mode  Kerr medium \citep{Jex1994,Vaccaro1995,Miranowicz2001,Rohith2014} where there is no  question of entanglement.

The phase-rotated coherent states (given in Eq.~(\ref{ch3superpositionofcs})) at the instants of $k$-subpacket fractional revival can also be visualized in the phase space plot of the Husimi $Q$ function. The Husimi $Q$ function for a state $\ket{\psi}$
is defined as 
\begin{equation}
Q(x,p)=\frac{1}{\pi}\modu{\int_{-\infty}^{\infty}dx^\prime~\psi_{\beta}(x^\prime)~\psi(x^\prime)}^2,
\end{equation}
where
\begin{equation}
\psi_{\beta}(x^\prime)=\pi^{-1/4}\exp\left[-\frac{(x^\prime-x)^2}{2}+ip(x^\prime-\frac{x}{2})\right]
\end{equation}
and  $\psi (x^\prime)$ are the position representation of the coherent state $\ket{\beta}$ and $\ket{\psi}$, respectively. The Husimi $Q$ function of a coherent state $\ket{\alpha}$ is a Gaussian distribution given by
\begin{equation}
Q(x,p)=\frac{1}{\pi}\exp\left[-\frac{1}{2}\left(x-x_0\right)^2-\frac{1}{2}\left(p-p_0\right)^2\right], \label{ch6QfunctionCS}
\end{equation}
where $x_0=\sqrt{2}\,\Re(\alpha)$ and $p_0=\sqrt{2}\,\Im(\alpha)$. All the coherent states in the  superposition, given in Eq.~(\ref{ch3superpositionofcs}), have the same amplitude $\modu{\alpha}$ and the center of the Gaussian peaks of each of these coherent states fall regularly on a circle of radius $\modu{\alpha}$ in the phase-plane. If we assume that the states are well separated when the distance between their Gaussian peaks in the phase-plane is equal to the diameter of the contour obtained when the section of the Gaussian bell is made at $0.1$ of its height, the maximum number of well-distinguished states that can be obtained for a given field strength $\modu{\alpha}^2$ is \citep{Miranowicz1990} 
	\begin{equation}
		N_{max}\cong 2\pi\modu{\alpha}/2\sqrt{\ln 10}.
		 \label{ch6Nmax}
	\end{equation}
For instance, for  $\modu{\alpha}^2=5$,  $N_{max}\cong 4.62$ and  maximum number of well-distinguished states is five (rounding $N_{max}$ to highest integer value). The contour plots of the Husimi $Q$ function   given in Fig.~\ref{ch6Husimim0} at fractional revival times $t=T_{\rm rev}/4$, $t=T_{\rm rev}/5$ and $t=T_{\rm rev}/6$ verifies this result. We find that highest order of fractional  revival that can be observed in the entropy plot is related  to the value of $N_{max}$. For example, the highest order of fractional revival that can be seen in the entropy plot is five for  $\modu{\alpha}^2=5$ and it is evident  from the dotted curve in Fig.~\ref{ch6m0nu510}. There are  $9$ well-distinguished local minima in this case and they corresponds to five-, four-, three-, and two-subpacket fractional revival times. When the   field strength $\modu{\alpha}^2$ increases  the radius of the circle in phase-plane increases and  higher-order fractional revivals are captured in Fig.~\ref{ch6m0nu510}. This is  evident in  the dashed and dotted curves  corresponds to field strengths $\modu{\alpha}^2=10$ and $20$, respectively,  in Fig.~\ref{ch6m0nu510}. 
\begin{figure}[h]
\centering
\includegraphics[height=6cm,width=\textwidth]{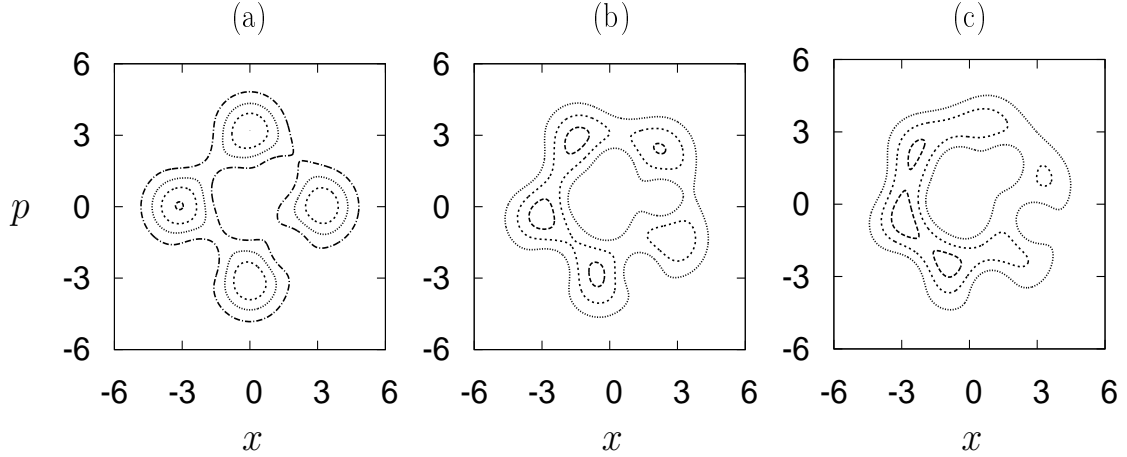}
\caption{Husimi $Q$ function of the state $\ket{\psi(t)}$ at times (a) $t=T_{\rm rev}/4$, (b) $t=T_{\rm rev}/5$, and (c) $t=T_{\rm rev}/6$ for an initial coherent state with $\modu{\alpha}^2=5$.}
\label{ch6Husimim0}
\end{figure}

\begin{figure}[h]
\centering
\includegraphics[scale=0.7]{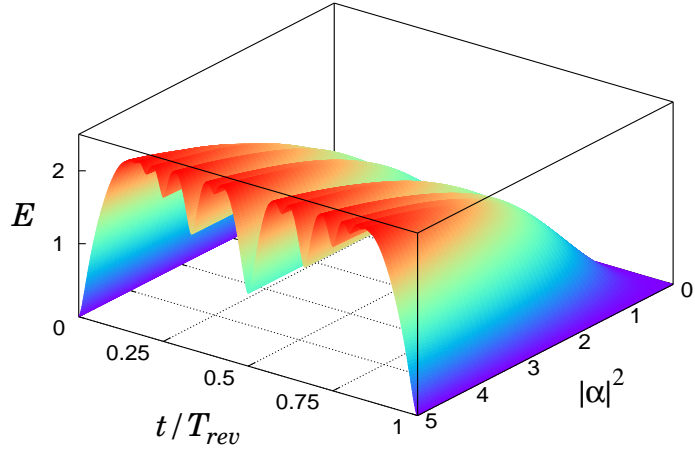}
\caption{Entanglement as a function of time $t/\trev$ and field strength $\modu{\alpha}^2$ for initial coherent state. Entanglement goes to zero at any instant when $\modu{\alpha}^2\rightarrow 0$.  For a given value of $\modu{\alpha}^2$, the entanglement attains a constant highest value   during collapse of wave packets and the highest  value of entanglement increases with increase in the field strength $\modu{\alpha}^2$.}
\label{ch6surf-m0}
\end{figure} 
Figure~\ref{ch6m0nu510} shows maxima in between  the fractional revivals  and the values of entropy $E$ at these  maxima are same. These maxima occur at the instants of  collapses  of the initial wave packet during the evolution in the medium.  
It is clear from the Fig.~\ref{ch6m0nu510} that the collapsed states are more nonclassical than the states at the instants of $k$-subpacket fractional revivals for a given value of $\modu{\alpha}^2$ \citep{Rohith2016b}.  
We denote the maximum value of entanglement by $E_{max}$ and  $E_{max}\cong 2.37$ ebits for $\modu{\alpha}^2=5$ (see Fig.~\ref{ch6m0nu510}).   For $\modu{\alpha}^2=10$ and $20$, $E_{max}$ is $2.90$ and $3.42$ ebits, respectively. Increase in the value of  $E_{max}$ with the increase in $\modu{\alpha}^2$ is justified because  the number, $n$, of Fock states contributing to the collapsed state  increases for larger $\modu{\alpha}^2$ which in turn increase the entropy.   
At revival times, the  entanglement  returns to its initial value of zero. 
Figure~\ref{ch6surf-m0} shows the variation of entanglement as a function time $t/\trev$ and field strength $\modu{\alpha}^2$. This figure clearly shows the  increase in  $E_{max}$   with an increase in the value of $\modu{\alpha}^2$. The value of entanglement becomes zero at any instants of time when $\modu{\alpha}^2\rightarrow 0$ because in this case  both the arms of the beam splitter contains the vacuum state $\ket{0}$, which is a classical state. 

\section{Optical tomogram of the entangled state generated in the Kerr medium}
In this section, we calculate the optical tomogram of the entangled state $\ket{\Phi(t)}$ given in Eq.~(\ref{ch6PhiinitialCS}), and find the signatures of entanglement in the  optical tomogram of the state. Substituting Eq.~(\ref{ch6PhiinitialCS}) in Eq.~(\ref{ch52modeOTdefAlternate}), we get the optical tomgram of the state $\ket{\Phi(t)}$ as 
\begin{align}
		\omega_t\left(X_{\theta_1},\theta _1; X_{\theta_2},\theta _2\right)=&\frac{\exp\left[-\modu{\alpha}^2-X_{\theta_1}^2-X_{\theta_2}^2\right]}{\pi}\left|\sum_{n=0}^\infty \frac{\alpha^n e^{-i\chi t n(n-1)}}{2^{n}}\right. \nonumber\\
		&\times\left.\sum_{p=0}^{n} \frac{H_p(X_{\theta_1})H_{n-p}(X_{\theta_2})\,e^{-i\left[p(\theta_1-\theta_2)+n\theta_2\right]}}{p!\,(n-p)!}\right|^2.\label{ch6OTinitialCS}
	\end{align}
At $k$-subpacket fractional revival time $t=\pi/k\chi$, the above expression reduces to 
\begin{align}
\omega^{(k)}(X_{\theta_1},\theta _1;X_{\theta_2},\theta _2)=\frac{1}{\pi}\modu{\sum_{s=0}^{k-1}f_{s,k}\,\eta(X_{\theta_1},\theta_1,\beta_s)\,\eta(X_{\theta_2},\theta_2,\beta_s)}^2, \label{ch6OTentangledKerr}
\end{align}
where the quantities $f_{s,k}$ and $\eta$ are defined in Eqs.~(\ref{ch3FourierCoefficients}) and (\ref{ch5eta}), respectively. Obviously, the optical tomogram given above can not be written as the product of optical tomograms of the subsystems. Using Eq.~(\ref{ch6OTentangledKerr}), we analyze the optical tomogram of the state in mode $c$ for different quadrature measurement in mode $d$. A measurement of the quadrature $\hat{X}_{\theta_2}$ in mode $d$ project the state $\ket{\Phi^{(k)}}$ given in Eq.~(\ref{ch6entangledcs}) to the state $\ket{\phi^{(k)}}_c$ in mode $c$:
\begin{align}
\ket{\phi^{(k)}}_c=\tilde{N}_k\sum_{s=0}^{k-1}f_{s,k}\,\psi_{\beta_s}\left(X_{\theta_2},\theta_2\right)\ket{\beta_s},\label{ch6projection}
\end{align}
where $\psi_{\beta_s}\left(X_{\theta_2},\theta_2\right)$ is the quadrature representation of the coherent state $\ket{\beta_s}$ and the normalization constant
\begin{align}
\tilde{N}_k=\left[\sum_{s=0}^{k-1}\sum_{s^\prime=0}^{k-1}f_{s,k}f^{\ast}_{s^\prime,k}
\psi_{\beta_s}\left(X_{\theta_2},\theta_2\right)\,\psi^{\ast}_{\beta_{s^\prime}}\left(X_{\theta_2},\theta_2\right)\,\bra{\beta_{s^\prime}}\ket{\beta_{s}}\right]^{-1/2}.
\end{align}
The value of the coefficients  $\psi_{\beta_s}\left(X_{\theta_2},\theta_2\right)$ of the states $\ket{\beta_s}$ in Eq.~(\ref{ch6projection}) changes depending upon the quadrature measurement (values of $X_{\theta_2}$ and $\theta_2$) in mode $d$, which gives different features for the state $\ket{\phi^{(k)}}_c$. These features will be reflected in the optical tomogram of the state in mode $c$. At two-subpacket fractional revival time ($k=2$), a measurement of $\hat{X}_{\theta_2}$ in mode $d$ project the state $\ket{\Phi^{(2)}}$ to the state
\begin{align}
\ket{\phi^{(2)}}_c=\tilde{N}_2\left[f_{0,2}\,\psi_{i\beta}\left(X_{\theta_2},\theta_2\right)\ket{i\beta}+f_{1,2}\,\psi_{-i\beta}\left(X_{\theta_2},\theta_2\right)\ket{-i\beta}\right],\label{ch6projectionK2}
\end{align}
where $f_{0,2}=(1-i)/\sqrt{2}$ and $f_{1,2}=(1+i)/\sqrt{2}$. Based on the relative strength of the coefficients $\psi_{i\beta}(X_{\theta_2},\theta_2)$ and $\psi_{-i\beta}(X_{\theta_2},\theta_2)$,  the state $\ket{\phi^{(2)}}_c$ can be one of following: $\ket{i\beta}$,  $\ket{-i\beta}$ and  a superposition of $\ket{i\beta}$ and  $\ket{-i\beta}$. All the Fourier coefficients $f_{s,k}$ have equal magnitude $1/\sqrt{k}$, and hence the probability for occurring the state $\ket{\pm i\beta}$ is proportional to 
$\modu{\psi_{\pm i\beta}(X_{\theta_2},\theta_2)}^2$. 

For $X_{\theta_2}\neq 0$, $\modu{\psi_{i\beta}(X_{\theta_2},\theta_2)}^2=\modu{\psi_{-i\beta}(X_{\theta_2},\theta_2)}^2$ for $\modu{\delta-\theta_2}=n\pi$, where $n=0,\,1$, and $2$. The states $\ket{\phi^{(2)}}_c$  for these three values of $\modu{\delta-\theta_2}$ are same. The probability for occurring $\ket{i\beta}$ and $\ket{-i\beta}$ in mode $c$ is $50:50$ and the optical tomogram in mode $c$ will display a double-stranded structure, in which, one strand corresponds to $\ket{i\beta}$ and the other corresponds to $\ket{-i\beta}$. The double-stranded structure of the optical tomogram in mode $c$ for $\modu{\delta-\theta_2}=0$ with $X_{\theta_2}=2.0$ is shown in Fig.~\ref{ch6fig:k2tomo}(a). 
 In the range $0\leq \modu{\delta-\theta_2}< \pi$, the state $\ket{\phi^{(2)}}_c$ can be approximated to the coherent state $\ket{i\beta}$ because  $\modu{\psi_{i\beta}(X_{\theta_2},\theta_2)}^2 \gg \modu{\psi_{-i\beta}(X_{\theta_2},\theta_2)}^2$, which gives a structure with a single strand for the optical tomogram in mode $c$. Figure~\ref{ch6fig:k2tomo}(c) shows single-stranded structure in the optical tomogram of the state in mode $c$ for  $\modu{\alpha}^2=10$, $\delta=0.2$,  $X_{\theta_2}=2.0$ and $\modu{\delta-\theta_2}=\pi/2$. 
In the vicinity of $\modu{\delta-\theta_2}=n\pi$, the state $\ket{\phi^{(2)}}_c$ is an intermediate state in which one of the coherent states in the superposition, given in Eq.~(\ref{ch6projectionK2}), is having high amplitude compared to the other. The optical tomogram of an intermediate state corresponding to $\modu{\delta-\theta_2}=\pi/60$ is shown in Fig.~\ref{ch6fig:k2tomo}(b). It shows the transition of the double-stranded structure to the single-stranded structure. When $X_{\theta_2}=0$, the both of the coherent states composing the superposition state $\ket{\phi^{(2)}}_c$ have same probability without any condition on  $\modu{\delta-\theta_2}$. This displays a structure with two sinusoidal strands in the optical tomogram.
\begin{figure}[h]
   \centering
	\includegraphics[height=6 cm,width=\textwidth]{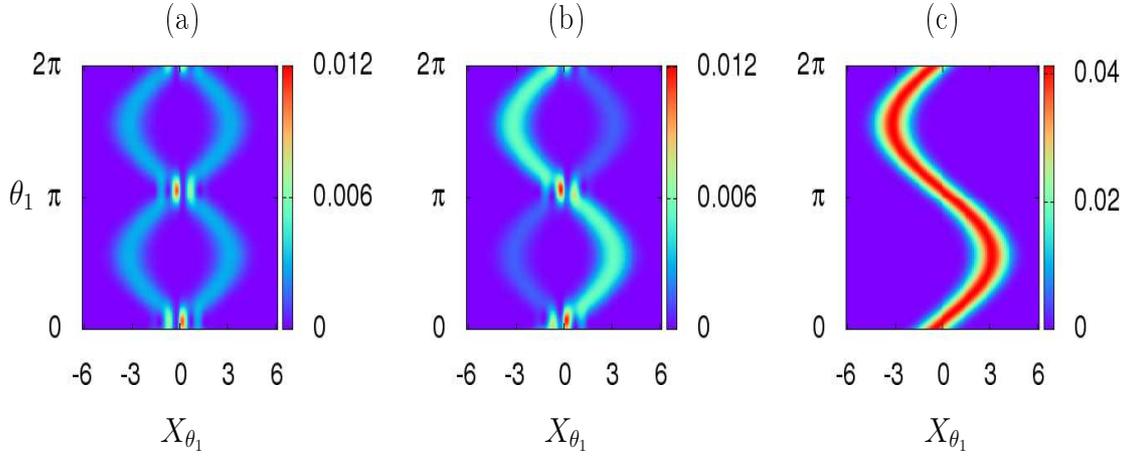}
   \caption {Optical tomograms $\omega^{(2)}(X_{\theta_1},\theta _1;X_{\theta_2},\theta _2)$ in mode $c$ for the entangled state $\ket{\Phi^{(2)}}$ with $\modu{\alpha}^2=10$, $\delta=0.2$, and $X_{\theta_2}=2.0$, for different relative phases $\modu{\delta-\theta_2}$ of the quadrature measurement in mode $d$: (a) $0$, (b) $\pi/60$ and (c) $\pi/2$.  The optical tomogram shows sinusoidal double-stranded structure for $\modu{\delta-\theta_2}=0$. The optical tomogram for $\modu{\delta-\theta_2}=\pi$ and $2\pi$ are exactly the  same as in the case of $\modu{\delta-\theta_2}=0$.   The plot in the second column is an optical tomogram of an intermediate case which shows the transition of the double-stranded structure to the single-stranded structure. The optical tomogram  shows  a sinusoidal single-stranded structure for $\modu{\delta-\theta_2}=\pi/2$.} 
\label{ch6fig:k2tomo}
   \end{figure}

At three-subpacket fractional revival time $t=\pi/3\chi$, the measurement of $\hat{X}_{\theta_2}$ in mode $d$ project the state $\ket{\Phi^{(3)}}$ to the state
\begin{align}
\ket{\phi^{(3)}}_c=&\tilde{N}_3\left[f_{0,3}\,\psi_{\beta}\left(X_{\theta_2},\theta_2\right)\ket{\beta}+f_{1,3}\,\psi_{\beta e^{-i2\pi/3}}\left(X_{\theta_2},\theta_2\right)\ket{\beta\,e^{-i2\pi/3}}\right.\nonumber\\
&\left.+f_{2,3}\,\psi_{\beta e^{-i4\pi/3}}\left(X_{\theta_2},\theta_2\right)\ket{\beta\,e^{-i4\pi/3}}\right],\label{ch6projectionK3}
\end{align}
where $f_{0,3}=(3-i\sqrt{3})/6$, $f_{1,3}=i/\sqrt{3}$ and $f_{2,3}=(3-i\sqrt{3})/6$. Depending on the relative strength of the coefficients of the coherent states composing the superposition state $\ket{\phi^{(3)}}_c$, the optical tomogram in mode $c$ shows different structures. Using the similar set of arguments as given in the $k=2$ case, we find that, with $X_{\theta_2}\neq 0$, the optical tomogram in mode $c$ shows a double-stranded structure for $\modu{\delta-\theta_2}=(2n-1)\pi/3$, where $n=1,\,2$, and $3$, and a single-stranded structure for all other values of $\modu{\delta-\theta_2}$. Figures~\ref{ch6fig:k3tomo}(a) displays the double-stranded structure of the optical tomogram of the state in mode $c$ for $\modu{\delta-\theta_2}=\pi/3$. These two sinusoidal strands corresponds to the coherent states  $\ket{\beta}$ and $\ket{\beta\,e^{-i4\pi/3}}$, for which the maximum intensities along the $X_{\theta_1}$ axis occur at $3.099$ and $-2.093$, respectively. This is due to the fact that, for $\modu{\delta-\theta_2}=\pi/3$, there is equal probability for occurring the  states $\ket{\beta}$ and $\ket{\beta\,e^{-i4\pi/3}}$ in mode $c$, that is $\modu{\psi_{\beta}\left(X_{\theta_2},\theta_2\right)}^2=\modu{\psi_{\beta e^{-i4\pi/3}}\left(X_{\theta_2},\theta_2\right)}^2$, which is very large compared to the probability for occurring the state $\ket{\beta\,e^{-i2\pi/3}}$.
\begin{figure}[h]
   \centering
	\includegraphics[scale=0.7]{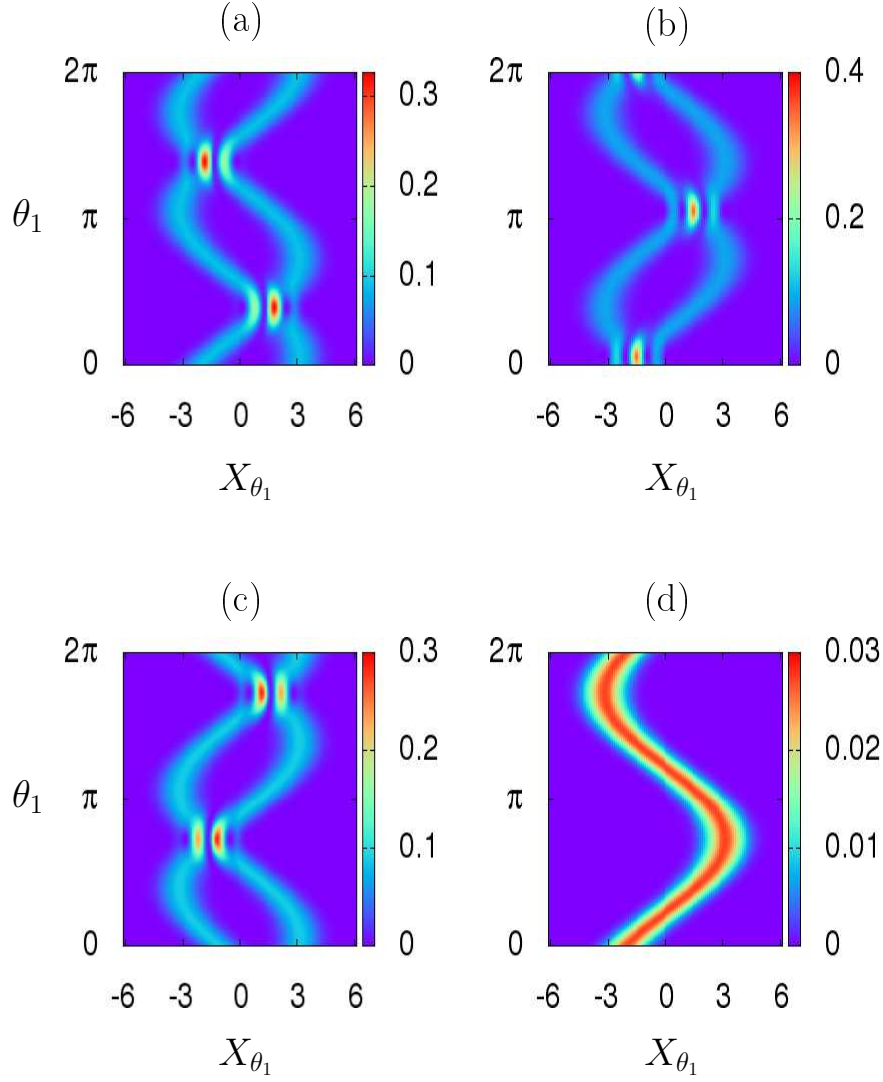}
   \caption {Optical tomograms $\omega^{(3)}(X_{\theta_1},\theta _1;X_{\theta_2},\theta _2)$ in mode $c$ for the entangled state $\ket{\Phi^{(3)}}$ with $\modu{\alpha}^2=10$, $\delta=0.2$, and $X_{\theta_2}=2.0$, for different relative phases $\modu{\delta-\theta_2}$ of the quadrature measurement in mode $d$: (a) $\pi/3$, (b) $\pi$, (c) $5\pi/3$, and (d) $2\pi/3$.} 
\label{ch6fig:k3tomo}
   \end{figure}

For $\modu{\delta-\theta_2}=\pi$, there is equal probability for occurring the  states $\ket{\beta\,e^{-i2\pi/3}}$ and $\ket{\beta\,e^{-i4\pi/3}}$ in mode $c$, that is $\modu{\psi_{\beta\,e^{-i2\pi/3}}\left(X_{\theta_2},\theta_2\right)}^2=\modu{\psi_{\beta e^{-i4\pi/3}}\left(X_{\theta_2},\theta_2\right)}^2$, which is very large compared to the probability for occurring the state $\ket{\beta}$. The two sinusoidal strands in the optical tomogram of the state in mode $c$ for $\modu{\delta-\theta_2}=\pi$, shown in Fig.~\ref{ch6fig:k3tomo}(b), correspond to the coherent states $\ket{\beta\,e^{-i2\pi/3}}$ and $\ket{\beta\,e^{-i4\pi/3}}$, for which the maximum intensities along the $X_{\theta_1}$ axis occur at $-1.005$ and $-2.093$, respectively.  Figures~\ref{ch6fig:k3tomo}(c) shows the double-stranded structure of the optical tomogram of the state in mode $c$ for $\modu{\delta-\theta_2}=5\pi/3$. These two sinusoidal strands belongs to the coherent states $\ket{\beta}$ and $\ket{\beta\,e^{-i2\pi/3}}$, because there is equal probability for occurring the  states $\ket{\beta}$ and $\ket{\beta\,e^{-i2\pi/3}}$ in mode $c$, that is $\modu{\psi_{\beta}\left(X_{\theta_2},\theta_2\right)}^2=\modu{\psi_{\beta e^{-i2\pi/3}}\left(X_{\theta_2},\theta_2\right)}^2$, which is very large compared to the probability for occurring the state $\ket{\beta\,e^{-i4\pi/3}}$.  In Fig.~\ref{ch6fig:k3tomo}(d), we have shown the single-stranded structure of the optical tomogram of the state in mode $c$ for $\modu{\delta-\theta_2}=2\pi/3$.
In this case, the probability of occurring the state $\ket{\beta\,e^{-i4\pi/3}}$ in mode $c$ is very large compared to the probabilities of occurring the states $\ket{\beta\,e^{-i4\pi/3}}$ and $\ket{\beta\,e^{-i4\pi/3}}$. Therefore, the optical tomogram shown in Fig.~\ref{ch6fig:k3tomo}(d) corresponds to the coherent state $\ket{\beta\,e^{-i4\pi/3}}$. For $X_{\theta_2}= 0$, all the three coherent states composing the superposition state $\ket{\phi^{(3)}}_c$ have same probability without any condition on  $\modu{\delta-\theta_2}$. This displays a structure with three sinusoidal strands in the optical tomogram. The forgoing analysis can be repeated for entangled state $\ket{\Phi^{(k)}}$ at higher-order fractional revival times ($k> 3$) and the conditions for the values of  $\modu{\delta-\theta_2}$ under which different structures are displayed in the optical tomogram in mode $c$ can be found. 

\section{Conclusions}
\label{Sec_conclusion}
We have investigated the optical tomogram of the entangled states generated using a beam splitter with a Kerr medium in one of its  input modes. The entanglement dynamics of the initial coherent state captures the signatures of revival and fractional revivals. The dynamics of entanglement using von Neumann entropy plot shows local minima at the instants of fractional revivals. These minima correspond to the generation of two-component Schr\"odinger cat states or multi-component Schr\"odinger cat-like states. The maximum entanglement $E_{max}$ is obtained at the instants of collapses of wave packets during the evolution in the medium. The maximum value of entanglement $E_{max}$ increases with an increase in the field strength $\modu{\alpha}^2$.  
We have found the signatures of entanglement in the optical tomogram of the entangled states generated at the instants of two- and three-subpacket fractional revival times. We have shown that, with $X_{\theta_2}\neq 0$, the optical tomogram in mode $c$ for the entangled state, generated at two-subpacket fractional revival time, shows a structure with two sinusoidal strands if $\modu{\delta-\theta_2}=n\pi$, where $n=0,\,1$, and $2$, and a structure with a single sinusoidal strand for all other values of $\modu{\delta-\theta_2}$.  For the entangled state generated at three-subpacket fractional revival time, the optical tomogram in mode $c$ shows a structure with two sinusoidal strands for $\modu{\delta-\theta_2}=(2n-1)\pi/3$, where $n=1,\,2$, $3$, and a structure with single sinusoidal strand for all other values of $\modu{\delta-\theta_2}$. 

\chapter{CONCLUSION}\label{Ch_Conclusion}
\thispagestyle{plain}
In the preceding chapters, we have described our results in detail, and also summarized them at appropriate  places in the text. It remains to place the work in a broader perspective and to list interesting open problems for future work.

In general terms, the present work has demonstrated the possibility of using the optical tomogram of the state, which is a directly measurable quantity, to study the nonclassical properties of light, such as fractional revival and entanglement. Since our methods avoid the reconstruction of the density matrix or the quasiprobabilty distributions of the state from the optical tomogram, more comprehensive is the information about the state, and thus highly sophisticated nonclassical phenomena can be studied with greater accuracy. The expressions for  the optical tomogram of the states in the presence of decoherence (with an appropriate value for the decay constants) found in this thesis provide the possibility of direct comparison of the optical tomograms obtained from the homodyne measurements. 

The present work opens up a number of avenues for further exploration. We list some of the interesting problems here. 

We have examined the signatures of superposed coherent states, which are a superposition of classical states, in the optical tomogram. It would be interesting to study  how the optical tomogram of the superposition state changes when the constituent states show a departure from its coherent nature. 

In this thesis, we have shown the signatures of revivals and fractional revivals in the optical tomogram of the state for an initial coherent state as well as for an initial superposed coherent states. We have also found the signatures of revivals and fractional revivals in the optical tomogram for an initial photon-added coherent state \citep{Rohith2015}. These investigations can be extended to different kinds of initial states, such as the photon-subtracted states, qudit coherent states, and the deformed coherent states.
The deformed coherent state can mimic the quantum state of the light from a nonideal laser \citep{Katriel1994}.

We have identified the signatures of bipartite entanglement for the maximally entangled states in two and three dimensions in the single-mode optical tomogram of the state. Is it possible to generalize these studies and find the signatures of entanglement in higher dimensions? Moreover, a quantitative estimation of the entanglement between the modes directly from the optical tomogram of the state is also an open problem to explore.

It will be interesting to study the signatures of quadrature squeezing of the electromagnetic field directly in the optical tomogram. An ideal candidate for this purpose would be a squeezed coherent state, which is having a tunable degree of squeezing. Here an exciting question comes. Is it possible to define a  quantitative measure of the degree of squeezing of the state in terms of its optical tomogram? Such an investigation  will be greatly useful for the experimentalists to characterize the squeezed states directly from its optical tomogram. 

The sub-Plank structures, structures at a scale smaller than the Plank's constant ($\hbar$), in phase space \citep{Zurek2001}  are direct signatures of quantum coherence and are formed as a result of interference between the two superposed cat states. These structures are extensively investigated using the quasiprobability distributions of the state. However, highly sophisticated experimental arrangements  must be needed to generate and observe these structures. Even if this is achieved, a small error in the reconstruction of the quasiprobability distributions or the density matrix from the optical tomogram of the state can ruin the studies. Therefore, it is of great significance to study the sub-Plank structures directly from the optical tomogram of the state.

\newpage
\addcontentsline{toc}{chapter}{REFERENCES}
\bibliography{reference}
\newpage
\appendix
\addcontentsline{toc}{chapter}{APPENDICES}
\chapter{Wigner function of the state at time $t=\trev/k$ for the initial state $\ket{\psi_{l,h}}$ propagating in a Kerr medium}\label{Appendix_superposedCS}
\thispagestyle{plain}
Consider propagation of a macroscopic superposition state composing $l$ coherent states, defined as
\begin{equation}
\ket{\psi_{l,h}}=N_{l,h}\sum_{r=0}^{l-1} e^{-i2\pi r h/l}\ket{\alpha\,e^{i2\pi r /l}},
\end{equation}
through a Kerr-like medium. Here $h=0,1,2,\dots,(l-1)$. The Hamiltonian governing the dynamics is given by
\begin{equation}
H=\hbar \chi \rm{\bf N}(\rm{\bf N}-1),\label{KerrHamilton}
\end{equation}
where ${\rm{\bf N}}=a^\dag a$ and $\chi$ is the third order nonlinear susceptibility of the medium. At time $t=\trev/k=\pi/k\chi$ the state can be written as
\begin{equation}
\ket{\psi(t=\trev/k)}=\ket{\psi^{(k)}_{l,h}}=U(\pi/k\chi)\,\ket{\psi_{l,h}},
\end{equation}
where
\begin{equation}
U\left(\pi/k\chi\right)=\exp\left[-\frac{i\pi}{k} {\rm{\bf N}}({\rm{\bf N}}-1)\right].
\end{equation}
The periodicity properties of the unitary time evolution operator $U\left(\pi/k\chi\right)$ (given in Eq.~(\ref{ch3periodicity1}) and Eq.~(\ref{ch3periodicity2})) enable us to write the time-evolved state at time $t=\trev/k$ as
\begin{equation}
\ket{\psi(t=\trev/k)}=\ket{\psi^{(k)}_{l,h}}=\sum_{s=0}^{k-1}\sum_{r=0}^{l-1} f_{s,k}\,e^{-i2\pi r h/l}\ket{\alpha_{r,s}}, \label{APDX1psi_trevbyk}
\end{equation} 
where $f_{s,k}$ is defined in Eq.~(\ref{ch3FourierCoefficients}) and 
\begin{align}
\alpha_{r,s}&=\begin{cases}
\alpha\, e^{i\,2 \pi \left(r /l- s/k\right)} & \text{if $k$ is odd}\\
\alpha\,e^{i\,2 \pi \left(r /l- s/k\right)}\,e^{i\pi/k} & \text{if $k$ is even.}
\end{cases}
\label{FourierCoefficients}
\end{align}
The Wigner function of the state $\ket{\psi^{(k)}_{l,h}}$ is 
\begin{equation}
W_{l,h}^{(k)}(\beta)=\frac{2\,e^{2\modu{\beta}^2}}{\pi^2}\int d^2 z \langle -z\left|\psi\left(t=\trev/k\right)\rangle\langle \psi\left(t=\trev/k\right) \right| z\rangle\, e^{2\left(z^\ast \beta-z\beta^\ast\right)},
\end{equation}
where $\ket{z}$ is a coherent state and $\beta=(x+i\,p)/\sqrt{2}$. Inserting Eq.~(\ref{APDX1psi_trevbyk}) in above equation, we get
\bea
\begin{split}
W_{l,h}^{(k)}(\beta)=\frac{2\,e^{2\modu{\beta}^2}}{\pi^2} \sum_{s,s^\prime=0}^{k-1}\sum_{r,r^\prime=0}^{l-1} f_{s,k}\,f_{s^\prime,k}^\ast\,e^{-i2\pi (r-r^\prime) h/l}\\
\times\int d^2 z \,\langle -z\ket{\alpha_{r,s}}\bra{\alpha_{r^\prime,s^\prime}} z\rangle\, e^{2\left(z^\ast \beta-z\beta^\ast\right)}. \label{wignerFun}
\end{split}
\eea
The inner product between two coherent states $\ket{\alpha}$ and $\ket{z}$ is 
\begin{equation}
\langle \alpha \ket{z}=\exp\left[-\frac{\modu{\alpha}^2}{2}-\frac{\modu{z}^2}{2}+\alpha^\ast z\right]\label{innerproduct}
\end{equation}
We make use of Eq.~(\ref{innerproduct}) to reduce the Eq.~(\ref{wignerFun}) to the following form: 
\bea
\begin{split}
W_{l,h}^{(k)}(\beta)=\frac{2\,e^{2\modu{\beta}^2-\modu{\alpha}^2}}{\pi^2} \sum_{s,s^\prime=0}^{k-1}\sum_{r,r^\prime=0}^{l-1} f_{s,k}\,f_{s^\prime,k}^\ast\,e^{-i2\pi (r-r^\prime) h/l}\\
\times\int d^2 z \,e^{-\modu{z}^2}\, e^{z^\ast\left(2 \beta-\alpha_{r,s}\right)-z\left(2\beta^\ast-\alpha_{r^\prime,s^\prime}\right)}. \label{WignerFun2}
\end{split}
\eea
It is straight forward to calculate the integral in above equation as
\begin{equation}
\frac{1}{\pi}\int d^2 z \,e^{-\modu{z}^2}\, e^{z^\ast\left(2 \beta-\alpha_{r,s}\right)-z\left(2\beta^\ast-\alpha_{r^\prime,s^\prime}\right)}= e^{-\left(2 \beta-\alpha_{r,s}\right)\left(2\beta^\ast-\alpha_{r^\prime,s^\prime}\right)}
\end{equation}
Therefore, it follows from Eq.~(\ref{WignerFun2}) that 
\begin{align}
W_{l,h}^{(k)}(\beta)=\frac{2\,e^{2\modu{\beta}^2-\modu{\alpha}^2}}{\pi} \sum_{s,s^\prime=0}^{k-1}\sum_{r,r^\prime=0}^{l-1} f_{s,k}\,f_{s^\prime,k}^\ast\,e^{-i2\pi (r-r^\prime) h/l}\,e^{-\left(2 \beta-\alpha_{r,s}\right)\left(2\beta^\ast-\alpha_{r^\prime,s^\prime}\right)}. \label{APDX1WignerFunfor_psilh_evolution}
\end{align}
\chapter{Calculation of entanglement of the state $\ket{\Phi}_{h}$ in the presence of amplitude damping}\label{Appendix_signatureEntanglement}
\thispagestyle{plain}
The beam splitting action on the even coherent state $\ket{\psi_{2,h}}$ with the vacuum state $\ket{0}$ generates the entangled state
\begin{align}
\ket{\Phi}_{h}=N_{2,h}\left[\ket{\beta}_c\ket{\beta }_d+e^{i\pi h}\ket{-\beta}_c\ket{-\beta }_d\right].\label{APPDX2entstate}
\end{align}
Evolution of this state in the presence amplitude decoherence is given by Eq.~(\ref{ch5rho_cd(t)}):
\begin{align}
\rho_{cd}(\tau)=&N^2_{2,h}\sum_{r,r^\prime=0}^{1} e^{i\pi h(r-r^\prime)}\exp\left[-2\modu{\beta}^2\left(1-e^{i\pi (r-r^\prime)}\right)\left(1-e^{-2\gamma \tau}\right)\right]\nonumber\\
&\times \ket{\beta_r\, e^{-\gamma \tau}}_{c} {\ket{\beta_{r^\prime}\, e^{-\gamma \tau}}_{d}}\,\,{_{c}\bra{\beta_r\, e^{-\gamma \tau}} _{d}\bra{\beta_{r^\prime}\, e^{-\gamma \tau}}},\label{APPDXrho_cd(t)}
\end{align}
where $\gamma$ is the rate of decay and $\tau$ is the time, $\beta_r=\beta\,e^{i\,\pi r}$ and $\beta_{r^\prime}=\beta\,e^{i\,\pi r^\prime}$ . The state $\rho_{cd}(\tau)$ is a mixed state for all the time $\tau >0$. The logarithmic negativity $E_N$ is a computable measure of entanglement for the mixed states \citep{Vidal2002}. In order to calculate the logarithmic negativity of the state $\rho_{cd}(\tau)$, we first have to express it in an orthogonal basis. This can be achieved by expressing the state $\rho_{cd}(\tau)$ in Fock basis. The Fock state representation of the state $\rho_{cd}(\tau)$ is given by
\begin{align}
\rho_{cd}(\tau)=&N^2_{2,h} e^{-2\modu{\beta}^2\,e^{-2\gamma\tau}}\sum_{r,r^\prime=0}^{1} e^{i\pi h(r-r^\prime)}\exp\left[-2\modu{\beta}^2\left(1-e^{i\pi (r-r^\prime)}\right)\left(1-e^{-2\gamma \tau}\right)\right]\nonumber\\
&\times \sum_{m_1=0}^{\infty}\sum_{m_2=0}^{\infty}\sum_{n_1=0}^{\infty}\sum_{n_2=0}^{\infty} \frac{{\beta}_{r}^{m_1}\,{\beta}_{r^\prime}^{m_2}\,{\beta^{\ast}_r}^{n_1}\,{\beta^{\ast}_{r^\prime}}^{n_2}\,e^{-\gamma\tau(m_1+m_2+n_1+n_2)}}{\sqrt{m_1!\,m_2!\,n_1!\,n_2!}}\nonumber\\
&\times\ket{m_1}_{c} {\ket{m_2}_{d}}\,\,{_{c}\bra{n_1} _{d}\bra{n_2}}.\label{APPDXrho_cd(t)}
\end{align}
The elements of this matrix can be written as
\begin{align}
\left[\rho_{cd}(\tau)\right]_{m_1 m_2;n_1 n_2}=&N^2_{2,h} e^{-2\modu{\beta}^2\,e^{-2\gamma\tau}}\sum_{r,r^\prime=0}^{1} e^{i\pi h(r-r^\prime)}\exp\left[-2\modu{\beta}^2\left(1-e^{i\pi (r-r^\prime)}\right)\left(1-e^{-2\gamma \tau}\right)\right]\nonumber\\
&\times \frac{{\beta}_{r}^{m_1}\,{\beta}_{r^\prime}^{m_2}\,{\beta^{\ast}_r}^{n_1}\,{\beta^{\ast}_{r^\prime}}^{n_2}\,e^{-\gamma\tau(m_1+m_2+n_1+n_2)}}{\sqrt{m_1!\,m_2!\,n_1!\,n_2!}}.\label{APPDXrho_cd_elements}
\end{align}
The partial transpose of $\rho_{cd}(\tau)$ with respect to the first mode is denoted by $\left[\rho_{cd}(\tau)\right]^{T_1}$ and is defined through the element-wise operation
\begin{align}
\left[\rho_{cd}(\tau)\right]_{m_1 m_2;n_1 n_2}^{^{T_1}}=\left[\rho_{cd}(\tau)\right]_{n_1 m_2;m_1 n_2}.
\end{align}
The logarithmic negativity of the state $\rho_{cd}(\tau)$ can be calculated as
\begin{equation}
E_N=\log_2 \parallel \rho_{cd}^{T_k}(\tau) \parallel, \label{APPDXlogNeg}
\end{equation}
where $\parallel \cdot \parallel$ denotes the trace norm operation, which  is equal to the sum of the absolute values of eigenvalues  for a Hermitian operator. The Eq.~(\ref{APPDXlogNeg}) can be evaluated numerically in a straightforward way by using standard linear algebra. In the case of infinite-dimensional matrices, convergence in numerical computation is provided by the factorials in the denominator of the summand in the expression derived above for the matrix elements. We use double precision arithmetic with an accuracy of $1$ part in $10^6$. We check the condition ${\rm Tr}\left[\rho_{cd}(\tau)\right]=1$, for the numerical computations. 

\chapter{Solution of the two-mode phase damping master equation}\label{Appendix_signatureEntanglementII}
\thispagestyle{plain}
Consider the zero-temperature phase damping master equation, given in Eq.~(\ref{ch5TwomodePhaseDampingMasterEq}), for the two-mode density matrix $\rho_{cd}$ of field modes at the output of the beam splitter  
\begin{equation}
		\pdv{\tau}\,{\rho_{cd}}=\sum_{s=1}^{2}\kappa_s \left(2 A_s\, \rho_{cd}\, A_s^\dag-A_s^\dag\, A_s \,\rho_{cd} -\rho_{cd} \,A_s^\dag\, A_s  \right),\label{APPDXTwomodePhaseME}
	\end{equation}
where $\kappa_1$ ($\kappa_2$) is the coupling strength of the mode $c$ (mode $d$) with the external environment, $\tau$ is the time, $A_1=c^\dag c$, and $A_2=d^\dag d$. The solution of Eq.~(\ref{APPDXTwomodePhaseME}) can be written in the Fock basis as 
\begin{align}
\rho_{cd}(\tau)=\sum_{m_1=0}^{\infty}\sum_{m_2=0}^{\infty}\sum_{n_1=0}^{\infty}\sum_{n_2=0}^{\infty} {\left[\rho_{cd}(\tau)\right]_{m_1 m_2;n_1 n_2}}\ket{m_1}_{c} {\ket{m_2}_{d}}\,\,{_{c}\bra{n_1}}\,{ _{d}\bra{n_2}}. \label{APPDXtwomoderho}
\end{align}
Substituting Eq.~(\ref{APPDXtwomoderho}) in Eq.~(\ref{APPDXTwomodePhaseME}), we get
\begin{align}
\pdv{\tau}\,{\left[\rho_{cd}(\tau)\right]_{m_1 m_2;n_1 n_2}}=&\sum_{s=1}^{2}\kappa_s\left\{2\,n_s m_s\,\,{\left[\rho_{cd}(\tau)\right]_{m_1 m_2;n_1 n_2}} -m_s^2 \,{\left[\rho_{cd}(\tau)\right]_{m_1 m_2;n_1 n_2}}\right.\nonumber\\
&\left.-n_s^2\,\,{\left[\rho_{cd}(\tau)\right]_{m_1 m_2;n_1 n_2}}\right\}.
\end{align}
This expression can be rewritten as
\begin{align}
\pdv{\tau}\,{\left[\rho_{cd}(\tau)\right]_{m_1 m_2;n_1 n_2}}=-\sum_{s=1}^{2}\kappa_s\left(n_s-m_s \right)^2\,{\left[\rho_{cd}(\tau)\right]_{m_1 m_2;n_1 n_2}}.
\end{align}
The solution of this equation can be easily found as
\begin{align}
{\left[\rho_{cd}(\tau)\right]_{m_1 m_2;n_1 n_2}}=&\exp\left[-\sum_{s=1}^{2}\kappa_s\tau\left(n_s-m_s \right)^2\right]\,{\left[\rho_{cd}(\tau=0)\right]_{m_1 m_2;n_1 n_2}}.
\end{align}
For the initial state $\ket{\Phi}_{h}$, given in Eq.~(\ref{APPDX2entstate}), the above equation becomes
\begin{align}
{\left[\rho_{cd}(\tau)\right]_{m_1 m_2;n_1 n_2}}=&\exp\left[-\sum_{s=1}^{2}\kappa_s\tau\left(n_s-m_s \right)^2\right]\,\bra{m_1,m_2}\ket{\Phi}_{h} {_{h}{\bra{\Phi}\ket{n_1,n_2}}}.
\end{align}
Simplifying the above expression, we get 
\begin{align}
{\left[\rho_{cd}(\tau)\right]_{m_1 m_2;n_1 n_2}}=&N^2_{2,h} e^{-2\modu{\beta}^2}\exp\left[-\sum_{s=1}^{2}\kappa_s\tau\left(n_s-m_s \right)^2\right]\nonumber\\
&\times\sum_{r,r^\prime=0}^{1} \frac{e^{-i\pi h(r-r^\prime)}\,{\beta}_{r}^{m_1}\,{\beta}_{r^\prime}^{m_2}\,{\beta^{\ast}_r}^{n_1}\,{\beta^{\ast}_{r^\prime}}^{n_2}}{\sqrt{m_1!\,m_2!\,n_1!\,n_2!}},
\label{APPDXelements}
\end{align}
where $\beta_r=\beta\,e^{i\,\pi r}$ and $\beta_{r^\prime}=\beta\,e^{i\,\pi r^\prime}$. It is clear from above equation  that the diagonal elements of the two-mode matrix $\rho_{cd}(\tau)$ do not decay due to phase damping. In the long time limit, the density matrix  give in Eq.~(\ref{APPDXtwomoderho}) reduces to
\begin{align}
\rho_{cd}(\tau\rightarrow\infty)=&2\,N^2_{2,h} e^{-2\modu{\beta}^2}\,\left[1+(-1)^h\right]\sum_{m=0}^{\infty}\sum_{n=0}^{\infty} \frac{e^{-2\kappa\tau(n-m)^2\,\modu{{\beta}}^{2m}\,\modu{{\beta}}^{2n}}}{m!\,n!}\nonumber\\
&\times\ket{m}_{c} {\ket{n}_{d}}\,\,{_{c}\bra{m}}\,{ _{d}\bra{n}}, \label{APPDXtwomoderholongtime}
\end{align}
 where we have set the coupling constants $\kappa_1=\kappa_2=\kappa$.

\newpage
\addcontentsline{toc}{chapter}{LIST OF PUBLICATIONS BASED ON THE THESIS}
\listofpapers
\section*{Papers in refereed international journals}
	\begin{enumerate}
 		\item Rohith, M., and Sudheesh, C. (2014). Fractional revivals of superposed coherent states.
\textit{Journal of Physics B: Atomic, Molecular and Optical Physics}, 47(4): 045504. 
 		
 		\item Rohith, M., and Sudheesh, C. (2015). Visualizing revivals and fractional revivals in a Kerr medium using an optical tomogram. \textit{Physical Review A}, 92(5): 053828.
 		
 		\item Rohith, M., and Sudheesh, C. (2016). Signatures of entanglement in optical tomogram. \textit{Journal of the Optical Society of America B}, 33(2): 126-133.
 		
		\item Rohith, M., Sudheesh, C., and Rajeev, R. (2016). Entanglement dynamics of quantum states generated by a Kerr medium and a beam splitter. \textit{Modern Physics Letters B}, 30(2): 1550269.

		 \end{enumerate}
\section*{Papers in conferences} 
		\begin{enumerate}
		\item Rohith, M., and Sudheesh, C. (2012). Multidimensional Entangled Photon-added Coherent States. {\it DAE-BRNS Symposium on Atomic, Molecular and Optical Physics}, IISER-Kolkata, Kolkata, Dec. 14-17, pp. 177.
		 
		\item Rohith, M., and Sudheesh, C. (2012). R\'{e}nyi uncertainty relations in Kerr-like media.  {\it DAE-BRNS Symposium on Atomic, Molecular and Optical Physics}, IISER-Kolkata, Kolkata, Dec. 14-17, pp. 85.
					
		\item Rohith, M., and Sudheesh, C. (2014). Entanglement dynamics of $m$-photon-added coherent states in a beam splitter. {\it Recent Trends in Information Optics \& Quantum Optics}, IIT-Patna, Patna, Nov. 07-08, pp. 64-65. ({\bf Best poster award})
		\item Rohith, M., and Sudheesh, C. (2014). Decoherence of superposed photon-added coherent states. {\it 20th National Conference on Atomic and Molecular Physics}, IIST, Thiruvananthapuram, India, Dec. 09-12, pp. 33.  
		
		\item Rohith, M., and Sudheesh, C. (2015). Quantum optical tomography of entangled states. ICOP2015, {\it International Conference on Optics and Photonics}, University of Calcutta, Kolkata, India, Feb. 20-22, P-273, pp. 90. 
		
		\item Rohith, M., and Sudheesh, C. (2015). Optical tomography of superposed photon-added coherent states. ICOPMA 2015, FOP 27, {\it $1^{\text{st}}$ International Conference on Opto-Electronics and Photonic Materials}, Sastra University, Thanjavur, Tamilnadu, India, Feb. 27-28, FOP 27, pp. 25. ({\bf Best poster award})
		 \end{enumerate}
\section*{Manuscript under preparation} 
		\begin{enumerate}
\item Rohith, M., and Sudheesh, C. Optical tomograms of the entangled states generated by a Kerr medium and a beam splitter.
	\end{enumerate}
\newpage
\begin{center}
{\large \bf CURRICULUM VITAE}
\end{center}
\begin{tabbing}
xxxxxxxxxxxxxxxx \= x\= xxxxxxxxxxxxxxxxxxxxxxxxxxxxxxxxxxxxxxxxxxxxxx \kill
Name \> :\> Rohith M.\\
Date of birth \> :\> 30.05.1988\\
Nationality \> :\> Indian\\
Permanent Address \> :\> Kozhiparambil House\\
		\> \>	V. K. Road, Nilambur (P. O.)\\
		\>	\> Malappuram, Kerala- 679 329\\
Present Address\> :\> Assistant Professor \\
				\> \> Department of Physics\\
				\> \> Government College Kasaragod\\
				\> \> Vidyanagar, Kerala- 671 123\\
Email ID \> :\> rohith.manayil@gmail.com\\
Phone number \> :\> +91 94 96 84 29 40\\
\end{tabbing}

\begin{flushleft}
{\bf Academic Record}
\end{flushleft}

\begin{itemize}
\item BSc. Physics, $4^{\rm th}$ Rank, Marthoma College Chungathara, University  of Calicut, (2008).
\item MSc. Physics, $1^{\rm st}$ Rank, University Campus, University of Calicut, (2010).
\item Qualified for Junior Research Fellowship in CSIR-UGC National Eligibility Test, (December 2010).
\end{itemize}

\begin{flushleft}
{\bf Papers in International Journals}
\end{flushleft}

\begin{itemize}
\item Rohith, M., and Sudheesh, C. (2014). Fractional revivals of superposed coherent states.
\textit{Journal of Physics B: Atomic, Molecular and Optical Physics}, 47(4): 045504. 
 		
 		\item Rohith, M., and Sudheesh, C. (2015). Visualizing revivals and fractional revivals in a Kerr medium using an optical tomogram. \textit{Physical Review A}, 92(5): 053828.
 		
 		\item Rohith, M., and Sudheesh, C. (2016). Signatures of entanglement in optical tomogram. \textit{Journal of the Optical Society of America B}, 33(2): 126-133.
 		
		\item Rohith, M., Sudheesh, C., and Rajeev, R. (2016). Entanglement dynamics of quantum states generated by a Kerr medium and a beam splitter. \textit{Modern Physics Letters B}, 30(2): 1550269.
\end{itemize}

\begin{flushleft}
{\bf Conferences Attended}
\end{flushleft}

\begin{itemize}
\item Rohith, M., and Sudheesh, C. (2012). Multidimensional Entangled Photon-added Coherent States. {\it DAE-BRNS Symposium on Atomic, Molecular and Optical Physics}, IISER-Kolkata, Kolkata, Dec. 14-17, pp. 177.
		 
		\item Rohith, M., and Sudheesh, C. (2012). R\'{e}nyi uncertainty relations in Kerr-like media.  {\it DAE-BRNS Symposium on Atomic, Molecular and Optical Physics}, IISER-Kolkata, Kolkata, Dec. 14-17, pp. 85.
					
		\item Rohith, M., and Sudheesh, C. (2014). Entanglement dynamics of $m$-photon-added coherent states in a beam splitter. {\it Recent Trends in Information Optics \& Quantum Optics}, IIT-Patna, Patna, Nov. 07-08, pp. 64-65. ({\bf Best poster award})
		\item Rohith, M., and Sudheesh, C. (2014). Decoherence of superposed photon-added coherent states. {\it 20th National Conference on Atomic and Molecular Physics}, IIST, Thiruvananthapuram, India, Dec. 09-12, pp. 33.  
		
		\item Rohith, M., and Sudheesh, C. (2015). Quantum optical tomography of entangled states. ICOP2015, {\it International Conference on Optics and Photonics}, University of Calcutta, Kolkata, India, Feb. 20-22, P-273, pp. 90. 
		
		\item Rohith, M., and Sudheesh, C. (2015). Optical tomography of superposed photon-added coherent states. ICOPMA 2015, FOP 27, {\it $1^{\text{st}}$ International Conference on Opto-Electronics and Photonic Materials}, Sastra University, Thanjavur, Tamilnadu, India, Feb. 27-28, FOP 27, pp. 25. ({\bf Best poster award})
\end{itemize}
\end{document}